\documentclass[aps,twocolumn,prb,amsmath,amssymb,superscriptaddress]{revtex4-1}
\usepackage{graphicx}
\usepackage{color}
\usepackage{natbib}
\usepackage{hyperref}
\usepackage{notes2bib}
\usepackage{mathbbol}
\usepackage{float}
\usepackage{ulem}
\usepackage{tikz}
\usepackage{xcolor}
\usepackage{svg}

\newcommand{\bpm}{\begin{pmatrix}}
\newcommand{\epm}{\end{pmatrix}}

\newcommand{\be}{\begin{equation}}
\newcommand{\ee}{\end{equation}}
\newcommand{\beq}{\begin{eqnarray}}
\newcommand{\eeq}{\end{eqnarray}}

\newcommand{\up}{\uparrow}
\newcommand{\down}{\downarrow}

\begin{document}

\title{Superconductivity in monolayer and few-layer graphene: \\ I. Review of possible pairing symmetries and basic electronic properties}

\author{Emile Pangburn}
\affiliation{Institut de Physique Th\'eorique, Universit\'e Paris Saclay, CEA
CNRS, Orme des Merisiers, 91190 Gif-sur-Yvette Cedex, France}
\author{Louis Haurie}
\affiliation{Institut de Physique Th\'eorique, Universit\'e Paris Saclay, CEA
CNRS, Orme des Merisiers, 91190 Gif-sur-Yvette Cedex, France}
\author{Adeline Cr\'epieux}
\affiliation{Aix Marseille Univ, Universit\'e de Toulon, CNRS, CPT, Marseille, France}
\author{Oladunjoye A.~Awoga}
\affiliation{Solid State Physics and NanoLund, Lund University, Box 118, S-221 00 Lund, Sweden}
\author{Annica M.~Black-Schaffer}
\affiliation{Department of Physics and Astronomy, Uppsala University, Box 516, S-751 20 Uppsala, Sweden}
\author{Catherine P\'epin}
\affiliation{Institut de Physique Th\'eorique, Universit\'e Paris Saclay, CEA
CNRS, Orme des Merisiers, 91190 Gif-sur-Yvette Cedex, France}
\author{Cristina Bena}
\affiliation{Institut de Physique Th\'eorique, Universit\'e Paris Saclay, CEA
CNRS, Orme des Merisiers, 91190 Gif-sur-Yvette Cedex, France}

\date{\today}

\begin{abstract}
We review all symmetry-allowed spin-singlet and spin-triplet superconducting (SC) order parameters in graphene ($s$-wave, $d$-wave, $p$-wave, and $f$-wave) generated by generic onsite, nearest-neighbor (NN), and next-to-nearest-neighbor (NNN) pairing interactions in a tight-binding model. For each pairing channel, we calculate both the band structure and the dependence of the density of states on energy, chemical potential, and on the pairing strength. In particular, we distinguish between nodal superconducting states and fully gapped states and study the dependence of gap closing points on the chemical potential and the superconducting pairing strength. We further investigate the difference between mono-, bi-, and tri-layer ABC and ABA graphene, including accounting for the effects of trigonal warping.
\end{abstract}

\maketitle

\section{Introduction}

The recent experimental discoveries of superconductivity in rhombohedral, or ABC-stacked, trilayer graphene\cite{Zhou2021SuperconductivityIR} and in twisted graphene bilayer systems \cite{cao2018unconventional} have received a lot of attention. A vast range of theoretical proposals has already emerged exploring the mechanisms and symmetries of the superconducting (SC) state in these all-carbon systems \cite{cea2021coulomb}, including the SC pairing mechanism being both phonon-mediated \cite{wu2018theory,peltonen2018mean,chou2021Acoustic} and electron-interaction mediated \cite{kennes2018strong} and various SC spin-singlet and spin-triplet order parameters, ranging from $s$-wave and $d$-wave to $p$-wave and $f$-wave symmetries \cite{chou2021correlation,chatterjee2022intervalley,ghazaryan2021unconventional,szabo2021parent,cea2021superconductivity,you2021kohnluttinger,lothman2022nematic,thingstad2020phonon,alidoust2019symmetry,alidoust2020josephson,vuvcivcevic2012d,hosseini2012model,hosseini2012unconventional,hosseini2015inhomogeneous,vahid2012graphene}. However, at present, there exist no experimental definite confirmation of a specific mechanism or pairing symmetry, nor an emerging consensus concerning these issues. 

While currently achieved critical temperatures in graphene-based systems are only of the order of a few Kelvin \cite{cao2018unconventional}, they host some tantalizing  similarities to the high-temperature cuprate superconductors \cite{kerelsky2019maximized}, in particular a similar phase diagram with multiple regions reminiscent of strongly correlated electron physics, such as Mott insulating \cite{saito2020independent} and strange metal behavior \cite{cao2020strange,lyu2021strange}. The importance of strong electron correlations are to be expected due to the normal state hosting low-energy flat energy bands \cite{Wang2013Flat,heikkila2016flat}, which effectively make even very small electron interactions dominate the kinetic energy. Taken together, understanding the underlying physics of carbon-based superconductors may help unveiling the mechanisms at the root of high-temperature SC and thereby also eventually increasing the presently accessible SC critical temperatures, which is crucial for technological developments.

In this work we focus on revealing all possible symmetry-allowed SC order parameters in a tight-binding model, without focusing on their possible origins nor on estimating the exact values that could be reached in realistic systems. We instead use a priori generic values for the pairing amplitudes and review the resulting basic electronic properties, such as the band structure and the density of states. 
We  start for simplicity with monolayer graphene, but the goal of our analysis is to also access multilayer systems, with both ABA- and ABC-stacking, particularly focusing on the latter, since it has already been demonstrated to exhibit superconducting properties \cite{Zhou2021SuperconductivityIR}. We claim that by comparing our results with experimental measurements one can identify the underlying SC order parameter in a graphitic system. 

Overall, our project consists of three parts: in the present work we focus on calculations of the band structure and density of states, with a particular focus on the gap closing points in the energy spectrum, as well as on evaluating the effects of trigonal warping present in bi- and trilayer graphene. In the upcoming two works we will focus first on the topological properties of the SC state in monolayer and multilayer graphene and on the correspoinding edges states, and secondly on the Yu-Shiba-Rusinov \cite{Yu1965,Shiba1968,Rusinov1969} (or in-short Shiba) subgap states appearing in the presence of impurities. We show that the results of these calculations depend strongly on the underlying SC order parameter, and thus a comparison with experimental measurements would allow to determine the SC symmetry as well as type of SC pairing.

By using a tight-binding formalism and modeling the SC pairing in real space as on-site (ON), nearest-neighbor (NN) and next-nearest-neighbor (NNN) couplings, in both the spin-singlet and spin-triplet channels, we capture all relevant symmetry possibilities, as classified by group theory~\citealp{Sigrist1991}. In particular, the possible SC order parameters can be split into spin-singlet pairing with $s$-wave (both a constant gap and extended $s$-wave) and $d$-wave ($d_{xy}$, $d_{x^2-y^2}$, and $d+id\,'$) spatial symmetry and into spin-triplet pairings with $p$-wave ($p_x$, $p_y$, and $p+ip\,'$) and $f$-wave spatial symmetry. 
We calculate the full band structure for each pairing possibility and symmetry, in particular, we focus on the dependence of the energy of the lowest (positive energy) band as a function of momentum and distinguish between nodal superconducting states ($d_{xy}$, $d_{x^2-y^2}$, $p_x$, and $p_y$), which all break the rotational symmetry of the normal state around both the Dirac points and the Brillouin zone center,  and fully gapped states (usually onsite, $p+ip\,'$, $d+id\,'$ and $f$), which in general preserve the spatial symmetries of the normal state. We note that such differences should be feasible to detect using, e.g., angle-resolved photoemission spectroscopy (ARPES), which would help identifying the underlying order parameter.

Moreover, we find that the density of states (DOS) for the nodal states has a linear dependence of energy at low energies (V-shaped), while the fully gapped states produce an U-shaped behavior. However both exhibit a gap-edge coherence peak in the DOS. We analyze how the DOS evolves for each state as a function of both the chemical potential and the SC pairing strength. In particular, we identify the position of the gap closing points in the parameter space. We find that the spin-triplet $p_x$, $p_y$, and $p+ip\,'$ states are the most peculiar by exhibiting most of these gap closing points. Among the spin-singlet states, only the $d_{xy}$ state exhibits gap closing points; for this nodal state we denote the point at which the gap edges merge at zero energy a gap closing point.

Finally, we find that most of the overall physical features are preserved when moving from monolayer to multilayer graphene. The notable difference is a doubling (for bilayer) or tripling (for trilayer) of the number of nodal points, all appearing in close proximity to each other in the Brillouin zone. This also leads to an increase in the number of gap closing points as a function of the chemical potential and the SC pairing strength. We note that, interestingly enough, trigonal warping has a significant effect on the number of gap closing points and can in fact also greatly reduce the number of gap closing points.

The remainder of this work is organized as follows. In Section II we present the tight-binding model for monolayer graphene and all possible SC order parameters up to NNN pairing, as well as the resulting low-energy band structure and DOS. In Section III we present the equivalent model for multilayer graphene and the corresponding modifications to the low energy band structure and the gap closing points in the DOS. We summarize our results in Section IV. Extra information and more calculational details are presented in the Appendices.

\section{Superconducting monolayer graphene}
\subsection{Tight-binding Hamiltonian}
\label{2a}
\begin{figure}[!t]
\centering
\includegraphics[width=8cm]{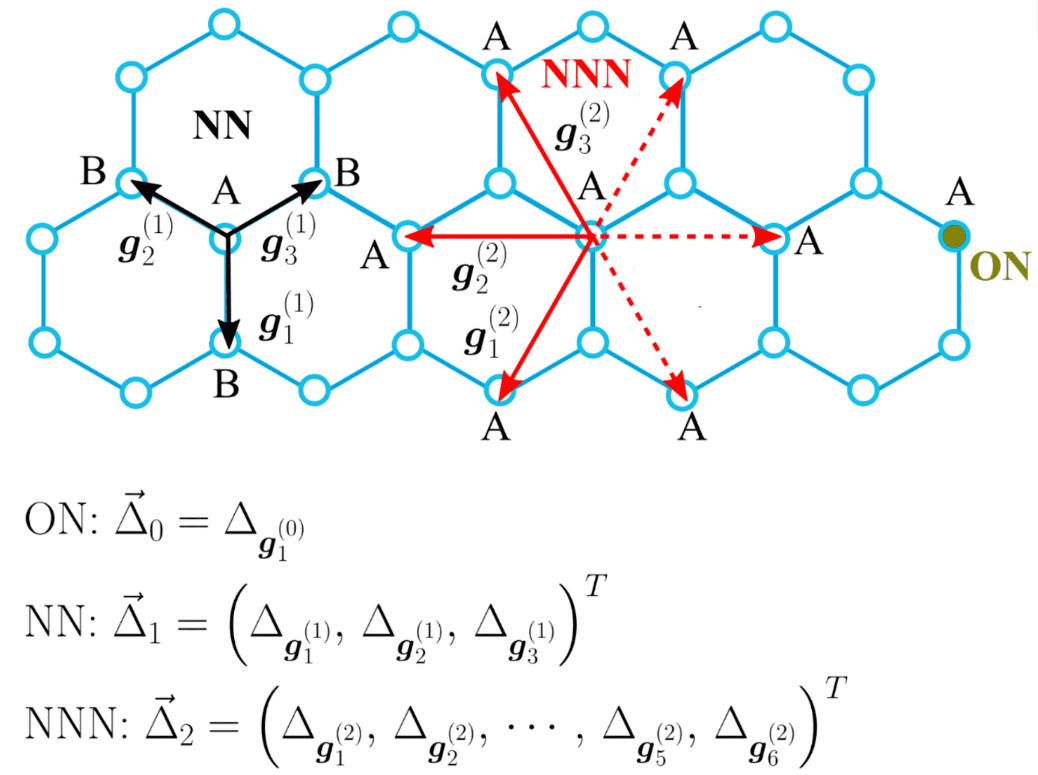}
\caption{Honeycomb lattice for monolayer graphene with ${\bf{g}}^{(1)}_{i}$ and ${\bf{g}}^{(2)}_{i}$ corresponding to the NN and NNN vectors. Real space SC order parameters $\vec{\Delta}_{0,1,2}$ for ON (green), NN (black), and NNN (red) couplings are also indicated in the figure and given in vector form.  Honeycomb lattice and its bond vectors are adapted from Ref.~[\onlinecite{AwogaABC}].}
\label{figure2a}
\end{figure}

Graphene has a honeycomb hexagonal lattice with two atoms per unit cell, here denoted A and B. 
We take the three vectors connecting the sites A to the NN sites B to be ${\bf{g}}^{(1)}_{1}=(0,-1)$, ${\bf{g}}^{(1)}_{2}=(-\dfrac{\sqrt{3}}{2},\dfrac{1}{2})$ and ${\bf{g}}^{(1)}_{3}=(\dfrac{\sqrt{3}}{2},\dfrac{1}{2})$, see Fig.\ref{figure2a}. 
For simplicity, we assume here that the distance between two carbon atoms, $a_0$, is equal to $1$.
%
%
The standard non-interacting Hamiltonian of graphene involves only NN hopping and can be written as
\begin{align}
&H_{0}=-t \sum \limits_{\langle i,j \rangle \sigma} \left[ a^\dagger_{i \sigma} b_{j \sigma} + b^\dagger_{j\sigma} a_{i\sigma}\right]- \mu \sum \limits_{i\sigma} \left[a^{\dagger}_{i\sigma}a_{i\sigma}+b^{\dagger}_{i\sigma}b_{i\sigma}\right],&
\end{align} 
where $t$ is the NN hopping parameter between A and B sites and $\mu$ is the chemical potential. The operators $a^\dagger_{i\sigma} (b^\dagger_{i\sigma})$ create an electron on site $i$ on sublattice A (B) with spin $\sigma$. 
Unless otherwise specified, we will consider for simplicity $t=1$.
Going to momentum space we arrive at the well-established Bloch Hamiltonian 
\begin{eqnarray}
\label{equation1}
H_0(\mathbf{k})&=&
\begin{pmatrix}
\mu&h_0(\mathbf{k})\\
h_0^{\ast}(\mathbf{k})&\mu
\label{h00}
\end{pmatrix}, \\
h_{0}(\mathbf{k})&=&-te^{-ik_{y}}\left(1+2e^{3ik_y/2}\cos\left(\dfrac{\sqrt{3}}{2}k_{x}\right)\right),
\end{eqnarray}
which for $\mu=0$ has a vanishing gap at two non-equivalent Dirac points in the Brillouin zone, $\mathcal{K}=(\dfrac{4\pi}{3\sqrt{3}},0)$ and $\mathcal{K}'=(-\dfrac{4\pi}{3\sqrt{3}},0)$. Near these two Dirac points, the electrons obey a linear dispersion relation leading to the famous Dirac cones that control the transport properties of undoped graphene. 

\subsection{Superconducting term}
\label{2b}

We next consider an interacting term $\text{H}_{I}$, such that the total Hamiltonian is $\text{H}=\text{H}_{0}+\text{H}_{I}$.
Here we consider generic four-body interactions extending as far as NNN, which we can write as\cite{Lothman2017,AwogaABC}
\begin{eqnarray}
H_{I}= \frac{1}{2} \sum\limits_{ij,\alpha \beta \delta \gamma}  \Gamma_{ij}^{\alpha \beta \delta \gamma}f^{\dagger}_{i\alpha}f^{\dagger}_{j\beta} f^{}_{i\delta} f^{}_{j\gamma},
\end{eqnarray}
with
\begin{eqnarray}
 \Gamma_{ij}^{\alpha \beta \delta \gamma}=\dfrac{1}{2}\sum\limits_{\eta}\left[\text{U}_{ij}\sigma_{\alpha \gamma}^{0} \sigma_{\beta \delta}^{0}+\text{J}_{ij} \sigma_{\alpha \gamma}^{\eta} \sigma_{\beta \delta}^{\eta} \right].
\label{eq:Gamma}
\end{eqnarray}
Here $\sigma^0$ and $\sigma^{\eta}$, $\eta \in \{x,y,z\}$ are the Pauli matrices acting in spin space, while $j$ is constrained to be equal to $i$, $i+\vec{\delta}_\text{NN}$ or  $i+\vec{\delta}_\text{NNN}$, respectively, for ON, NN ($\vec{\delta}_\text{NN}={\bf g}^{(1)}_{1,2,3}$), and NNN  interactions ($\vec{\delta}_\text{NNN}={\bf g}^{(2)}_{1,2,3,4,5,6}$).  For notational simplicity, we use here the operators $f_{i \sigma}$ which correspond to either $a_{i,\sigma}$ or  $b_{i,\sigma}$ depending on whether $i$ corresponds to an electron on an $A$ or $B$ sublattice. 
By providing the decomposition of the interaction terms in Eq.~\eqref{eq:Gamma}, we separate already at this level the effective Coulomb interactions $\text{U}_{ij}$ from the the effective spin-spin interactions~$\text{J}_{ij}$

To study superconductivity we need to carry out a mean-field decoupling of $\text{H}_{I}$ into the SC Cooper channels. Following Ref.~[\onlinecite{AwogaABC}], we introduce the SC order parameters, or equivalently the mean-field decoupling fields and  arrive at
\begin{eqnarray}
&&H_{I}^\text{MF}= \sum \limits_{\langle i j \rangle, \eta}\left[ \Delta_{ji}^\eta (g_{ij}^\eta)^\dagger + (\Delta_{ji}^\eta)^* g_{ij}^\eta\right]\nonumber\\
&&+ \sum \limits_{\langle \langle i j \rangle \rangle, \eta}\left[ \Delta_{ji}^\eta (g_{ij}^\eta)^\dagger + \Delta_{ji}^\eta (g_{ij}^\eta)^\dagger
+ (\Delta_{ji}^\eta)^* g_{ij}^\eta +  (\Delta_{ji}^\eta)^* g_{ij}^\eta\right]\nonumber\\
&&+ 2\sum \limits_{i,\sigma} \left[\Delta^\eta_{ii} (g_{ii}^0)^\dagger + (\Delta^\eta_{ii})^* g_{ii}^0\right],
\end{eqnarray}
where the SC order parameters are given by the self-consistent equations
\begin{align}
\Delta_{ji}^{\eta}=\Gamma_{ij}^{\eta,+}\langle g_{ij}^{\eta} \rangle~,
\label{eq:selfcons}
\end{align}
with
\begin{eqnarray}
  \hspace{0.2 cm} g_{ij}^{\eta}=\dfrac{1}{2}\sum\limits_{\alpha \beta}f_{j\alpha}[i\sigma^y\sigma^{\eta}]_{\alpha \beta} f'_{i\beta}.
\end{eqnarray}
In the above equations we have an on-site (ON) pairing determined by $\Delta^\eta_{ii}$, as well as pairings between nearest neighbors (NN) $\langle i j \rangle$, and next nearest neighbors (NNN) $\langle \langle i j \rangle \rangle$, both denoted by $\Delta_{ji}^\eta$.
As no spin-orbit coupling is present in graphene (carbon atoms are light with an atomic number of $Z=6$), we can  separate the treatment of spin-singlet SC, where $\eta =0$ so that we have the interactions $\Gamma_{ij}^{0,+}=U_{ij}-3 J_{ij}$, from spin-triplet SC, where $\eta = x,y,z$ and $\Gamma_{ij}^{\eta,+}=U_{ij}+ J_{ij}$. We use here the notation $\Gamma_{ij}^{\eta,+}$ for interactions acting in the SC channels, but a similar treatment can be done for $\Gamma_{ij}^{\eta,-}$ to capture putative mean-field magnetic or charge-ordered states.
As we will shortly see, this model captures all reasonable pairing symmetries, including phonon-mediated superconductivity as well as states very reminiscent of the cuprate high-temperature superconductors\cite{Lothman2017}. More specifically, an ON coupling necessarily implies spin-singlet superconductivity and $i=j$, such that $\Delta_{ji}^{\eta}=\vec{\Delta}_\text{ON}$ is a 1-component vector. For NN pairing we can have both spin-singlet $\eta =0$ and spin-triplet $\eta = x,y,z$ superconductivity. Here $\Delta_{ji}^{\eta}$ has three non-zero components, one on each NN bond, and can thus be written compactly as a 3-component vector $\vec{\Delta}_1$, see Fig.~\ref{figure2a}. Similarly, for NNN we create a 6-component vector  $\vec{\Delta}_2$, see Fig.~\ref{figure2a}, but where only three components are technically independent. 
Because the spin-orbit coupling is negligible in graphene\cite{Neto2009}, $H^{\text{MF}}$ does not break spin rotation symmetry. Consequently, the three spin-triplet order parameters are degenerate. Therefore, when it comes to the basic electronic properties, such as the band structure and density of states, it is enough to focus on only one triplet channel. In what follows we will generally use the $\eta=x$ channel. 

If we know the interaction parameters $U$ and $J$ we could use Eq.~\eqref{eq:selfcons} to calculate self-consistently all SC order parameters $\vec{\Delta}$. But we very rarely do, and currently we have very little notion of these interaction parameters in SC graphene systems. Instead, we concentrate in this work on the symmetries and overall properties of all relevant SC states. 
As such, we assume a given overall amplitude~$\Delta_0$, for either $\vec{\Delta}_\text{ON}$, $\vec{\Delta}_1$, or $\vec{\Delta}_2$. 
Moreover, we know that at the critical temperature, the SC order parameter is forced to belong to one of the irreducible representations of the crystal symmetry group \cite{Sigrist1991}, in this case the hexagonal lattice group $D_{6h}$. This means we can readily quantify all allowed symmetries of the SC state by using the irreducible representations of the crystal symmetry group.
Starting with ON pairing we further can have only spin-singlet $s$-wave symmetry as the pairing is fully localized in real space, and hence uniform in momentum space. This state belongs to the trivial irreducible representation. For NN and NNN pairing the possibilities include both spin-singlet and spin-triplet states, with the lowest spatial harmonics going up to $f$-wave symmetry (higher harmonics induces  more nodes and will thus not generally be energetically favored). 
In Table~\ref{Table1} we summarize the allowed $\vec{\Delta}$ and their symmetries for ON, NN and NNN pairing for spin-singlet pairing ($\eta=0$) and in Table~\ref{Table2} for spin-triplet pairing ($\eta \in \{x,y,z\})$ \cite{BlackSchafferHonerkamp14,AwogaABC}. We note that the extracted symmetries are those obtained for the intraband pairing in reciprocal space and around the Brillouin zone center, see further Section \ref{sec:intraband}.

\begin{table}
\renewcommand*\arraystretch{2.2}
\begin{tabular}{c|c|c}
\label{TableDelta}
Range & $\Vec{\Delta}$ & Symmetry\\
\hline
ON & 1 & $s_\text{ON}$\\
\hline
NN &$\dfrac{(1,1,1)^T}{\sqrt{3}}$ & $s_\text{ext}$\\
NN &$\dfrac{(2,-1,-1)^T}{\sqrt{6}}$ & $d_{x^2-y^2}$\\
NN & $\dfrac{(0,-1,1)^T}{\sqrt{2}}$& $d_{xy}$\\
\hline
 NNN &$\dfrac{(1,1,1,1,1,1)^T}{\sqrt{6}}$ & $s_\text{ext}$\\
 NNN &$\dfrac{(-1,2,-1,-1,2,-1)^T}{\sqrt{12}}$ & $d_{x^2-y^2}$\\
 NNN & $\dfrac{(-1,0,1,-1,0,1)^T}{2}$& $d_{xy}$\\
\end{tabular}
\caption{Spin-singlet SC symmetries ($\eta=0$) for ON, NN, and NNN pairing in the form of the $\vec{\Delta}$ order parameter (up to an overall amplitude $\Delta_0$) and their spatial symmetries in reciprocal space.}\label{Table1}
\end{table}

\begin{table}
\renewcommand*\arraystretch{2.2}
\begin{tabular}{c|c|c}
\multicolumn{1}{c|}{Range} & \multicolumn{1}{c|}{$\Vec{\Delta}$} & \multicolumn{1}{c}{Symmetry}\\
\hline
NN &$\dfrac{(2,-1,-1)^T}{\sqrt{6}}$ & $p_y$\\
NN & $\dfrac{(0,-1,1)^T}{\sqrt{2}}$& $p_x$\\
\hline
NNN &$\dfrac{(1,-1,1,-1,1,-1)^T}{\sqrt{6}}$ & $f_{x(x^2-3y^2)}$\\
NNN &$\dfrac{(-1,-2,-1,1,2,1)^T}{\sqrt{12}}$ & $p_x$\\
NNN & $\dfrac{(-1,0,1,1,0,-1)^T}{2}$& $p_y$\\
\end{tabular}
\caption{Spin-triplet SC symmetries ($\eta \in \{x,y,z\})$ for ON, NN, and NNN pairing in the form of the $\vec{\Delta}$ order parameter (up to an overall amplitude $\Delta_0$) and their spatial symmetries in reciprocal space.}\label{Table2}
\end{table}

\subsection{Bogoliubov-de-Gennes Hamiltonian}
\label{2c}
In order to proceed we rewrite the total Hamiltonian $H=H_0 + H_I^{\textrm MF}$ in momentum space in the compact Bogoliubov-de-Gennes (BdG) matrix form. Writing the momentum-space Hamiltonian in a matrix form will allow us later to diagonalize $H$ and obtain the band structure. 
For this purpose we introduce the spinor
\begin{equation}
\label{equation3}
\psi_{\mathbf{k}} = \{ a_{\mathbf{k} \uparrow},b_{\mathbf{k} \uparrow},a_{\mathbf{k} \downarrow},b_{\mathbf{k} \downarrow}, a_{-\mathbf{k} \uparrow}^{\dagger},b_{-\mathbf{k}\uparrow}^{\dagger},a_{-\mathbf{k} \downarrow}^{\dagger},b_{-\mathbf{k} \downarrow}^{\dagger}\}^{T},
\end{equation}
with $a_{\mathbf{k}\sigma}$ and $b_{\mathbf{k}\sigma}$ the usual annihilation operators, but now $\mathbf{k}$ belonging to the first Brillouin zone.
With this notation we express the total Hamiltonian
\begin{align}
H=\sum\limits_{\mathbf{k}} \psi_{\mathbf{k}}^{\dagger}H_{\text{BdG}}(\mathbf{k})\psi_{\mathbf{k}}.
\end{align}
Note that the dimension of the matrix associated to the BdG Hamiltonian is doubled compared to the standard BdG Hamiltonian, since we take into account separately the up and down spins, as well as the electrons and holes. 

The normal-state terms within the BdG form can be written as (see Appendix \ref{appendix1} for details)
\begin{eqnarray}
&&H_{0}= \sum \limits_{{\mathbf{k}},\sigma} h_0({\mathbf{k}})( a^\dagger_{{\mathbf{k}} \sigma} b_{{\mathbf{k}} \sigma} - b_{-{\mathbf{k}},\sigma}a^\dagger_{{\mathbf{k}},\sigma})\nonumber\\
&&+ \sum \limits_{{\mathbf{k}},\sigma} h^*_0({\mathbf{k}})( b^\dagger_{{\mathbf{k}} \sigma} a_{{\mathbf{k}} \sigma} - a_{-{\mathbf{k}},\sigma}b^\dagger_{-{\mathbf{k}},\sigma})\nonumber\\
&& -\mu \sum \limits_{{\mathbf{k}} \sigma} ( a^\dagger_{{\mathbf{k}},\sigma} a_{{\mathbf{k}},\sigma} - a_{-{\mathbf{k}},\sigma} a^\dagger_{-{\mathbf{k}},\sigma}
+ b^\dagger_{{\mathbf{k}},\sigma} b_{{\mathbf{k}},\sigma} - b_{-{\mathbf{k}},\sigma} b^\dagger_{-{\mathbf{k}},\sigma}).\nonumber\\
\end{eqnarray}
We next  write all the SC terms within the BdG form. The details of the calculations are given in Appendix \ref{appendix1}. Writing out everything explicitly for NN pairing we have in real space for spin-singlet pairing ($\eta =0$):
\begin{align}
&H^0_\text{NN}=\sum\limits_{\langle ij \rangle}\Delta_{ij}^{\eta=0}(a^\dagger_{i\uparrow}b^\dagger_{j\downarrow}-a^\dagger_{i\downarrow}b^\dagger_{j\uparrow})+\text{h.c},&
\end{align}
and for spin-triplet pairing ($\eta =x,y,z$) we have
\begin{align}
H^x_\text{NN}=&\sum \limits_{\langle i,j \rangle}  \Delta_{ij}^{\eta=x} (a^\dagger_{i\up} b^\dagger_{j\up} -  a^\dagger_{i\down} b^\dagger_{j\down}) + h.c. \\
H^y_\text{NN}=&\sum \limits_{\langle i,j \rangle}  \Delta_{ij}^{\eta=x} (a^\dagger_{i\up} b^\dagger_{j\up} + a^\dagger_{i\down} b^\dagger_{j\down} )+ h.c. \\
H^z_\text{NN}=& \sum \limits_{\langle i,j \rangle} \Delta^{\eta =z}_{ij} (a^\dagger_{i\up} b^\dagger_{j\down}  + a^\dagger_{i\down}b^\dagger_{j\up}) + h.c.
\end{align}
By Fourier transforming these into reciprocal space we obtain
\begin{equation}
H^0_\text{NN}= \sum \limits_{{\mathbf{k}}}h_\text{NN}^{0}({\mathbf{k}}) ( a^\dagger_{{\mathbf{k}}\up} b^\dagger_{-{\mathbf{k}}\down}  - a^\dagger_{{\mathbf{k}}\down}b^\dagger_{-{\mathbf{k}}\up}) + h.c.
\end{equation}
and 
\begin{align}
H^x_\text{NN}=&\sum \limits_{{\mathbf{k}}} h_\text{NN}^x ({\mathbf{k}})(a^\dagger_{{\mathbf{k}}\up} b^\dagger_{-{\mathbf{k}}\up} -  a^\dagger_{{\mathbf{k}}\down} b^\dagger_{-{\mathbf{k}}\down} )+ h.c. \\
H^y_\text{NN}=& i\sum \limits_{{\mathbf{k}}}h_\text{NN}^y({\mathbf{k}})( a^\dagger_{{\mathbf{k}}\up} b^\dagger_{-{\mathbf{k}}\up}+ a^\dagger_{{\mathbf{k}},\down} b^\dagger_{-{\mathbf{k}}\down}) + h.c.  \\
H^z_\text{NN}=& \sum \limits_{{\mathbf{k}}}h_\text{NN}^z({\mathbf{k}})(a^\dagger_{{\mathbf{k}}\up}b^\dagger_{-{\mathbf{k}}\down}+ a^\dagger_{{\mathbf{k}}\down}b^\dagger_{-{\mathbf{k}}\up})+ h.c. 
\end{align}
Here $h_\text{NN}^{\eta}({\mathbf{k}})$ are the overall form factors whose expressions depend on both the spin channel and the spatial symmetry of the order parameter. Their general expression is

\begin{eqnarray}
h_\text{NN}^{\eta}({\mathbf{k}})&=&\Delta_\text{NN}^{\eta,d=1}e^{-ik_y}+\Delta_\text{NN}^{\eta,d=2} e^{\frac{i}{2}k_y -\frac{\sqrt{3}i}{2}k_x}\nonumber\\
&&+ \Delta_\text{NN}^{\eta,d=3} e^{\frac{i}{2}k_y + \frac{\sqrt{3}i}{2}k_x},
\end{eqnarray}
where $d=1,2,3$ correspond to the three NN bonds, following the convention of Fig.~\ref{figure2a} and Tables \ref{Table1} and \ref{Table2}. We summarize $h_\text{NN}^{\eta}({\mathbf{k}})$ for each symmetry in Table \ref{Table3} for spin-singlet pairing and in Table \ref{Table4} for spin-triplet pairing. In these Tables we also present the corresponding results for $h_\text{NNN}^{\eta}({\mathbf{k}})$ (see Appendix \ref{appendix1} for details).
\begin{table}
\renewcommand*\arraystretch{1.8}
\begin{tabular}{|l|l|c|}
\hline
   Sym.& Range & Form factor  \\
  \hline
  $s_\text{ext}$& NN & $ h_\text{NN}^{0,s_\text{ext}}({\mathbf{k}})= \frac{\Delta_0}{\sqrt{3}} \tilde{h}_0({\mathbf{k}})$  \\
  $d_{x^2-y^2}$& NN & $ h_\text{NN}^{0,d_{x^2-y^2}}({\mathbf{k}})= \frac{2\Delta_0}{\sqrt{6}}e^{-ik_y}\left[1-e^{\frac{3i}{2}k_y}\cos(\frac{\sqrt{3}}{2}k_x)\right]$ \\
  $d_{xy}$& NN & $ h_\text{NN}^{0,d_{xy}}({\mathbf{k}})=  \Delta_{0}\sqrt{2}  i \ e^{\frac{i}{2}k_y}\sin(\frac{\sqrt{3}}{2}k_x)$ \\ 
  \hline
  $s_\text{ext}$& NNN & $ h_\text{NNN}^{0,s_\text{ext}}({\mathbf{k}})=\frac{ 2\Delta_0}{\sqrt{6}}\Big[\cos(\sqrt{3}k_x)  $ \\
  & & $+ 2 \cos(\frac{\sqrt{3}}{2}k_x)\cos(\frac{3}{2}k_y)\Big]$\\
  $d_{x^2-y^2}$& NNN &  $ h_\text{NNN}^{0,d_{x^2-y^2}}({\mathbf{k}})=\frac{4\Delta_0}{\sqrt{12}}\Big[\cos(\sqrt{3}k_x)$ \\
    & & $-\cos(\frac{\sqrt{3}}{2}k_x) \cos(\frac{3}{2}k_y)\Big]$ \\
  $d_{xy}$& NNN & $ h_\text{NNN}^{0,d_{xy}}({\mathbf{k}})= - 2 \Delta_{0} \sin(\frac{3}{2}k_y) \sin(\frac{\sqrt{3}}{2}k_x)$ \\
  \hline
\end{tabular} 
\caption{Form factors $h_\text{NN}^{0}({\mathbf{k}})$ and $h_\text{NNN}^{0}({\mathbf{k}})$ for each spin-singlet SC symmetry ($\eta =0$) for NN and NNN pairing. Here $\tilde{h}_0(\mathbf{k})=\frac{h_{0}(\mathbf{k})}{t}$.
}\label{Table3}
\end{table}

\begin{table}
\renewcommand*\arraystretch{1.8}
\begin{tabular}{|l|l|c|}
  \hline
   Sym.& Range & Form factor  \\
  \hline
  $p_y$& NN & $h_\text{NN}^{\eta,p_y}({\mathbf{k}})= \frac{2 \Delta_0}{\sqrt{6}}e^{-ik_y}\left[1-e^{\frac{3i}{2}k_y}\cos(\frac{\sqrt{3}}{2}k_x)\right]$  \\
  $p_x$& NN & $h_\text{NN}^{\eta,p_x}({\mathbf{k}})= i\sqrt{2}\Delta_0 e^{\frac{i}{2}k_y}\sin(\frac{\sqrt{3}}{2}k_x)$  \\
 \hline
  $f_x$& NNN &  $ h_\text{NNN}^{\eta,f_x}({\mathbf{k}})=\frac{2i\Delta_0}{\sqrt{6}}\Big[\sin(\sqrt{3}k_x)$ \\
  &&$-2\sin(\frac{\sqrt{3}}{2}k_x) \cos(\frac{3}{2}k_y)\Big]$\\
    $p_x$&NNN &  $ h_\text{NNN}^{\eta,p_x}({\mathbf{k}})=\frac{4 i \Delta_0}{\sqrt{12}}\Big[\cos(\sqrt{3}k_x)$ \\
    &&$+\cos(\frac{3}{2}k_y)\Big] \sin(\frac{\sqrt{3}}{2}k_x]$\\
  $p_y$& NNN& $h_\text{NNN}^{\eta,p_y}({\mathbf{k}})= -2 i \Delta_{0} \sin(\frac{3}{2}k_y) \cos(\frac{\sqrt{3}}{2}k_x)$ \\
  \hline
\end{tabular} 
\caption{Form factors $h_\text{NN}^{\eta}({\mathbf{k}})$ and $h_\text{NNN}^{\eta}({\mathbf{k}})$ for each spin-triplet SC symmetry ($\eta =x,y,z$) for NN and NNN pairing.}
  \label{Table4}
\end{table}

For completeness we write out the full BdG Hamiltonian in each spin channel in their matrix form.
For the spin-singlet pairing ($\eta=0$) we obtain
\begin{equation}
	\begin{pmatrix}
		H_0({\mathbf{k}}) & \mathbf{0}_{2\times 2} & \mathbf{0}_{2\times 2}  & -h_\Delta^0({\mathbf{k}}) \\
		\mathbf{0}_{2\times 2} & H_0({\mathbf{k}}) & h_\Delta^0({\mathbf{k}}) & \mathbf{0}_{2\times 2} \\
		\mathbf{0}_{2\times 2} &  \left( h_\Delta^0({\mathbf{k}})\right)^\dagger & - H_0({\mathbf{k}}) & \mathbf{0}_{2\times 2}  \\
		-\left(h_\Delta^0({\mathbf{k}})\right)^\dagger & \mathbf{0}_{2\times 2} &  \mathbf{0}_{2\times 2} & - H_0({\mathbf{k}})
	\end{pmatrix}
	\label{eq:matrixsinglet}
\end{equation}
and for $\eta=x$ spin-triplet pairing we have
\begin{equation}
	\begin{pmatrix}
		H_0({\mathbf{k}}) & \mathbf{0}_{2\times 2}  & -h_\Delta^x({\mathbf{k}})  & \mathbf{0}_{2\times 2}\\
		\mathbf{0}_{2\times 2} & H_0({\mathbf{k}}) & \mathbf{0}_{2\times 2}  & h_\Delta^x({\mathbf{k}})  \\
		-\left( h_\Delta^x\right)^\dagger &	\mathbf{0}_{2\times 2}  & - H({\mathbf{k}}) & \mathbf{0}_{2\times 2}  \\
		\mathbf{0}_{2\times 2} &	\left(h_\Delta^x\right)^\dagger   &  \mathbf{0}_{2\times 2} & - H({\mathbf{k}})
	\end{pmatrix}
	\label{eq:matrixtriplet}
\end{equation}
where $	\mathbf{0}_{2\times 2}$ is a $2\times 2$ null matrix, $H_0({\mathbf{k}})$ is the normal state Hamiltonian matrix given in Eq.~\ref{equation1}, and

\begin{align}
	h_\Delta^0{\mathbf{k}}) & =\frac{1}{2} \begin{pmatrix}
		\Delta_{\rm ON}+{h}_\text{NNN}^{0}(-{\mathbf{k}})& {h}_\text{NN}^{0}({\mathbf{k}}) \\	{h}_\text{NN}^{0}(-{\mathbf{k}})& \Delta_{\rm ON}+{h}_\text{NNN}^{0}(-{\mathbf{k}})
	\end{pmatrix},\nonumber\\	
	h_\Delta^x({\mathbf{k}}) & =\frac{1}{2} \begin{pmatrix}
		{h}_\text{NNN}^{x}(-{\mathbf{k}})& -{h}_\text{NN}^{x}({\mathbf{k}}) \\	{h}_\text{NN}^{x}(-{\mathbf{k}})& {h}_\text{NNN}^{x}(-{\mathbf{k}})\nonumber
	\end{pmatrix}
\end{align}
are the corresponding superconducting order parameter matrices.
The superconducting order parameter  matrices for the $\eta=y,z$ spin-triplet pairing are given in Appendix \ref{Appendix2}. Diagonalizing the matrices in Eqs.~\eqref{eq:matrixsinglet}-\eqref{eq:matrixtriplet} yields both the full energy spectrum and the eigenstates of the various SC states.

\subsection{Intraband pairing symmetries}
\label{sec:intraband}
While the matrix forms in Eqs.~\eqref{eq:matrixsinglet}-\eqref{eq:matrixtriplet} allow for straightforward numerical diagonalization to easily find e.g.~the energy spectrum, it is still useful to first analyze the order parameter in some more detail. For this purpose it is beneficial to not work in the sublattice basis with the operators $a_{\mathbf{k}\sigma}, b_{\mathbf{k}\sigma}$, but to instead in the band basis with the operators $c_{\mathbf{k}\sigma}, d_{\mathbf{k}\sigma}$ for the two bands, where the normal-state Hamiltonian $H_0$ is diagonal, with the band energies $\epsilon_c (\mathbf{k}), \epsilon_d(\mathbf{k})$ as diagonal entries. In particular, in the band basis we know that $\epsilon_{c}(\mathbf{k})=0$ (or $\epsilon_{d}(\mathbf{k})=0$ depending on the value of $\mu$) on the Fermi surface. In graphene the Fermi surface forms circles around the Dirac points at $\mathcal{K}$ and $\mathcal{K}'$ with increasing radius as the chemical potential increases from 0. At $|\mu|=1$ these two Dirac Fermi surfaces meet and instead form a separatrix line joining the $M$ points in the Brillouin zone. Finally, for $|\mu| >1$, the Fermi surface becomes centered around the Brillouin zone center $\Gamma$.

Setting $\alpha_{\mathbf{k}} = ( a_{\mathbf{k} \sigma},b_{\mathbf{k} \sigma})^{T}$ and $\chi_{\mathbf{k}} = (c_{\mathbf{k} \sigma},d_{\mathbf{k} \sigma})^{T}$, we find that the basis change can be expressed as
\begin{equation}
	\alpha_{\mathbf{k}}=\hat{U}\left(\mathbf{k}\right) \chi_{\mathbf{k}}
\end{equation}
with the unitary matrix
\begin{equation}\label{eq:uslg}
	\hat{U}\left(\mathbf{k}\right)=\frac{1}{\sqrt{2}}\left(\begin{array}{cc}
		- e^{-i \varphi_{\boldsymbol{k}}} & e^{-i \varphi_{\boldsymbol{k}}}  \\
		1 & 1
	\end{array}\right), 
\end{equation}
where $\varphi_{\boldsymbol{k}}=\arg\left(h_0(\mathbf{k})\right)$. 
Using $\hat{U}\left(\mathbf{k}\right)$ to transform also the superconducting terms from the sublattice basis of the previous subsections into the band basis, we arrive for spin-singlet ON pairing at
\begin{equation}\label{eq:swaveBand}
	H^0_\text{ON}=\Delta_0\sum_{\mathbf{k}}\left(c_{\mathbf{k} \uparrow}^{\dagger} c_{-\mathbf{k} \downarrow}^{\dagger} + d_{\mathbf{k} \downarrow}^{\dagger} d_{-\mathbf{k} \uparrow}^{\dagger}\right)+h.c.,
\end{equation}
while for spin-singlet NN pairing we get 
\begin{equation}
	\begin{split}
		H_\text{NN}^{0} & =\Delta_0\sum_{\mathbf{k},d}\Delta_d\left[ \cos\left(\mathbf{k}\cdot \mathbf{g}_d^{(1)}-\varphi_{\boldsymbol{k}}\right) \left(d_{\mathbf{k} \downarrow}^{\dagger} d_{-\mathbf{k} \uparrow}^{\dagger} -c_{\mathbf{k} \uparrow}^{\dagger} c_{-\mathbf{k} \downarrow}^{\dagger}\right) \right. \\
		& + \left. i\sin\left(\mathbf{k}\cdot\mathbf{g}_d^{(1)}-\varphi_{\boldsymbol{k}}\right) \left(c_{\mathbf{k} \uparrow}^{\dagger} d_{-\mathbf{k} \downarrow}^{\dagger} - d_{\mathbf{k} \uparrow}^{\dagger} c_{-\mathbf{k} \downarrow}^{\dagger}\right)\right] + h.c.,
	\end{split}
	\label{eq:intrasinglet}
\end{equation}
and for spin-triplet NN pairing with $\eta =z$
\begin{equation}
	\begin{split}
		H_\text{NN}^z & =\Delta_0\sum_{\mathbf{k},d}\Delta_d\left[ i\sin\left(\mathbf{k}\cdot\mathbf{g}_d^{(1)}-\varphi_{\boldsymbol{k}}\right) \left(d_{\mathbf{k} \downarrow}^{\dagger} d_{-\mathbf{k} \uparrow}^{\dagger} -c_{\mathbf{k} \uparrow}^{\dagger} c_{-\mathbf{k} \downarrow}^{\dagger} \right) \right. \\
		& + \left.  \cos\left(\mathbf{k}\cdot\mathbf{g}_d^{(1)}-\varphi_{\boldsymbol{k}}\right) \left(c_{\mathbf{k} \uparrow}^{\dagger} d_{-\mathbf{k} \downarrow}^{\dagger} - d_{\mathbf{k} \uparrow}^{\dagger} c_{-\mathbf{k} \downarrow}^{\dagger}\right)\right] + h.c..
	\end{split}
	\label{eq:intratriplet}
\end{equation}
Here $d = 1,2,3$ mark the three NN bonds, following the convention of Fig.~\ref{figure2a}, and $\Delta_d$ is the $d$th component of the NN $\vec{\Delta}$ bond order parameter, given in Table \ref{Table1}.
 
  \begin{figure*}[!t]
	\centering
	\includegraphics[width=1\textwidth]{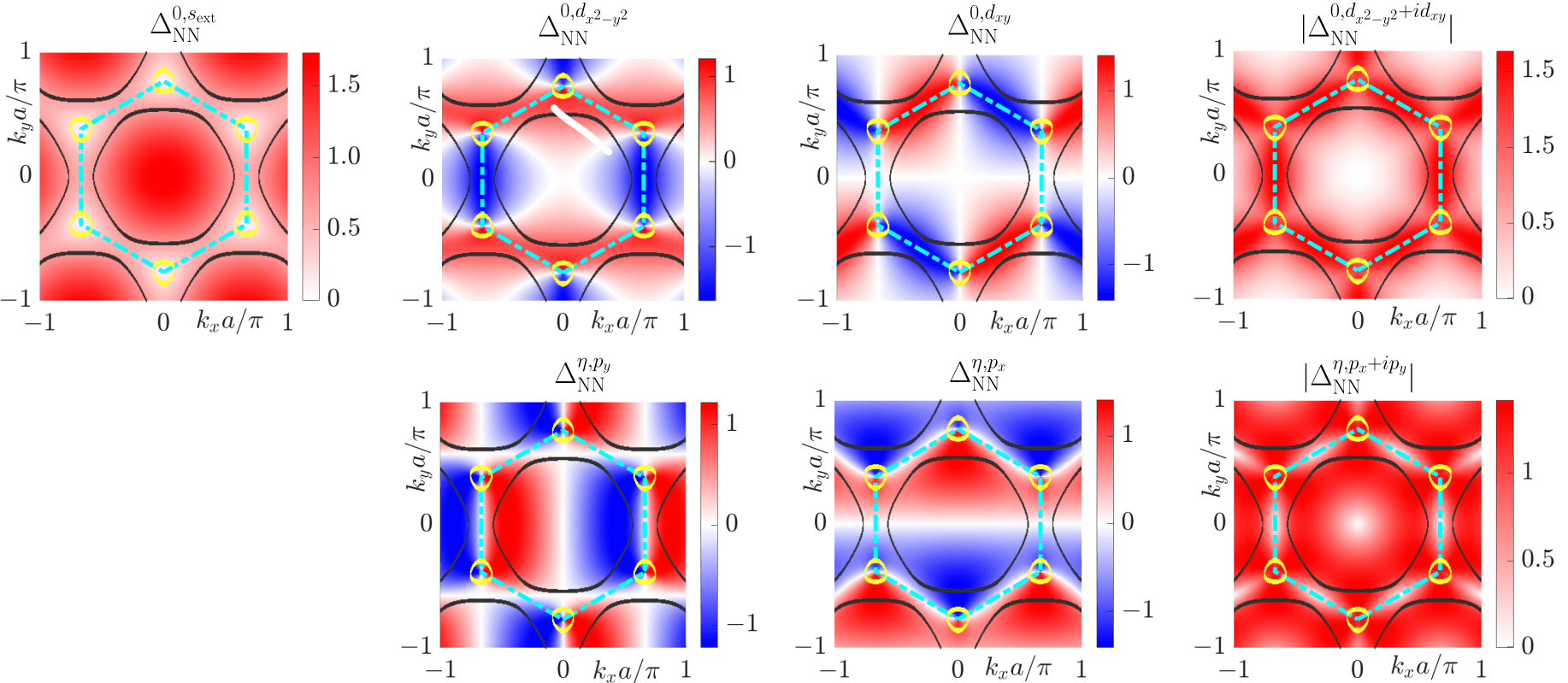}
	\caption{Intraband SC order parameter form factors for NN ($s_{\rm ext}$-, $d$-, $p$-wave) pairing. Top (bottom) row corresponds to spin-singlet (spin-triplet) pairing. Cyan hexagon denotes the first Brillouin zone, while yellow (black) circles centered around the $\mathcal{K},\mathcal{K}'$-points ($\Gamma$-point) are the normal-state Fermi surfaces for $\mu=0.4t, \left(\mu=1.2t\right)$. White regions correspond to nodes, or zeroes, of the SC order parameter.}
	\label{fig:figureOPs}
\end{figure*}

From Eqs.~\eqref{eq:intrasinglet}-\eqref{eq:intratriplet} we see that in the band basis superconductivity consists of both {\it intraband} pairing through the terms $c_{\mathbf{k} \uparrow}^{\dagger} c_{-\mathbf{k} \downarrow}^{\dagger}$ and $d_{\mathbf{k} \uparrow}^{\dagger} d_{-\mathbf{k} \downarrow}^{\dagger}$ for bands $c$ and $d$, respectively, and {\it interband} pairing though terms such as $c_{\mathbf{k} \uparrow}^{\dagger} d_{-\mathbf{k} \downarrow}^{\dagger}$. 
Generally, the existence of both intra- and interband pairing means that the energy spectrum of the BdG Hamiltonian cannot simply be expressed as $E = \pm \sqrt{\epsilon^2 + |\Delta|^2}$, but instead both the intra- and interband pairing terms contribute to the energy $E$. Here we are not concerned with deriving the exact expressions, but instead note that for a qualitative understanding it has in the past often been enough to only consider the intraband pairing and ignore the interband term \cite{BlackSchaffer07, BlackSchafferHonerkamp14, Black-SchafferLeHur14,Awoga2018}. This can be understood by noticing that only one of the bands crosses zero energy (except at the Dirac point), and therefore contributes to the formation of the Fermi surface. Thus, pairing within that band is the most important contribution, while  pairing in the other band and in-between bands are often of less importance. It is also the symmetry of the intraband order parameters in the Brillouin zone that we use for naming the different possible symmetries as $s,d,p,f$-waves in Tables \ref{Table1}-\ref{Table4}.

Based on the above arguments, we analyze the intraband SC order parameters for both NN and NNN pairing.
In Fig.~\ref{fig:figureOPs} we plot the intraband SC order parameter symmetry for NN pairing, or equivalently the form factors  in front of the intraband terms in Eqs.~(\ref{eq:intrasinglet})-(\ref{eq:intratriplet}).
We also mark the Fermi surface for both $\mu = 0.4$ (yellow lines) and $\mu =1.2$ (black lines). 
We first note that the extended $s$-wave follows the symmetry of the normal state, with nodal points only at the Dirac points. Assuming a simplified energy dispersion, $E = \pm \sqrt{\epsilon^2 + |\Delta|^2}$, with $\Delta$ only given by the intraband component, we would then expect a fully gapped energy spectrum for any finite $\mu$.
Moving on to the $d_{x^2-y^2}$- and $d_{xy}$-wave order parameters we note that, for large chemical potential $|\mu|>1$, when the Fermi surface is centered around $\Gamma$ (black lines), the SC state has $d$-wave symmetry, and we thus expect four nodal points in the band structure, one at each of the intersections between the Fermi surface and the zero-gap lines (white in Fig.~\ref{fig:figureOPs}). In contrast, around $\mathcal{K},\mathcal{K}'$, the $d_{x^2-y^2}$-wave order has an effective $p_y$-wave symmetry, while the $d_{xy}$-wave order has an effective $p_x$-wave symmetry. As such, for $|\mu|<1$, when the Fermi surface is centered around $\mathcal{K},\mathcal{K}'$ (yellow lines), one expects two nodal points per Dirac cone. 
Here we note that the two Dirac cones have opposite signs on their effective $p$-wave order, such that the spin-singlet state still has an overall even spatial symmetry, as required by Fermi-Dirac statistics.
We can also combine the two $d$-wave order parameters into a chiral combination $d_{x^2-y^2 + ixy} = d+id\,'$, resulting in a fully gapped order parameter with restored rotational symmetry as seen in Fig.~\ref{fig:figureOPs}. It is only at the $\Gamma,\mathcal{K},\mathcal{K}'$ points that the order parameter is equal to zero. This full gap is the reason why the chiral $d+id\,'$-wave combination is often found to be most stable among all the $d$-wave states \cite{BlackSchaffer07,Honerkamp2008Density, BlackSchaffer12PRL, Nandkishore12, Kiesel12,Lothman14,Awoga2017Domain}. 
Moving on to the NN spin-triplet order parameters, we find that the $p_y$ and $p_x$-wave order parameters have a $p$-wave symmetry both around the $\Gamma$ point and the $\mathcal{K},\mathcal{K}'$ Dirac points. This results in two nodal points per Fermi surface when it is centered around the Dirac points and also two nodal points for a Fermi surface centered around $\Gamma$. 
We also note that the $p$-wave states hosts additional nodal points, in particular at the $M$-points. This is a consequence of adopting a $p$-wave symmetry to the six-fold symmetric honeycomb lattice with its hexagonal first Brillouin zone. 
Finally, the chiral combination $p_x + ip_y = p+ip\,'$~\cite{Ma2014Possible}, is again fully gapped, except at the $\Gamma,\mathcal{K},\mathcal{K}'$, as well as at the $M$-points where both individual $p$-wave components have nodes. Thus, we expect that the $p+ip\,'$ SC to be fully gapped except when the chemical potential $|\mu|=0,1,3$.

 \begin{figure*}[!t]
	\centering
	\includegraphics[width=1\textwidth]{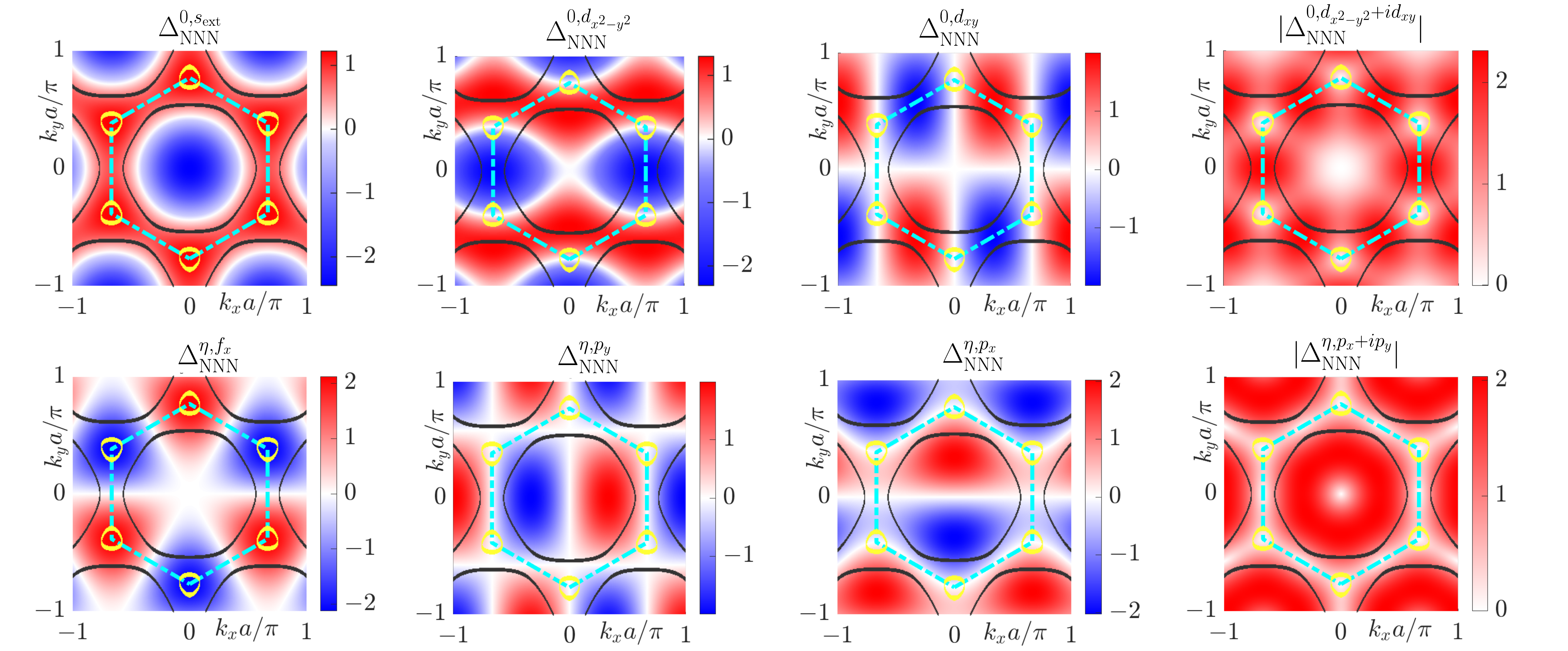}
	\caption{Intraband SC order parameter form factors for NNN ($s_{\rm ext}$-, $d$-, $p$-, $f_x$-wave) pairing. Top (bottom) row corresponds to spin-singlet (spin-triplet) pairing. Cyan hexagon denotes the first Brillouin zone, while yellow (black) circles centered around the $\mathcal{K},\mathcal{K}'$-points ($\Gamma$-point) are the normal-state Fermi surfaces for $\mu=0.4, \left(\mu=1.2\right)$. White region corresponds to the nodes, or zeroes, of the SC order parameter.}
	\label{fig:figureOPsNNN}
\end{figure*}

We also perform a similar analysis for NNN pairing, for which we in fact find that the interband pairing terms are identically zero. Thus, the energy $E$ in the SC state is exactly given by $E = \pm \sqrt{\epsilon^2 + |\Delta|^2}$, with $\epsilon$ the energy of the normal state, and $\Delta$ the intraband order parameter. This can be understood by noticing that the NNN pairing couples electrons on the same sublattice and thus there is no sublattice mixing or any interband pairing terms.
In Fig.~\ref{fig:figureOPsNNN} we plot the resulting NNN intraband order parameter symmetries. Overall we find very similar results to the NN pairing with only a few exceptions: The extended $s$-wave state has a peculiar nodal structure centered around $\Gamma$. However, unless $\mu$ is fine tuned such that the Fermi surface exactly hits this node, the SC state will be fully gapped. Also, for NNN pairing, an $f$-wave symmetry is allowed. This order parameter display the required $f_x = f_{x(x^2-3y^2)}$-wave symmetry with a total of six nodal points for a Fermi surface centered around $\Gamma$ (black line), but is notably fully gapped on Fermi surfaces centered around the $\mathcal{K},\mathcal{K}'$ Dirac points. Technically there also exists another, $f_y = f_{y(y^2-3x^2)}$-wave, state that has nodal lines going through the $\mathcal{K},\mathcal{K}'$ Dirac points, such that the energy for $|\mu|<1$ has six nodes per Dirac cone. However, due to its high number of nodes per Dirac cone, this state will be much less energetically favorable at all doping levels $|\mu| <1$ and we do not considering it in this work.

\subsection{Lowest energy bands}

Having analyzed the symmetry of the intraband order parameter in detail in the previous subsection, we now turn to the complete solution, attained by diagonalizing the BdG Hamiltonian in Eqs.~\eqref{eq:matrixsinglet} and \eqref{eq:matrixtriplet}, for spin-singlet and spin-triplet pairing, respectively. In this subsection we are interested in the low-energy band structure, which we analyze by plotting the lowest energy band ($E>0$) as a function of momentum. These results directly tell us about both the existence of nodal points in the energy spectrum and the overall symmetry of the SC state.
We focus primarily on ON $s$-wave, NN $s$-, $d$-, and $p$-wave, and NNN $f$-wave SC states. 
Due to the similarities between the remaining NNN order parameters and the corresponding NN ones, we expect to cover all relevant behavior with this selection.

\begin{figure*}[tbh]
\hspace{-0.5cm}
\includegraphics[width=5.2cm]{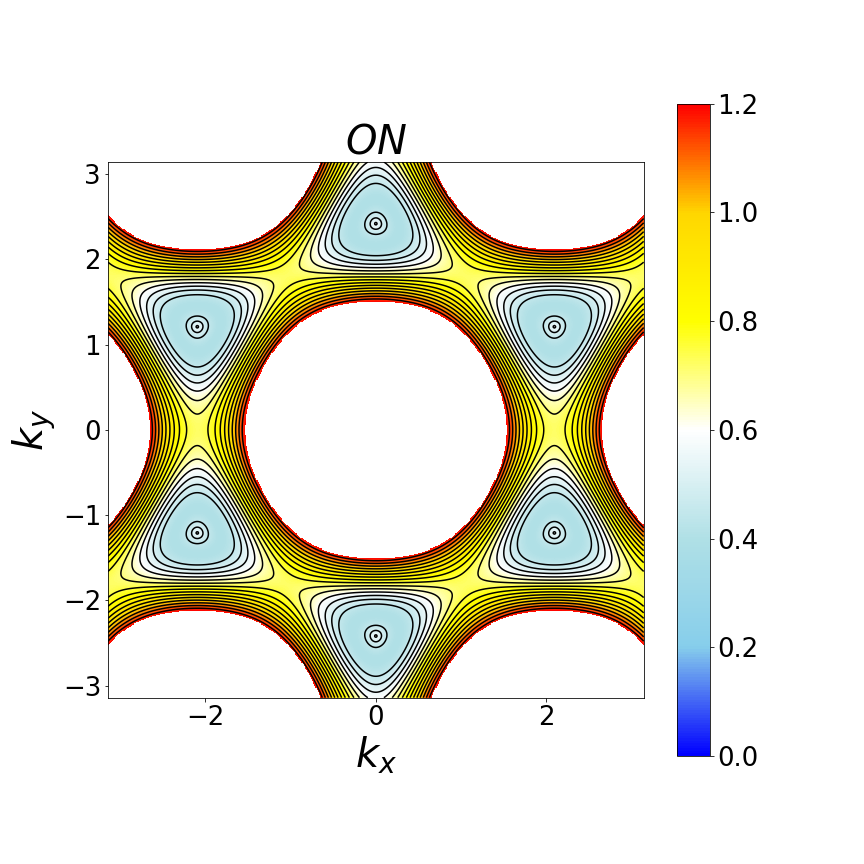}
\hspace{-0.5cm}
\includegraphics[width=5.2cm]{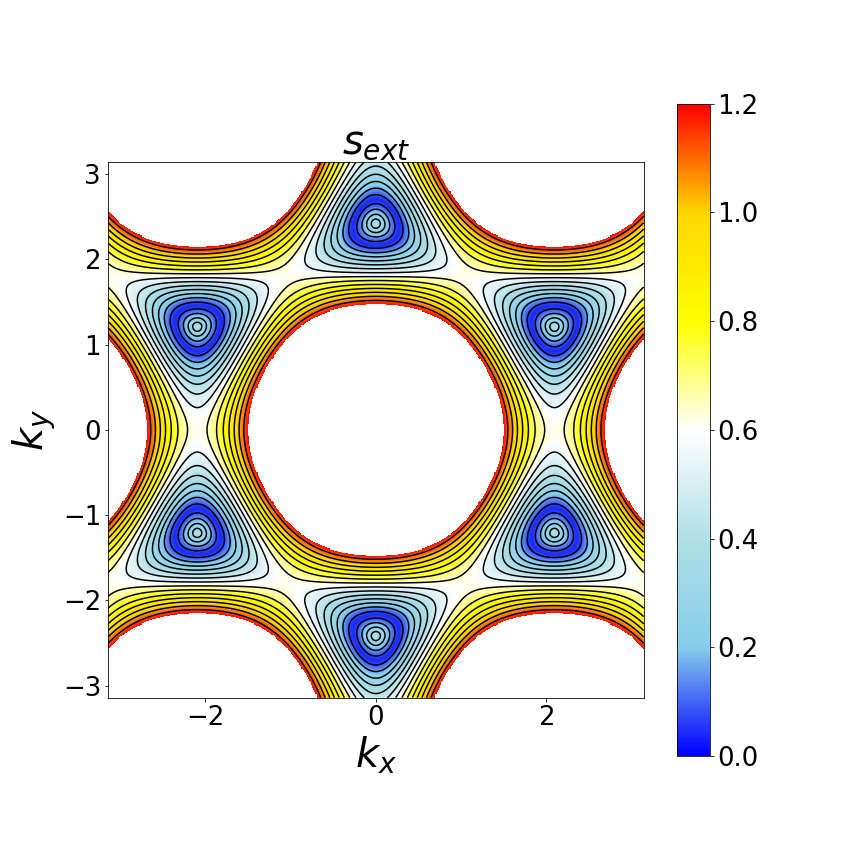}
\hspace{-0.5cm}
\includegraphics[width=5.2cm]{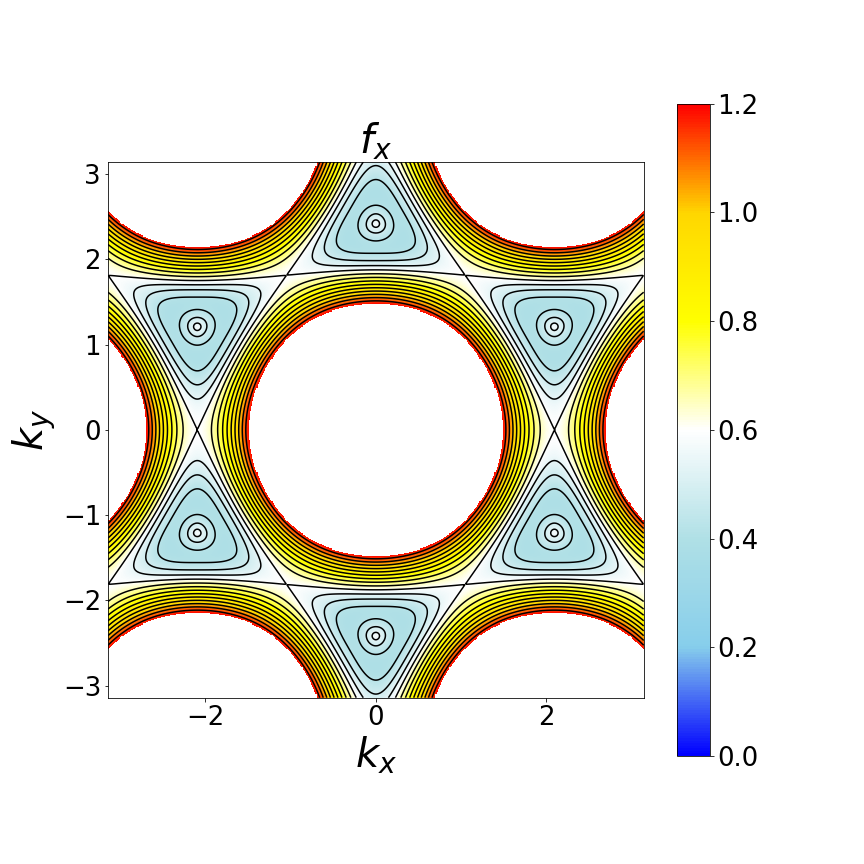}

\vspace{-0.5 cm}

\hspace{-0.5cm}
\includegraphics[width=5.2cm]{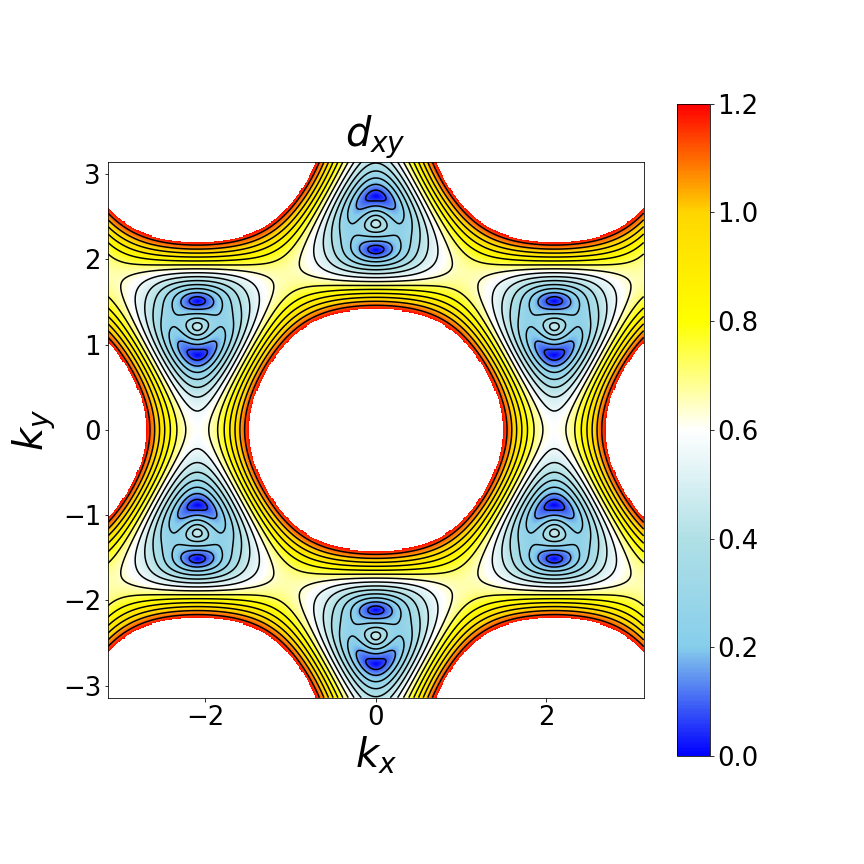}
\hspace{-0.5cm}
\includegraphics[width=5.2cm]{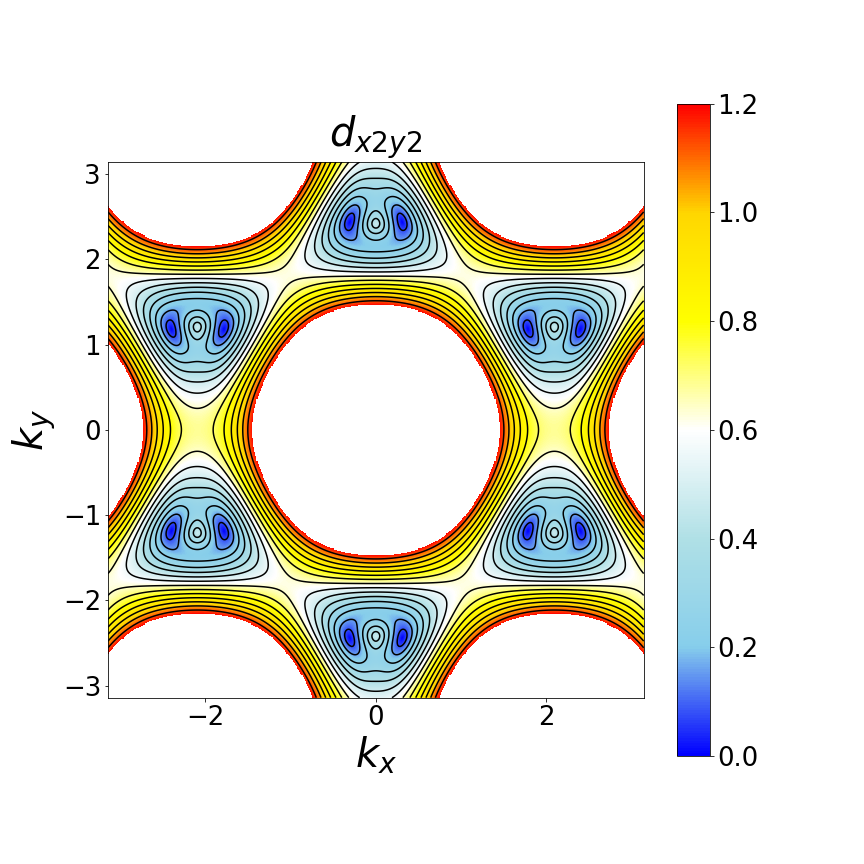}
\hspace{-0.5cm}
\includegraphics[width=5.2cm]{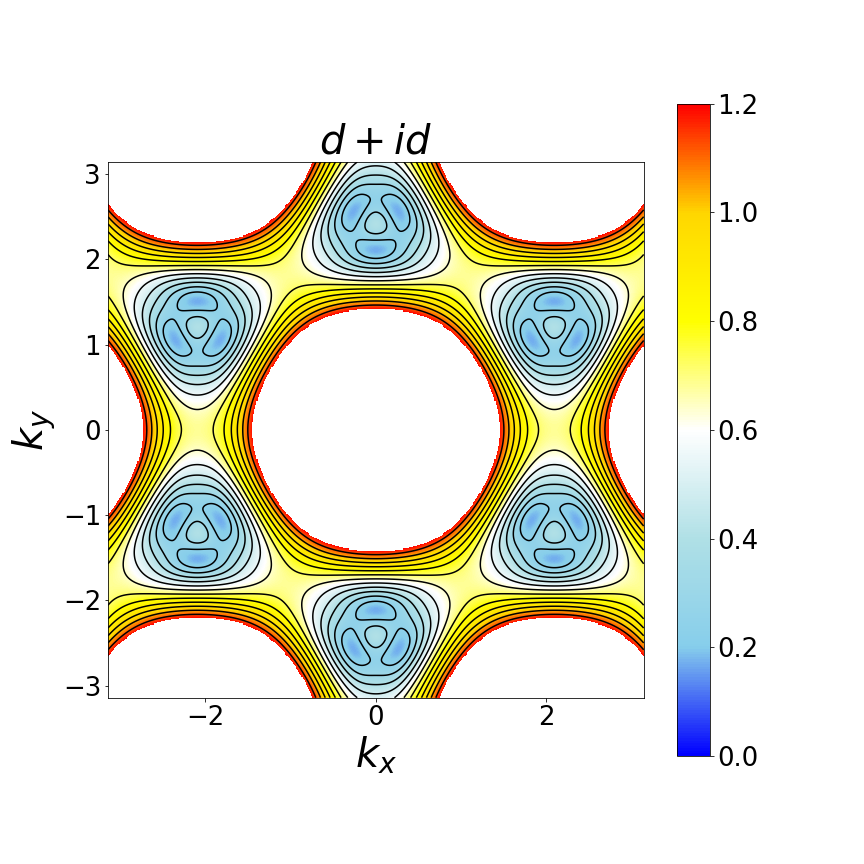}

\vspace{-0.5 cm}

\hspace{-0.5cm}
\includegraphics[width=5.2cm]{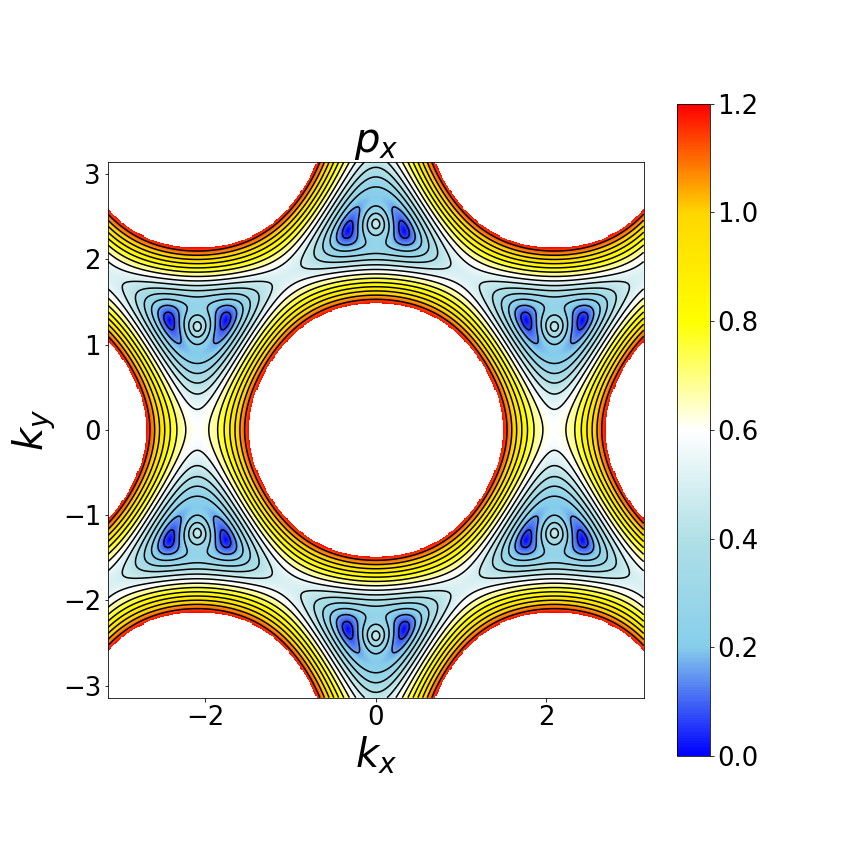}
\hspace{-0.5cm}
\includegraphics[width=5.2cm]{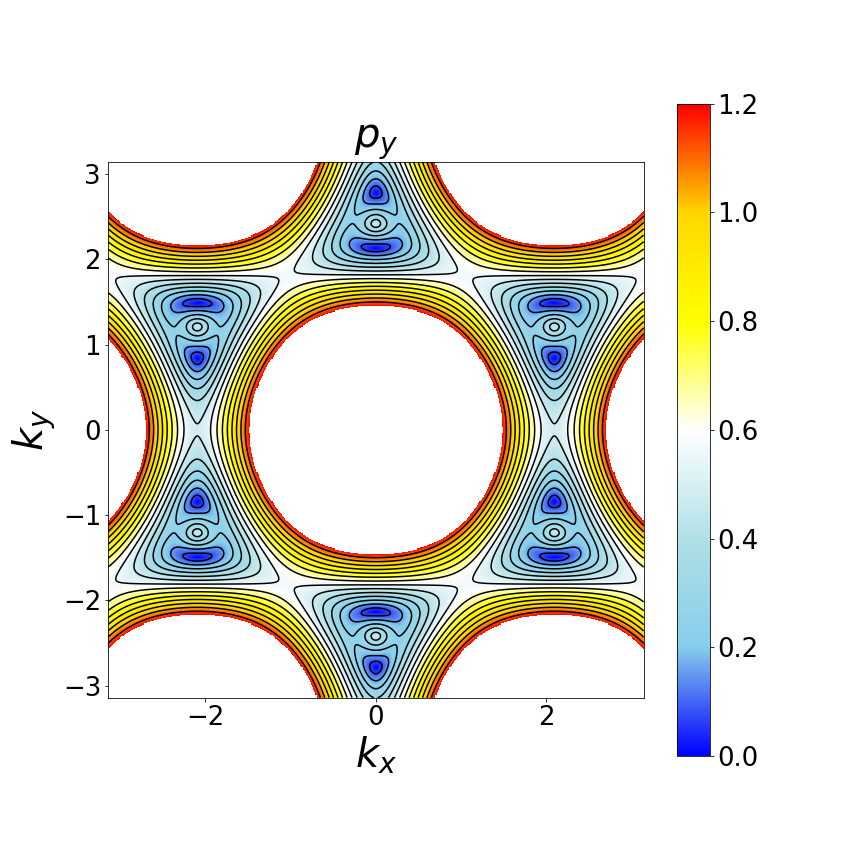}
\hspace{-0.5cm}
\includegraphics[width=5.2cm]{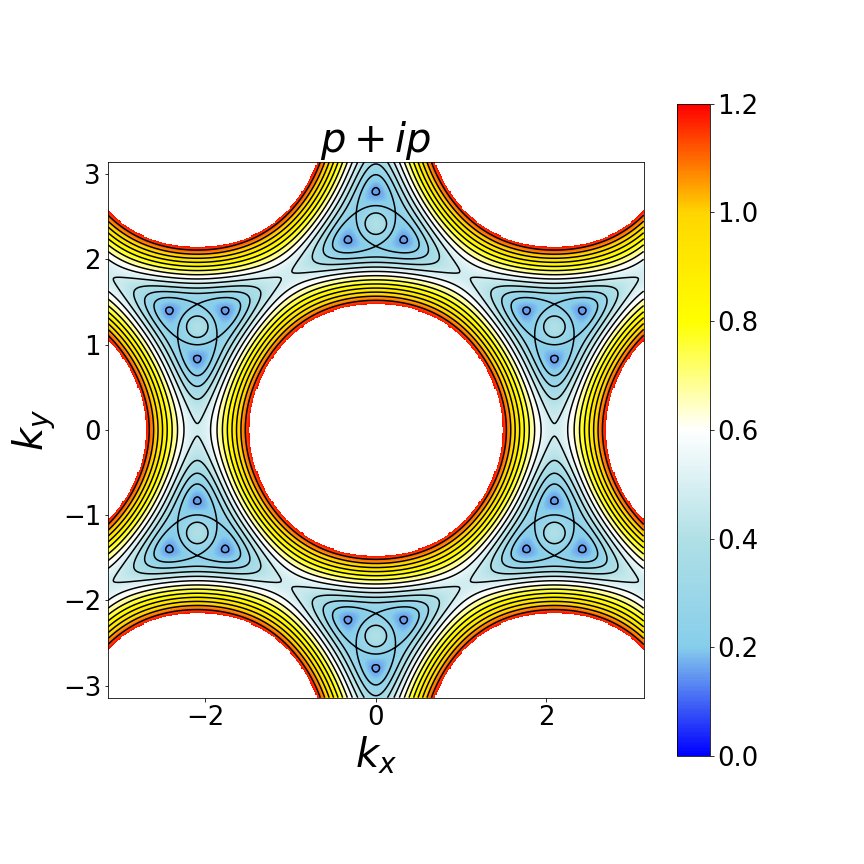}
\caption{Lowest energy band for $\mu=0.4$, $\Delta_0=0.4$ and for SC states with ON $s$-wave, NN $s_\text{ext}$-, $d_{xy}$-, $d_{x^2-y^2}$-, $p_x$-, $p_y$-, $p+ip\,'$-, $d+id\,'$-wave, as well as NNN $f_x$-wave symmetries. Dark blue corresponds to zero energy.}
\label{figurebands}
\end{figure*}

In Fig.~\ref{figurebands} we plot the lowest energy band as a function of 
$k_x$ and $k_y$ for $\mu=0.4$. This value corresponds to a normal-state Fermi surface that is a small circle centered around the Dirac points $\mathcal{K},\mathcal{K}'$, see Fig.~\ref{fig:figureOPs}. 
As expected, and consistent with the intuitive analysis in the previous subsection, we find that the ON $s$-wave and NN $s_\text{ext}$-wave SC states preserve the rotational symmetry in reciprocal space around the $\mathcal{K},\mathcal{K}'$ Dirac points. Also, these SC states give rise to a fully gapped band structure.
On the other hand, we note that the $d_{xy}$-, $d_{x^2-y^2}$- , $p_x$-, and $p_y$-wave SC states all break the rotation symmetry around the $\mathcal{K},\mathcal{K}'$ points. In particular, we observe a nodal energy spectrum with two nodal points per Fermi surface arranged in agreement with the nodal structure of the intraband order parameters in Fig.~\ref{fig:figureOPs}. These nodal points can be viewed as the split of the original normal-state Dirac point into two points in the SC state. 
Thus the $d_{xy}$-, $d_{x^2-y^2}$-, $p_x$-, and $p_y$-wave SC states all correspond to nodal superconductors.

By continuing with analyzing the chiral $d+id\,'$- and $p+ip\,'$-wave SC states, we find fully gapped energy spectra and a restored rotational symmetry around the $\mathcal{K},\mathcal{K}'$ points (up to the same three-fold symmetry as for the $s$-wave state). We note that the convention to generate these states is to consider both a $d_{xy}$ and $d_{x^2-y^2}$ amplitude of  $\Delta_0$ for the $d+id\,'$ state, and similarly for the $p+ip\,'$ state. This convention corresponds to an effective coupling $\sqrt{2}$ stronger for the chiral states than for the nodal ones. 
Finally, for the NNN $f_x$-wave state we find a fully gapped state with preserved rotation symmetry around the $\mathcal{K},\mathcal{K}'$ points. Again, all of these results are consistent with the intraband order parameter picture discussed in Fig.~\ref{fig:figureOPs}.

In Fig.~\ref{figurebands} we focus on Fermi surfaces centered around the $\mathcal{K},\mathcal{K}'$ points. For completeness we report in Appendix~\ref{appendix3} the equivalent results for $\mu =1$, where the Fermi level sits at the van Hove singularity with the Fermi surface forming a separatrix line connecting the $M$-points.  
Similarly, we report the results for NNN pairing in Appendix~\ref{appendix4}, with the conclusion that NN and NNN pairing hosts very similar low-energy band structures.

\subsection{Density of states and gap closing points}
Having studied the lowest energy bands we next turn to evaluating the density of states (DOS) at low energies. For this purpose, we define the Green's function of the Bloch Hamiltonian $H_{\mathrm{BdG}}(\mathbf{q})$ as 
\begin{align}
&G(\mathbf{q},E)=(E-H_{\mathrm{BdG}}(\mathbf{q})+i0^+)^{-1}&
\end{align}
We then integrate $G(\mathbf{q},E)$ over the first  Brillouin zone to get the density of states $\rho_0(E)$ \cite{Kaladzhyan2016CharacterizingUS}
\begin{align}
&\rho_{0}(E)=-\dfrac{1}{\pi} \text{Tr}_{\text{el}}\left[\text{Im}\int_{\text{BZ}}\text{d}\mathbf{q}G(\mathbf{q},E)\right],&
\end{align}\\
where the trace Tr$_{\text{el}}$ is performed over the electron modes only (first half of the diagonal of the Green's function matrix).

\begin{figure}[tbh]
\centering
\includegraphics[width=6cm]{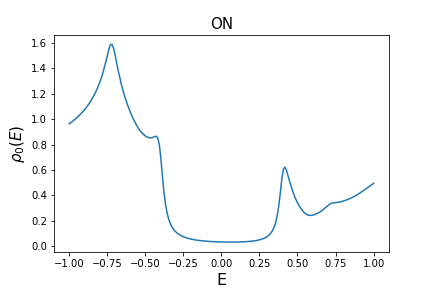}
\includegraphics[width=6cm]{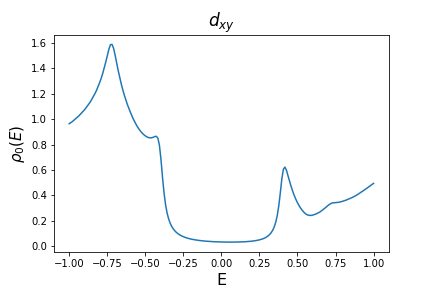}
\caption{DOS as a function of energy for ON $s$-wave and NN $d_{xy}$-wave SC states at $\Delta_0=0.4$ and $\mu=0.4$ illustrating the difference between the U-shaped and V-shaped behavior, corresponding to a full gap and a nodal spectrum, respectively.}
\label{figure4}
\end{figure}

We expect that the DOS for the nodal states to have a linear dependence of energy at low energies (V-shaped), while the fully gapped states to produce an U-shaped behavior. 
Indeed, this difference is visible in Fig.~\ref{figure4}, where we plot the DOS as a function of energy for  the ON $s$-wave and the NN $d_{xy}$-wave SC states.  The $s$-wave state displays a full gap giving rise to an overall U-type DOS profile. The nodal $d$-wave state on the other hand exhibits a V-type DOS near zero energy due to the existence of nodal quasiparticles. Moreover, we note  an additional near V-type feature at higher energies, centered around $E=\mu=0.4$. This corresponds to the normal-state Dirac point appearing at finite energy due to the finite $\mu$ \cite{Awoga2018}. We also note that both nodal and fully gapped profile exhibit a gap-edge coherence peak in the DOS.

\begin{figure*}[tbh]
\includegraphics[width=5.5cm]{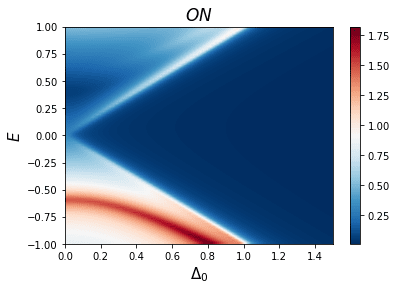}
\includegraphics[width=5.5cm]{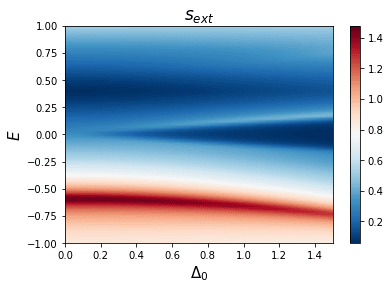} 
\includegraphics[width=5.5cm]{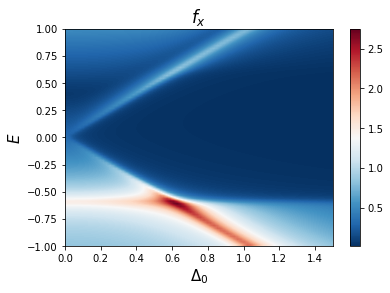}\\
\includegraphics[width=5.5cm]{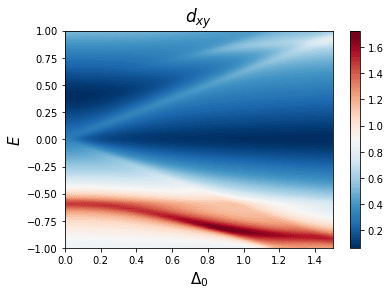} 
\includegraphics[width=5.5cm]{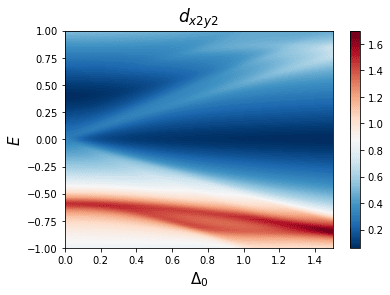}
\includegraphics[width=5.5cm]{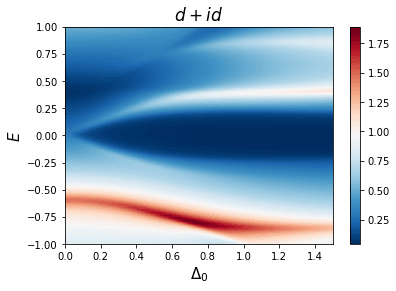}\\
\includegraphics[width=5.5cm]{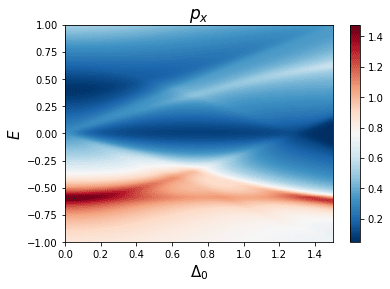}
\includegraphics[width=5.5cm]{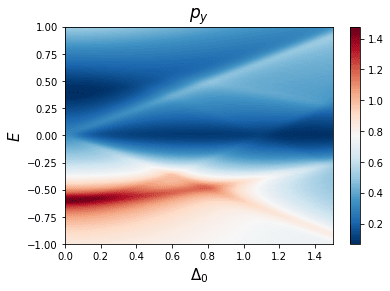}
\includegraphics[width=5.5cm]{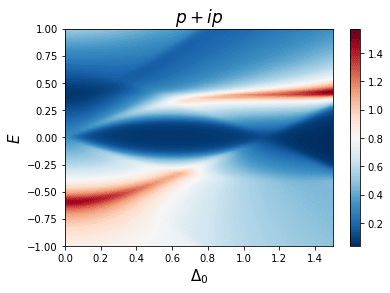}
\caption{DOS as a function of SC amplitude $\Delta_0$ for $\mu = 0.4$ for SC states with ON $s$-wave, NN $s_\text{ext}$-, $d_{xy}$-, $d_{x^2-y^2}$-, $p_x$-, $p_y$-, $p+ip\,'$-, $d+id\,'$-wave, as well as NNN $f_x$-wave symmetries. Dark blue represents zero DOS.
}
\label{figure3}
\end{figure*}

In Fig.~\ref{figure3} we plot the DOS as a function of the overall strength, or amplitude, of the SC order parameter $\Delta_0$ for the same chemical potential $\mu=0.4$ as in Fig.~\ref{figurebands}. 
The overall particle-hole asymmetry, especially strong at higher energies, is due to a pronounced particle-hole asymmetry in the normal state for finite $\mu$. 
Focusing primarily on low energies, as relevant for superconductivity, we note that for the ON $s$-wave state the energy gap is roughly linear in $\Delta_0$. We observe a similar linear dependence for the gap edges (or coherence peaks) as a function of $\Delta_0$ for the NN $s_\text{ext}$-, $d_{xy}$-, $d_{x^2-y^2}$-, $p_x$-, and $p_y$-wave SC states. Note, however, that for the $d_{xy}$-, $d_{x^2-y^2}$-, $p_x$-, and $p_y$-wave states the intensity is not zero inside the gap, which is consistent with the observations in the previous sections that all these states exhibit nodes in the band structure. 

In Fig.~\ref{figure3} we also note that the overall size of the gap, or equivalently half the distance between the coherence peaks for the nodal states, is no longer equal to the overall amplitude of SC order parameter $\Delta_0$, but generally lower, except for the ON $s$-wave state and the NNN $f$-wave state. The two exceptions have the normal-state Fermi surface being gapped out by an SC order parameter constant along the Fermi surface, which generates a gap equal to $\Delta_0$. For all the other states the low-energy spectrum is strongly modified by having the SC order parameter vary along the normal-state Fermi surface. Still, even in these cases we can identify a gap that is initially linearly increasing with $\Delta_0$, albeit with a smaller than 1 coefficient. At larger $\Delta_0$ this linearity breaks down for some of the symmetries. This is not surprising, as $\Delta_0 \sim 1$ corresponds to the energy scale of the van Hove singularity point, $\mu =1$, which marks the energy where the normal state band structure dramatically changes. For example in Fig.~\ref{figure3} we noticed a gap closing point for the $p_x$, $p_y$  and $p+ip\,'$ at $\Delta_0 \sim 1$. 

\begin{figure*}[tbh]
\includegraphics[width=5.5cm]{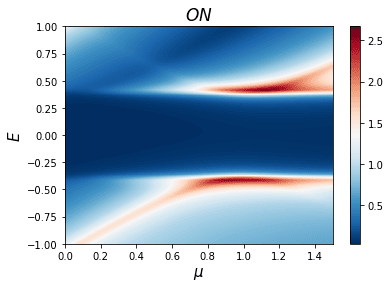}
\includegraphics[width=5.5cm]{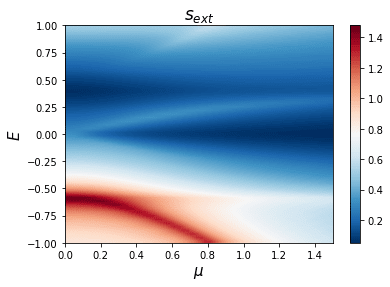} 
\includegraphics[width=5.5cm]{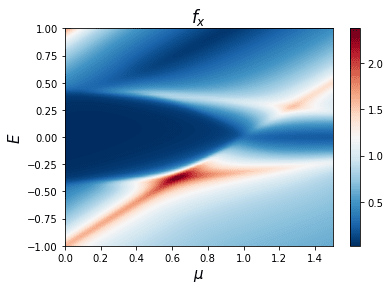}\\
\includegraphics[width=5.5cm]{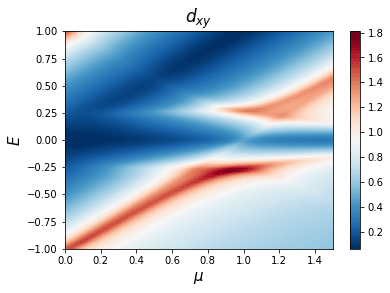} 
\includegraphics[width=5.5cm]{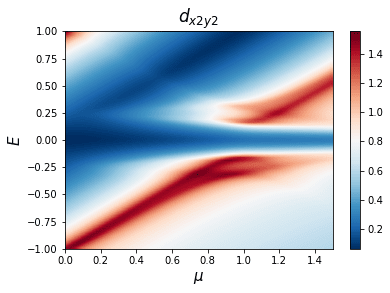}
\includegraphics[width=5.5cm]{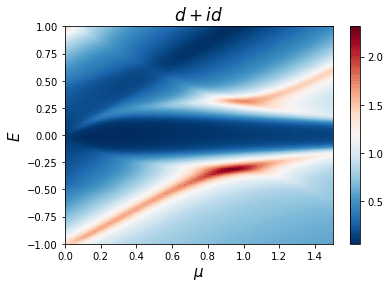} \\
\includegraphics[width=5.5cm]{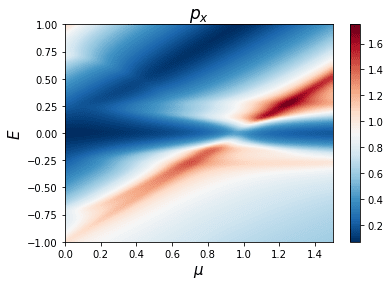}
\includegraphics[width=5.5cm]{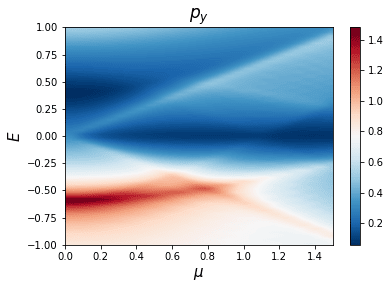}
\includegraphics[width=5.5cm]{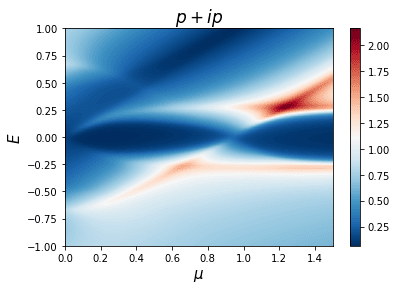}
\caption{DOS as a function of chemical potential $\mu$ for SC amplitude $\Delta_0=0.4$ for SC states with ON $s$-wave, NN $s_\text{ext}$-, $d_{xy}$-, $d_{x^2-y^2}$-, $p_x$-, $p_y$-, $p+ip\,'$-, $d+id\,'$-wave, as well as NNN $f_x$-wave symmetries. Dark blue represents zero DOS.
}
\label{figure5}
\end{figure*}

We also plot in Fig.~\ref{figure5} the DOS as a function of chemical potential $\mu$, while instead keeping the SC amplitude $\Delta_0$ fixed. Normally changing the chemical potential should not affect the DOS much as it only alters the underlying normal-state band structure.  Indeed, this is what we see for the ON $s$-wave which exhibits a constant gap for a large range of $\mu$. For some of the other states, however, the intraband SC order parameters exhibits nodes at the $\mathcal{K},\mathcal{K}'$ Dirac points, such that when $\mu =0$, both the SC order parameter and the normal state energy is zero at the Dirac points. This leads to a gapless system at $\mu=0$ and to a linear dependence of the gap for small $\mu$. 

\begin{figure}[tbh]
\begin{center}
\includegraphics[width=6cm]{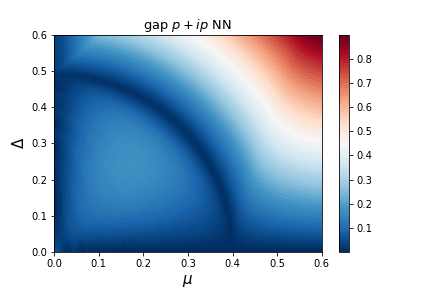}
\caption{DOS as a function of chemical potential $\mu$ and SC amplitude $\Delta_0$ for the NN $p+ip\,'$-wave state. Dark blue represents zero DOS and illustrates the gap closing points.}
\label{figure6}
\end{center}
\end{figure}

\begin{figure*}[tbh]
\centering
\includegraphics[width=4.9cm]{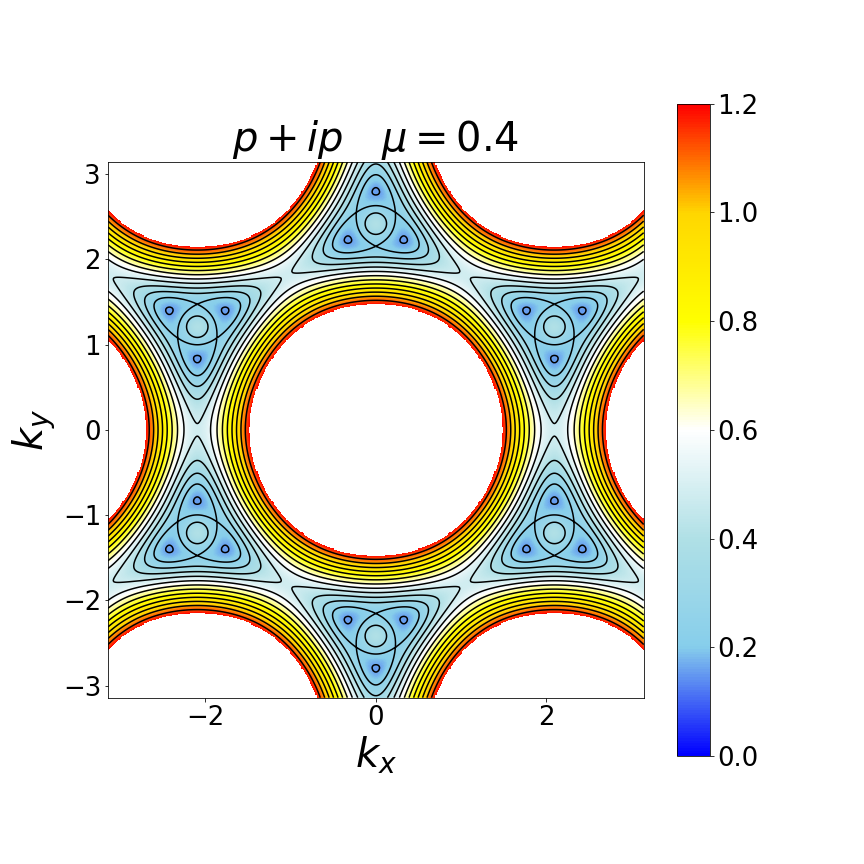}
\includegraphics[width=4.9cm]{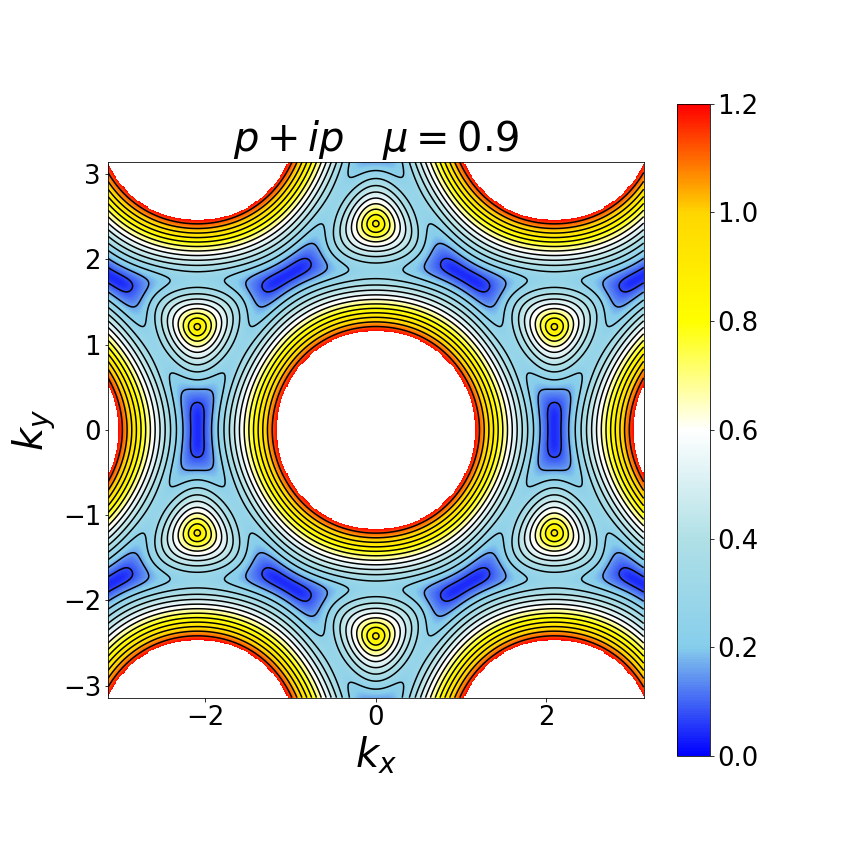}
\includegraphics[width=4.9cm]{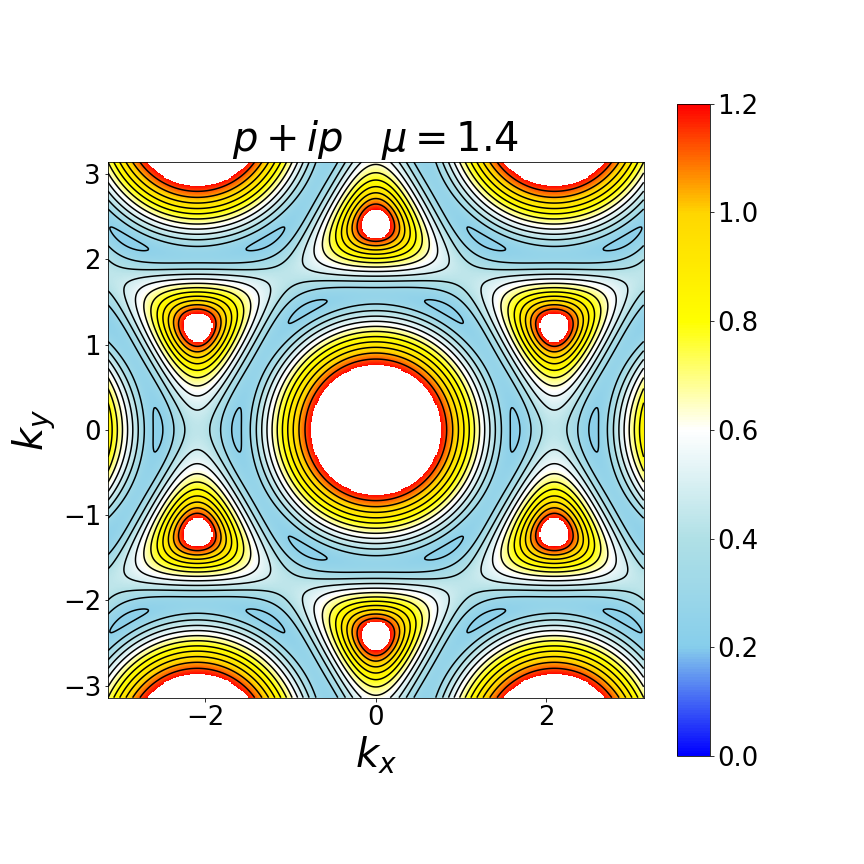}
\caption{Lowest energy band structure for SC amplitude $\Delta_0=0.4$ at three different values of the chemical potential, $\mu=0.4, 0.9$ and $1.4$ for the NN $p+ip\,'$-wave state. Only the middle panel hosts zero energy states.
}
\label{figure7}
\end{figure*}

Around $\mu=1$ we observe further substantial changes in the gap, or equivalently in the distance between the gap coherence peaks, for several SC states ($f_x$, $p_x$, $p_y$, $d_{xy}$, $p+ip'$). These changes consist primarily in the existence of points in parameter space where the energy gap is closing.
In order to track these points we plot in Fig.~\ref{figure6} the energy gap in the band structure for the chiral $p+ip\,'$-wave state as a function of both the chemical potential $\mu $ and the SC amplitude $\Delta_0$. We note that a continuous gap closing line indeed occurs connecting $\mu = 1$ with $\Delta_0 \sim 1$.
These gap closing points are important for experimentally distinguishing between different SC states. They are also important from the point of view of topology, as we will discuss in detail in the next part of this work. 
%
We can trace the gap closing points in all the $p$-wave states to the existence of nodes in the intraband SC order parameter at the $M$ points in the Brillouin zone, as depicted in Fig.~\ref{fig:figureOPs}. This leads to both a vanishing band energy and intraband order parameter at $\mu \approx 1$.  To verify this behavior we plot in Fig.~\ref{figure7} the lowest energy band for the chiral $p+ip\,'$-wave state at three different values of $\mu$. Keeping $\Delta_0 = 0.4$, we find that the gap closing occurs at the $M$ points of the Brillouin zone and at $\mu\approx 0.9$, corresponding to the middle panel of Fig.~\ref{figure7}. The difference from the intuitively expected value of $\mu \approx 1$  comes from interband pairing due to the sizable $\Delta_0$. For $\mu$ below this gap closing at $M$, the topology of the low-energy contours corresponds to having an origin at the $\mathcal{K},\mathcal{K}'$ Dirac points, while for larger $\mu$'s the contours are centered around the $\Gamma$ point. We also note that the $d_{xy}$-wave intraband order parameter has nodes at two $M$ points, which generates the gap closing point at $\mu =1$ in Fig.~\ref{figure5}. However, the $d_{x^2-y^2}$-wave state, and consequently the chiral $d+id\,'$-wave state, host no nodes as function of $\mu$. This different behavior offers a clear way to distinguish between all $p$- and $d$-wave states. Overall these results shows that, even if the total energy is also dependent on the interband pairing, many features can be understood simply by reasoning using the intraband order parameter and its nodes.

To provide a comprehensive treatment, we report also the DOS for the NNN pairing in Appendix \ref{appendix5}. Here we find that $s_\text{ext}$-wave and $d$-wave states have exactly the same type of behavior as their NN counterparts, while the NNN $p$-wave states only shows a gap closing as a function of $\mu$, at $\mu =1$ but not as a function of the SC pairing $\Delta_0$. The difference can be explained by the influence of the interband pairing term for NN pairing but not for NNN pairing. As such, the gap closing stays at $\mu=1$ for all values of $\Delta_0$.

\section{Superconducting multilayer graphene}

\begin{figure}[tbh]
\centering
\includegraphics[width=0.4\textwidth]{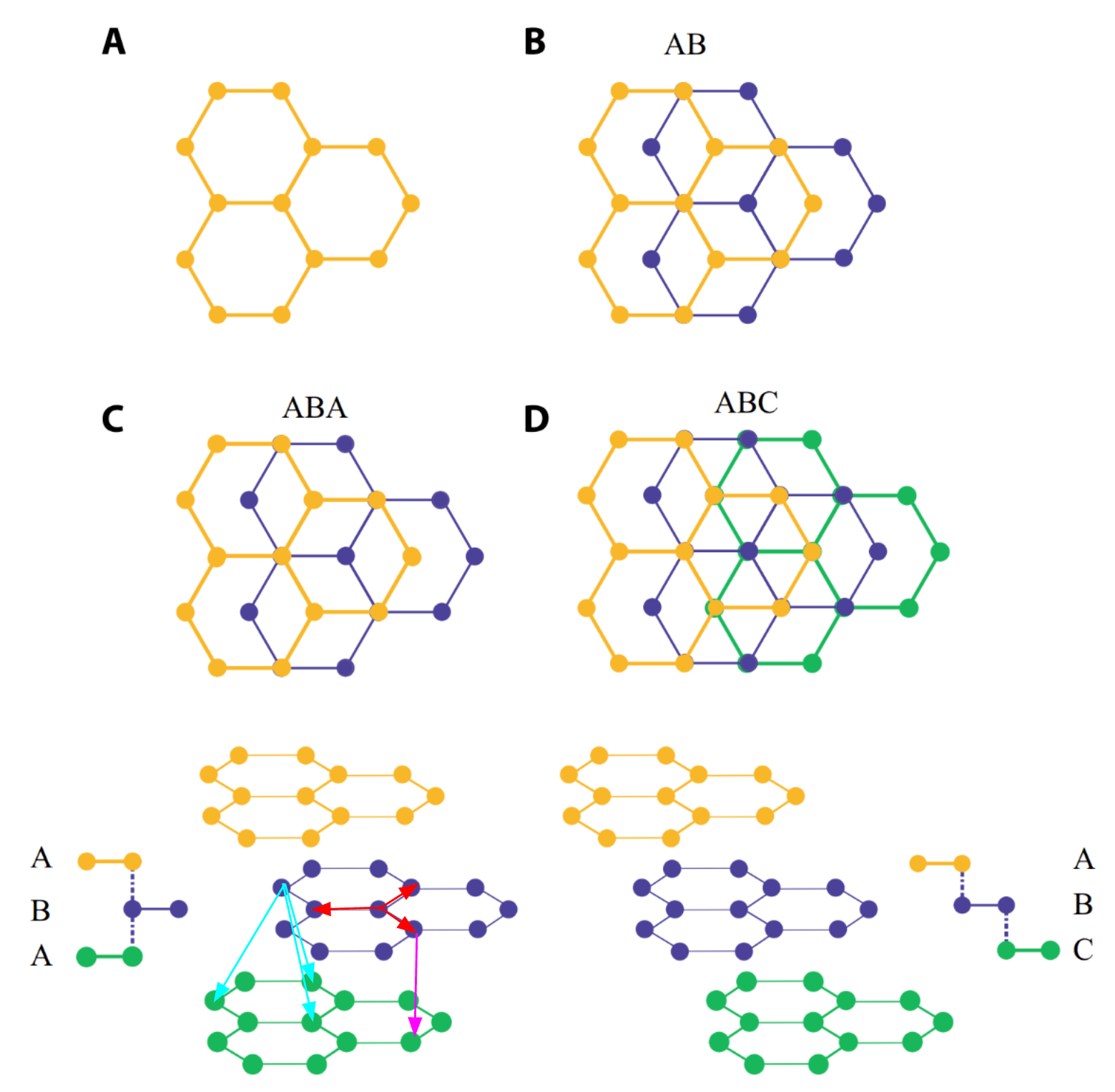}
\caption{Illustration of bilayer and trilayer graphene adapted from Ref.~[\onlinecite{Shan2018StackingSG}]. A: monolayer graphene. B: AB or Bernal-stacked bilayer graphene. C: ABA or Bernal-stacked trilayer graphene. D: ABC, or rhombohedral-stacked, trilayer graphene. Red arrows correspond to intralayer hopping $t$,  magenta to  interlayer hopping $\gamma_1$, and cyan to trigonal warping terms $\gamma_3$.}
\label{figure11}
\end{figure}

We next turn to the treatment of SC multilayer graphene. Here the 2D graphene sheets are stacked together with interlayer coupling primarily through van der Waals interaction. The naive configuration of placing two layers directly on top of each other is not energetically stable. The most stable form of stacking is instead known as AB, or Bernal stacking\cite{Charlier1991FirstprinciplesSO}. Figure~\ref{figure11}B illustrates this stacking: the two layers are stacked with a relative relative translation of $a_0\hat{x}$. For bilayer graphene, there is only this one choice. For trilayer graphene, we have two choices, we can either perform a $a_0\hat{x}$ translation between the first and second layer, then also a $a_0\hat{x}$ translation between the second and third layer, giving rise to a rhombohedral trilayer system, or ABC-stacking, see Fig.~\ref{figure11}D. We can also perform a translation of $-a_0\hat{x}$ for the third layer, which yields a Bernal trilayer system, or ABA-stacking, see Fig.~\ref{figure11}C. ABA is the energetically most stable stacking, while ABC is technically only metastable. Still, ABC-stacked multilayer graphene has been constructed using layer-by-layer deposition in multiple experiments \cite{Yan2011,Vervuurt2017}. ABC-stacking has also been produced by imposing mechanical constraints on Bernal graphite. \cite{Yankowitz2014}

To be able to model the normal-state in multilayer graphene we need to introduce interlayer hopping terms.
We denote the two closest hoppings between sites A and B in neighboring layers by $\gamma_1$ and $\gamma_3$. The parameter $\gamma_1$ is the direct hopping term between planes, coupling sites A and B that are spatially on top of each other. Since this is a purely vertical hopping, it appears as a constant in the tight-binding Hamiltonian. 
In constract, $\gamma_3$ parametrizes the non-direct interlayer hopping by coupling sites A and B that are not spatially on top of each other. Due to its finite intralayer reach it has a $k$-space modulation and act as a trigonal deformation term.
The values of these parameters have previously been determined, for example in Ref.~\onlinecite{Kuzmenko2009DeterminationOT} to be $\gamma_1=0.38$ eV, and $\gamma_3=0.38$ eV. Comparing this to the value of intralayer hopping $t=3.16$ eV, we note that the graphene layers are only weakly coupled to each other, as expected for van der Waals bonded layers. This weak interlayer coupling leads us in the following to consider only intralayer superconductivity, and thus disregard the possibility of interlayer SC bond pairing as that would require much stronger interlayer bonds than in the present van der Waals layered structures.

To write down the BdG Hamiltonian for SC bilayer and trilayer graphene, we use $\alpha \in \{1,2,3\}$ to describe the bottom (1), middle (2) or top (3) layers, respectively, and we write down
\begin{align}
&c^{\alpha}_{\mathbf{k}} =  \{ a_{{\mathbf{k}} \uparrow}^{\alpha},b_{{\mathbf{k}} \uparrow}^{\alpha},a_{{\mathbf{k}} \downarrow}^{\alpha},b_{{\mathbf{k}} \downarrow}^{\alpha} \}.&
\end{align}
Thus the bases to use for bilayer and trilayer graphene are, respectively,
\begin{align}
\label{equation23}
&\Psi_{\mathbf{k}}^{\text{bilayer}} = \{c^{1}_\mathbf{k},c^{2}_\mathbf{k},c^{1 \dagger}_{-\mathbf{k}},c^{2 \dagger}_{-\mathbf{k}}\}^T, \text{and} &
\end{align}
\begin{align}
\label{equation24}
&\Psi_{\mathbf{k}}^{\text{trilayer}} = \{c^{1}_\mathbf{k},c^{2}_\mathbf{k},c^{3}_\mathbf{k},c^{1 \dagger}_{-\mathbf{k}},c^{2 \dagger}_{-\mathbf{k}},c^{3 \dagger}_{-\mathbf{k}} \}^T.&
\end{align}
Below we treat each case separately.

\subsection{Bilayer Hamiltonian}
The BdG Hamiltonian for AB bilayer graphene can be expressed as 
\begin{align}
H_{AB}=
&\begin{pmatrix} 
H_{0}^{AB} & H_\text{SC} \\
H_\text{SC}^{\dagger} & -H_{0}^{AB} \\
\end{pmatrix},
\end{align}
with
\begin{align}
H_{0}^{AB}=
\begin{pmatrix}
H_{1} & H_{12} \\
H_{12}^{\dagger} & H_{2} \\
\end{pmatrix},
\hspace{0.2cm} 
H_\text{SC}=
\begin{pmatrix}
H_\text{SC}^{1} & 0 \\
0 & H_\text{SC}^{2} \\
\end{pmatrix}.
\end{align}
The components of the normal-state tight-binding Hamiltonian $H_{0}^{AB}$ are given by
\begin{align}
H_1=H_2=
\begin{pmatrix}
\mu & h_0({\mathbf{k}}) & 0 & 0 \\
h_0^{*}({\mathbf{k}}) & \mu & 0 & 0 \\
0 & 0 & \mu & h_0({\mathbf{k}}) \\
0 & 0 & h_0^{*}({\mathbf{k}}) & \mu \\
\label{h01}
\end{pmatrix},
\end{align}
which describe the intralayer hoppings in each layer 1 and 2, while the interlayer terms are given by
\begin{align}
&H_{12}=
\begin{pmatrix}
0 & \gamma_{3} \tilde{h}_{0}^{*}({\mathbf{k}}) & 0 & 0 \\
\gamma_1 & 0 & 0 & 0 \\
0 & 0 & 0 & \gamma_{3} \tilde{h}_{0}^{*}({\mathbf{k}}) \\ 
0 & 0 & \gamma_1 & 0 \\
\end{pmatrix},
\label{eqh12}
\end{align}
where $h_0({\mathbf{k}})$ and $\tilde{h}_0({\mathbf{k}})=h_0({\mathbf{k}})/t$ are introduced in Section~\ref{2c}.
The SC state is represented through $H_\text{SC}^{1,2}$, which are $4\times4$ matrices describing the intralayer superconducting pairing in layer 1 and 2, respectively. Due to the absence of inter-layer SC pairing, $H_\text{SC}$ is diagonal in the layer index. Furthermore, making the natural assumption of having the same superconducting mechanism in each layer, we consider that both layers have the same superconducting symmetry, leading to $H_\text{SC}^1=H_\text{SC}^2$, and $H_\text{SC}^1$ then also identical to the SC monolayer graphene matrix derived in Section~\ref{2c}.

\subsection{Trilayer Hamiltonian}
For ABA/ABC trilayer graphene the BdG Hamiltonian is given by
\begin{align}
H_\text{ABA/ABC}=
&\begin{pmatrix} 
H_{0}^\text{ABA/ABC} & H_\text{SC} \\
H_\text{SC}^{\dagger} & -H_{0}^\text{ABA/ABC} \\
\end{pmatrix},
\end{align}
with
\begin{align}
H_\text{SC}=
\begin{pmatrix}
H_\text{SC}^{1} & 0 & 0\\
0 & H_\text{SC}^{2} & 0 \\
0 & 0 & H_\text{SC}^{3} \\
\end{pmatrix}.&
\end{align}
As in the bilayer case, $H_\text{SC}$ does not couple the different layers, such that the matrix is diagonal in the layer space and is also characterized for each layer by the same superconducting Hamiltonian, $H_\text{SC}^{1} = H_\text{SC}^{2} = H_\text{SC}^{3}$.
For the normal-state Hamiltonian we have
\begin{align}
&H_0^\text{ABC}=
\begin{pmatrix}
H_1 & H_{12} & 0\\
H_{12}^{\dagger} & H_2 & H_{23}\\
0 & H_{23}^{\dagger} & H_3\\
\end{pmatrix},\\
 \; &H_{0}^\text{ABA}=
\begin{pmatrix}
H_1 & H_{12} & 0\\
H_{12}^{\dagger} & H_2 & H'_{23} \\
0 &(H'_{23})^{\dagger} & H_3\\
\end{pmatrix}.&
\end{align}
Here the intralayer terms are identical, $H_1=H_2=H_3$  and are given by Eq.~(\ref{h01}) above. Furthermore, for  the ABC trilayer $H_{23}=H_{12}$, with $H_{12}$ given by the bilayer expression in Eq.~\eqref{eqh12}. This is the consequence of having the same interlayer coupling between the bottom and middle layers as between the middle and top layers. In contrast, for the ABA trilayer, the type of pairs of sites connecting the top two layers is opposite to that connecting the bottom two layers: $\gamma_1$ connects a site A in the middle layer to a site B in the top layer, while $\gamma_3$ connects a site B in the middle layer to a site A in the top layer. Thus we need to swap the role of $\gamma_1$ and $\gamma_3$ in the interlayer Hamiltonian and we find $H'_{23}$ to have the form
\begin{align}
& H'_{23}=
\begin{pmatrix}
0 & \gamma_1 & 0 & 0 \\
\gamma_{3} \tilde{h}_{0}^{*}({\mathbf{k}}) & 0 & 0 & 0 \\
0 & 0 & 0 & \gamma_1 \\ 
0 & 0 & \gamma_{3} \tilde{h}_{0}^{*}({\mathbf{k}}) & 0 \\
\end{pmatrix}.&
\end{align}

\subsection{Lowest energy bands}
\begin{figure}[tbh]
\includegraphics[width=4.4cm]{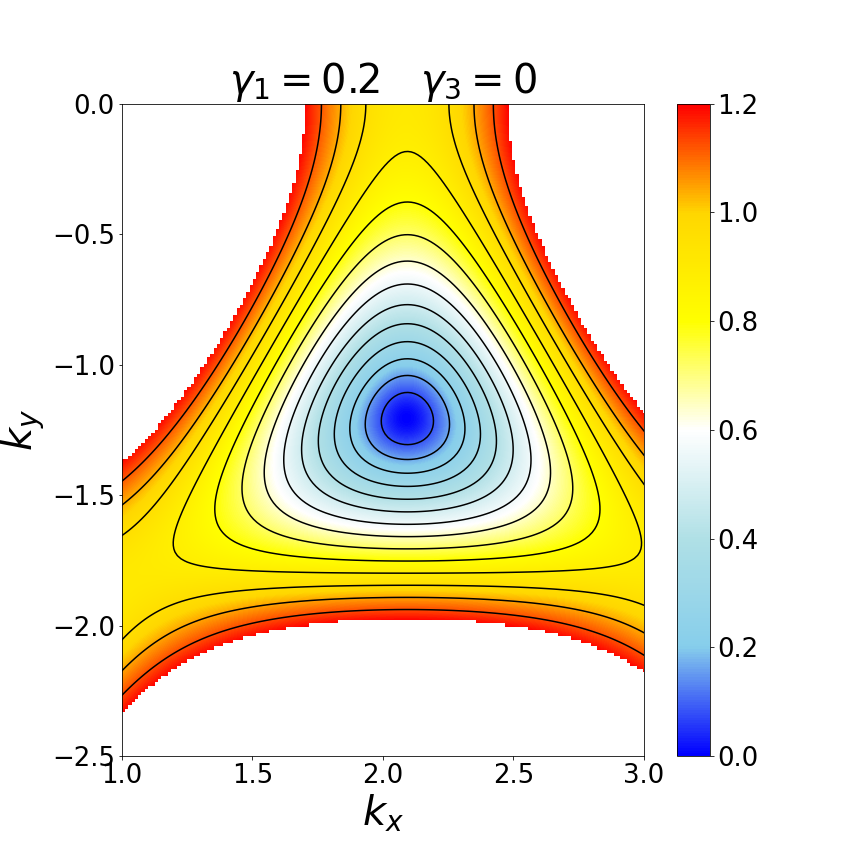}
\hspace{-0.5cm}
\includegraphics[width=4.4cm]{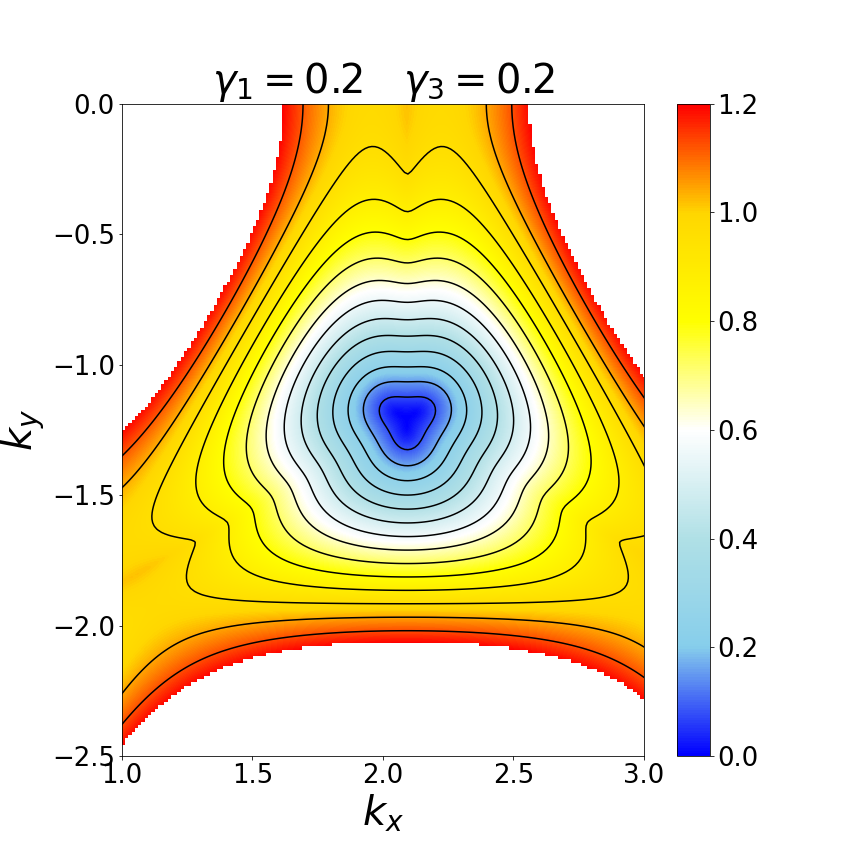}
\caption{Lowest energy band for undoped bilayer graphene ($\mu =0$) in the normal-state ($\Delta_0 = 0$) when $\gamma_3=0$ (left) and $\gamma_3=0.2$ (right). Trigonal warping gives rise to deformations in the lowest-energy iso-energy contours.}\label{figure9}
\end{figure}

Having derived the Hamlitonians for bilayer and trilayer graphene, we continue with exact diagonalization to find the energy spectrum.
In the following we focus mainly on bilayer graphene, aiming to extract the generic differences between the monolayer and any multilayer band structures. First we describe briefly  the effect of trigonal warping in the normal-state. To this end, we plot in Fig.~\ref{figure9} the lowest energy band in bilayer graphene for zero chemical potential and vanishing SC pairing. When the trigonal warping $\gamma_3$ is set to zero, we obtain at very low energy (dark blue region) a quadratic and isotropic dispersion relation with circular iso-energy contours around the $\mathcal{K},\mathcal{K}'$ Dirac points \cite{McCann2013TheEP}. When including also the interlayer $\gamma_3$ coupling we note that the behavior of the iso-energy contours around the $\mathcal{K},\mathcal{K}'$ points is significantly altered and acquires a trigonal distortion also at the lowest energies, hence the name trigonal warping term for $\gamma_3$.

\begin{figure*}[tbh]
\centering
\includegraphics[width=4.5cm]{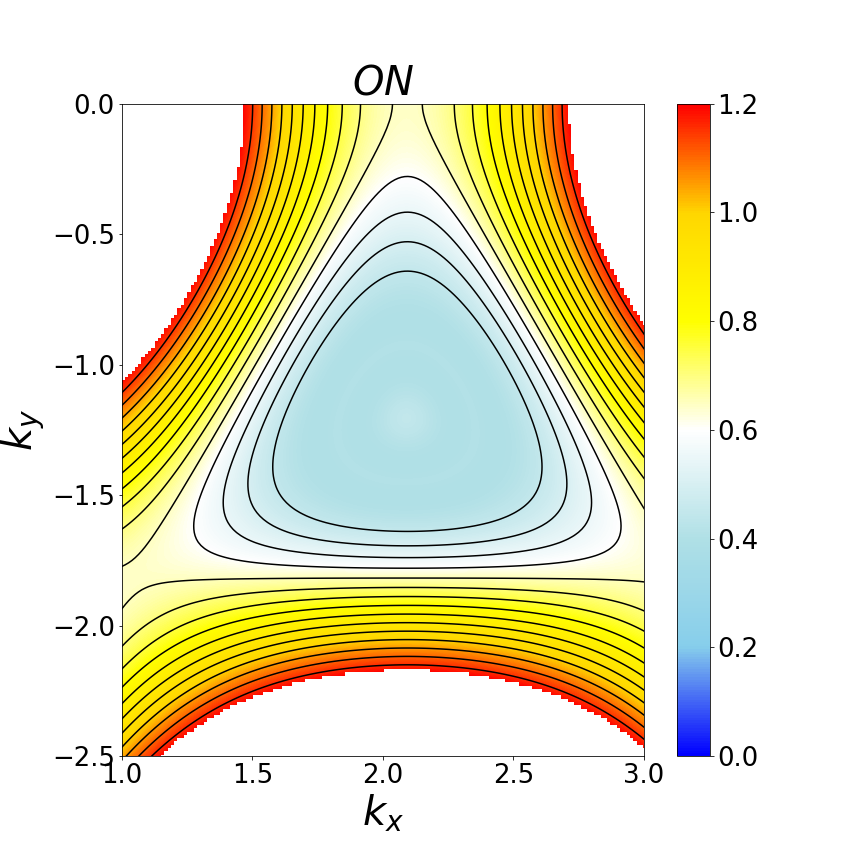}
\hspace{-0.5cm}
\includegraphics[width=4.5cm]{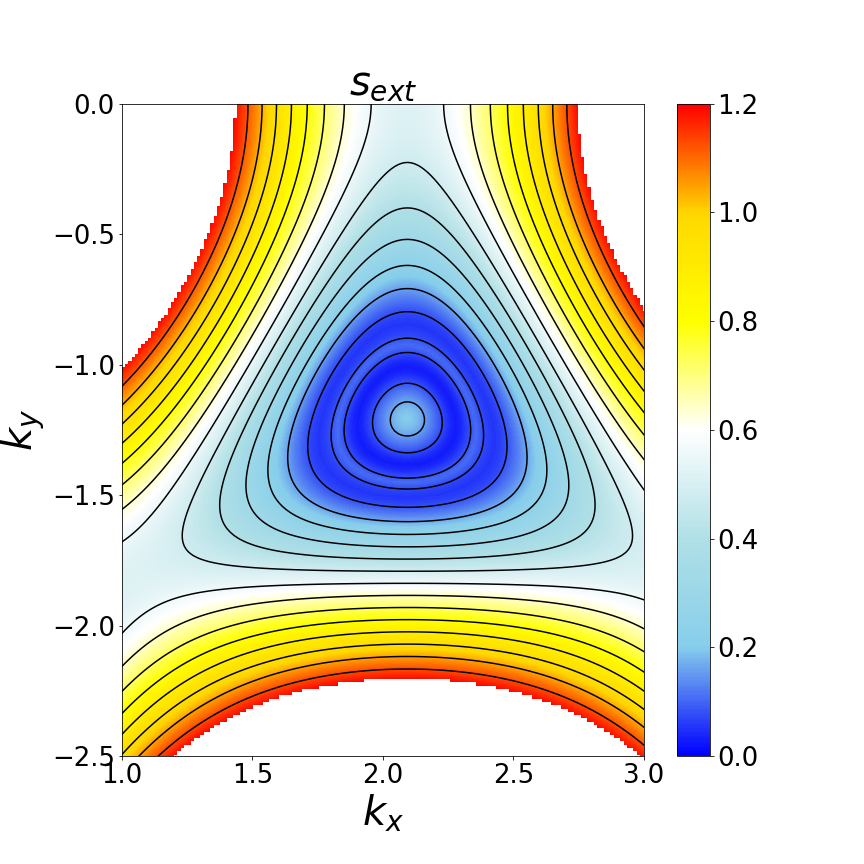}
\hspace{-0.5cm}
\includegraphics[width=4.5cm]{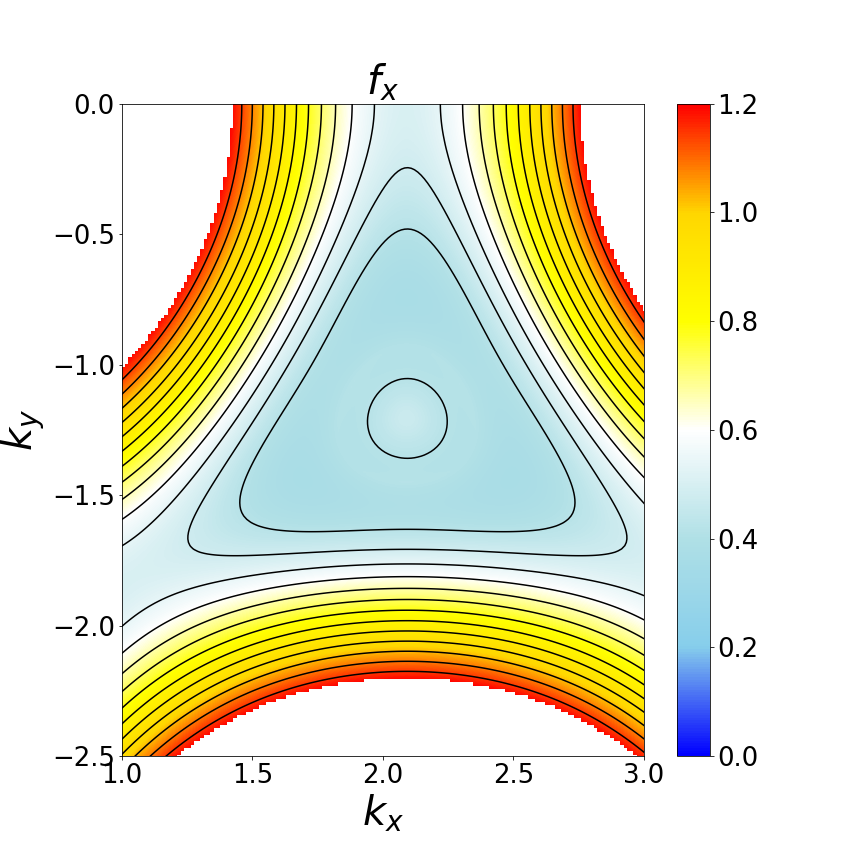}

\vspace{-0.2cm}

\includegraphics[width=4.5cm]{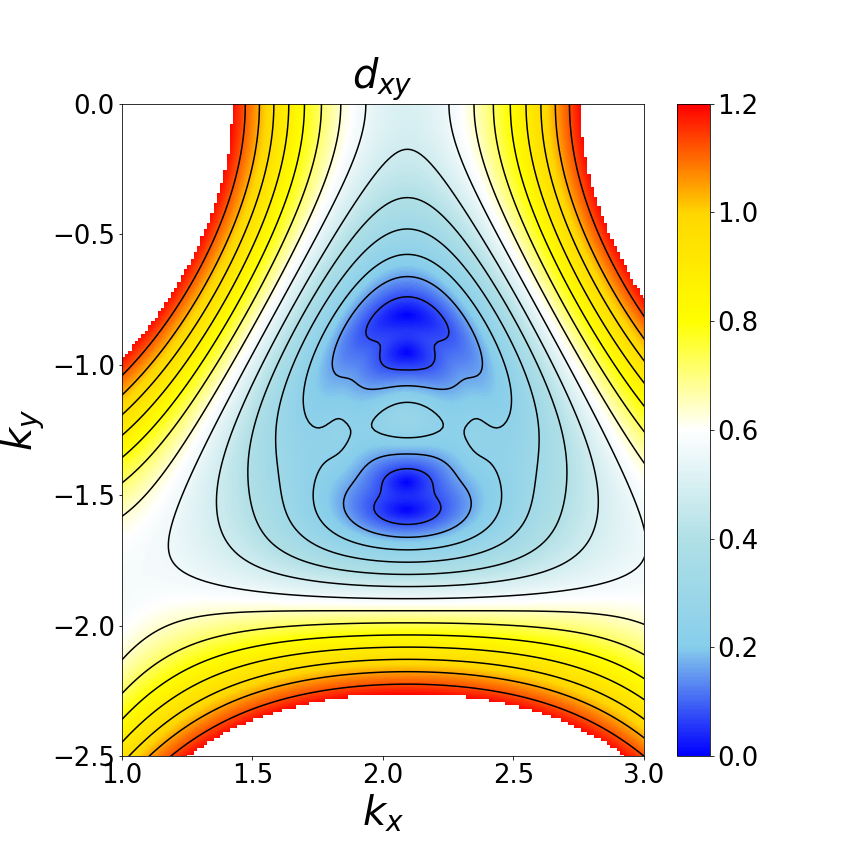}
\hspace{-0.5cm}
\includegraphics[width=4.5cm]{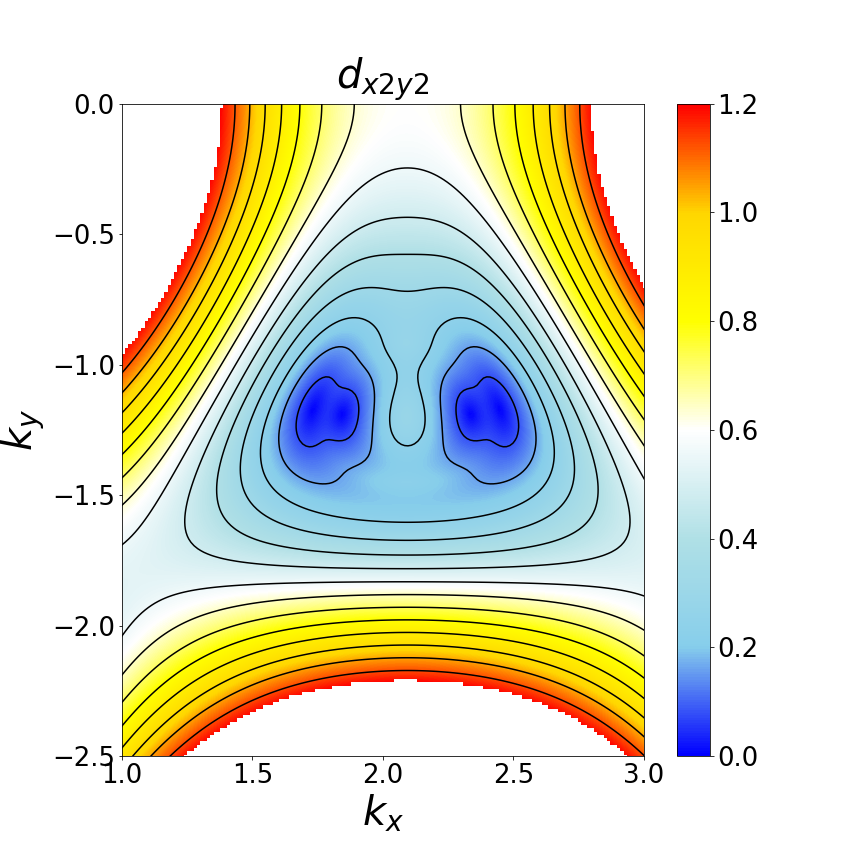}
\hspace{-0.5cm}
\includegraphics[width=4.5cm]{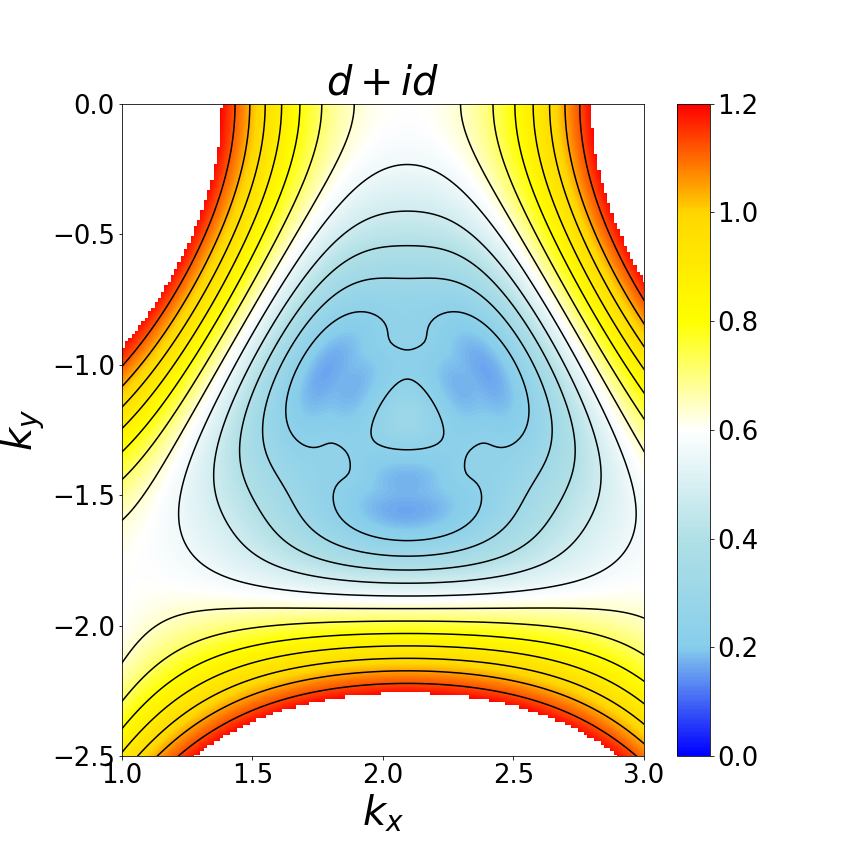}

\vspace{-0.2cm}

\includegraphics[width=4.5cm]{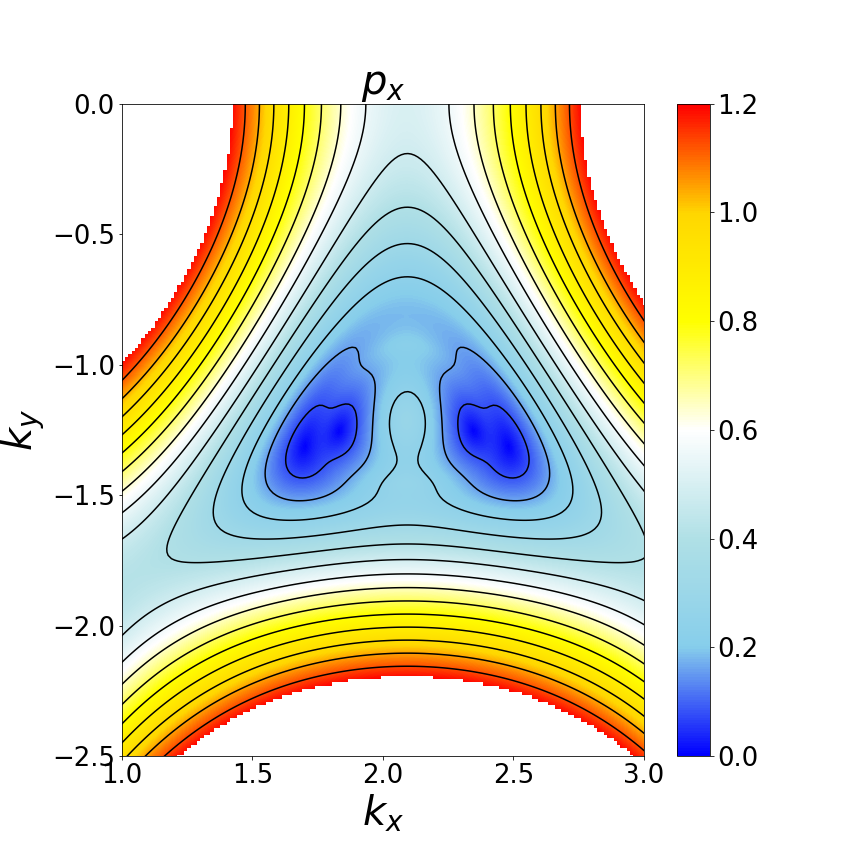}
\hspace{-0.5cm}
\includegraphics[width=4.5cm]{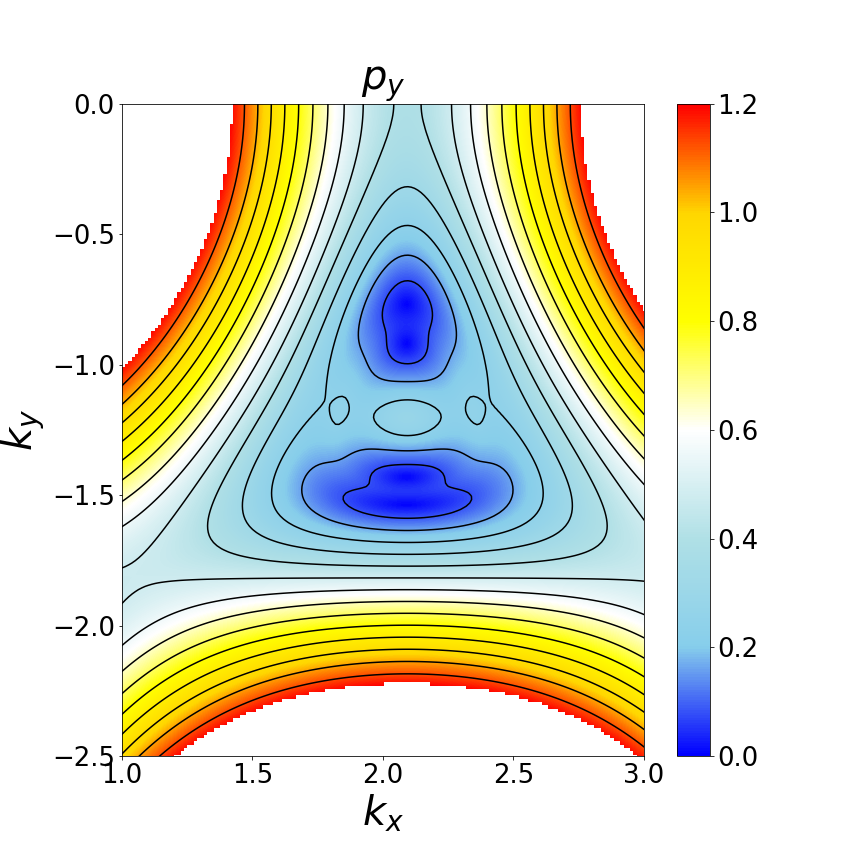}
\hspace{-0.5cm}
\includegraphics[width=4.5cm]{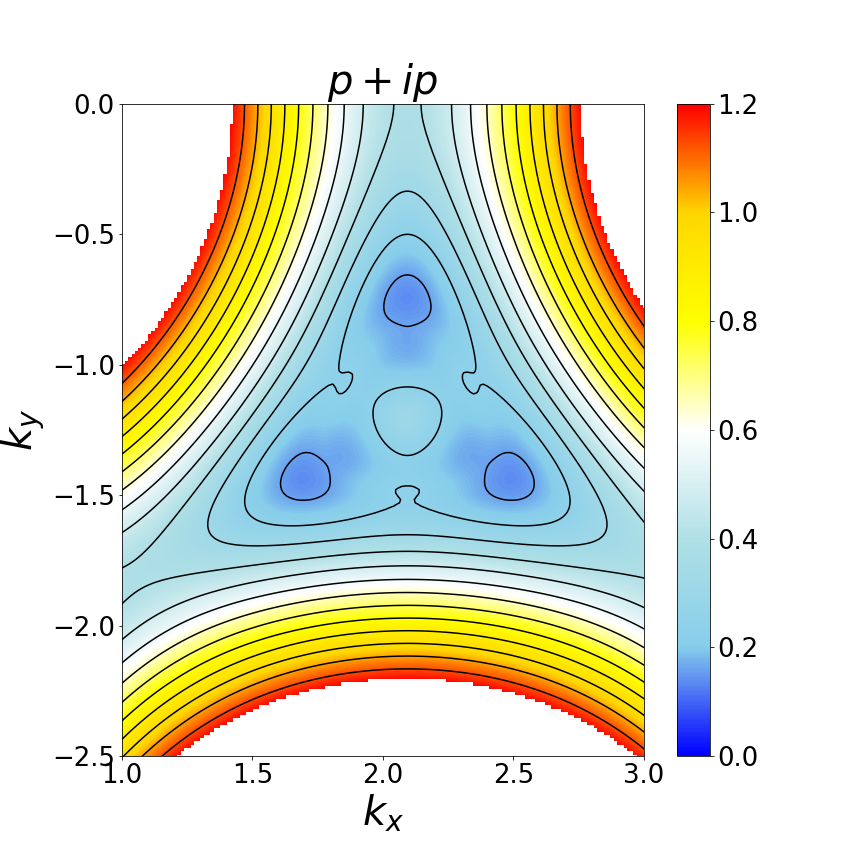}
\caption{Lowest energy band structure for $\mu=0.4$, $\Delta_0=0.4$, $\gamma_1=0.2$, $\gamma_3=0$,  and for SC states with ON $s$-wave, NN $s_\text{ext}$-, $d_{xy}$-, $d_{x^2-y^2}$-, $p_x$-, $p_y$-, $p+ip\,'$-, and $d+id\,'$-wave, as well as $f_x$-NNN symmetries. Only the region in the Brillouin zone close to the Dirac point $\mathcal{K}$ is displayed. Dark blue corresponds to zero energy.}
\label{figure10}
\end{figure*}

\begin{figure*}[tbh]
\includegraphics[width=4.8cm]{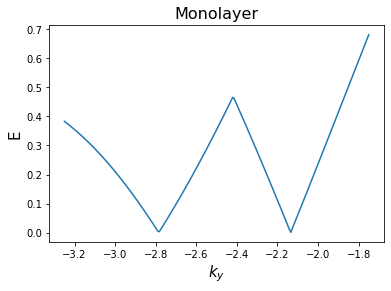}
\hspace{0.3cm}
\includegraphics[width=4.8cm]{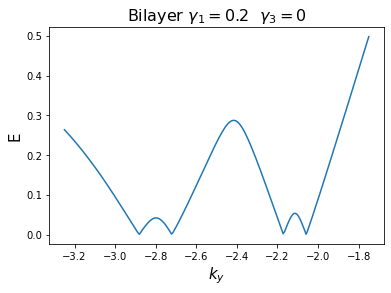}
\hspace{0.3cm}
\includegraphics[width=4.8cm]{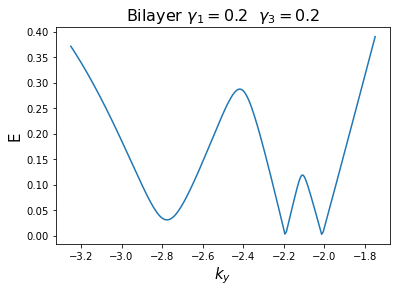}
\caption{Lowest energy band structure for the NN $p_y$-wave SC state as a function of $k_y$ for $k_x=0$. We compare monolayer graphene (left), bilayer graphene with no trigonal warping, $\gamma_1=0.2$ and $\gamma_3=0$ (middle), and bilayer graphene with finite trigonal warping, $\gamma_1=0.2$ and $\gamma_3=0.2$ (left). Here $\Delta_0=0.4$ and $\mu=0.4$.}
\label{np}
\end{figure*}

Moving on to the SC states, we plot in Fig.~\ref{figure10} the lowest energy bands for the same SC symmetries as in monolayer graphene  (c.p.~Fig.~\ref{figurebands}), but now zoomed-in around the Dirac point $\mathcal{K}$, to capture most clearly the effects of the interlayer coupling. We here first use $\gamma_3=0$. We find that for the fully gapped states, the lowest energy bands are similar to the monolayer case. However, for all  the nodal states we find a significant difference from the monolayer, in that all the nodal points double by exhibiting a small splitting in momentum space, see dark blue color gradient. Thus the nodal $d$- and $p$-wave states now have four nodal points, two on each side of the Dirac point. 
Next we turn on the trigonal warping and find that it has a significant effect on the formation of the nodes. In Fig.~\ref{np} we plot the lowest band as a function of $k_y$ for $k_x =0$ for the illustrative case of NN $p_y$-wave symmetry. We note that the two nodal points in monolayer graphene split and give rise to four nodal points for bilayer graphene with no trigonal warping. When turning on $\gamma_3$ we find that this term slightly gaps two of the four nodes.

\subsection{Density of states and gap closing points}
It is also interesting to study the DOS in bilayer and trilayer graphene, in particular elucidating the existence of gap closing points. Using a similar procedure as in the previous section we calculate the DOS as a function of energy, chemical potential, and SC order parameter amplitude. We find that it has a very similar dependence to that of monolayer graphene. For completeness we provide plots of the DOS for the most interesting case, the $p+ip\,'$-wave state, in Appendix \ref{appendix6} for both bilayer and trilayer graphene. The main features observed for the monolayer are preserved, except that the number of gap closing points is different for multilayer graphene. 
\begin{figure*}[tbh]
\hspace*{1.cm}\includegraphics[width=5.5cm]{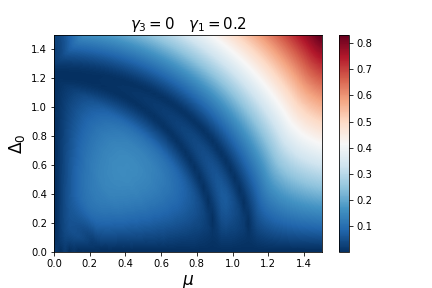}
\includegraphics[width=5.5cm]{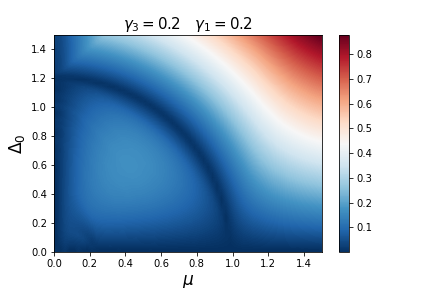}
\includegraphics[width=5.5cm]{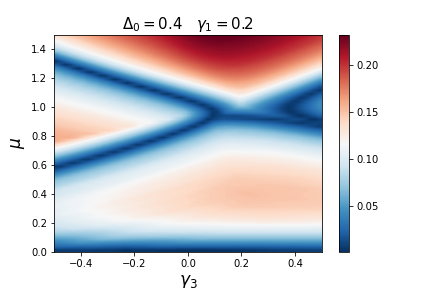}
\caption{DOS as a function of chemical potential $\mu$ and SC amplitude $\Delta_0$ for the NN $p+ip\,'$-wave state in bilayer graphene for $\gamma_3=0$ (left), $\gamma_3 = 0.2$ (middle) with $\gamma_1 = 0.2$. (Left): DOS as a function of $\gamma_3$ and $\mu$ for $\Delta_0 = 0.4$ and $\gamma_1 = 0.2$. Dark blue corresponds to a vanishing DOS and to the gap closing points. }
\label{figure12}
\end{figure*}
Focusing on these gap closing points for the chiral $p+ip\,'$-wave SC state, we plot in Fig.~\ref{figure12} the gap in the energy spectrum as a function of the chemical potential and the SC amplitude for bilayer graphene without trigonal warping ($\gamma_3=0$). We find that there are now two gap closing lines for each $\Delta_0$ as a function of $\mu$, compared to the single gap closing line in monolayer graphene in Fig.~\ref{figure6}. Adding a finite $\gamma_3$ changes this, and we find again only one gap closure line. Thus the trigonal warping has a strong influence on the number of gap closing points. Indeed this is confirmed in the right panel in Fig.~\ref{figure12}, where we plot the DOS as a function of both $\mu$ and $\gamma_3$. We find a similar effect in trilayer graphene, with the number of gap closing points oscillating between one and three. A more detailed analysis is presented in Appendix \ref{appendix6}.

\begin{figure}[tbh]
\includegraphics[width=5cm]{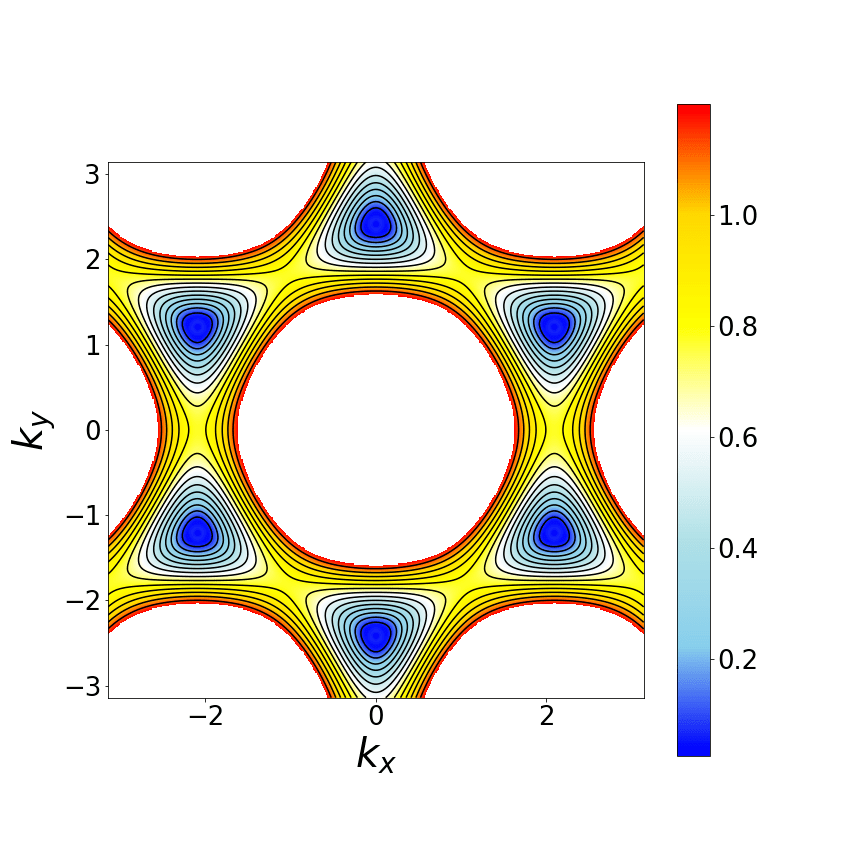}
\caption{Lowest energy band for bilayer graphene for the NN $p+ip\,'$-wave SC state for $\mu=0.1$, $\Delta_0=0.1$, $\gamma_1=0.2$ and $\gamma_3=0$. The dark blue regions correspond to zero energy.}
\label{fig16}
\end{figure}

Finally, we note that the formation of an interesting additional gap closing point at small $\mu$ and $\Delta_0$, not present in the monolayer case, see the dark blue line close to $\mu \approx 0.2$ in Fig.~\ref{figure12}. Since any realistic system will have a rather small $\Delta_0$, this is particularly alluring from an experimental point of view. 
It appears that this gap closing point is not overly sensitive to the value of $\gamma_3$. In Fig.~\ref{fig16} we plot the lowest energy band for the parameters corresponding to this gap closing point and establish that this interlayer-induced gap closing occurs at the $\mathcal{K},\mathcal{K}'$ Dirac points, in contrast to the previously described gap closing points that occur due to a nodal intraband pairing at the $M$ points. We attribute this gap closing to the combination of nodal points at the $\mathcal{K},\mathcal{K}'$ points in the intraband SC order parameter and the bottom of the second graphene band touching the $\mathcal{K},\mathcal{K}'$ points when the chemical potential $\mu$ is equal to the interlayer coupling $\gamma_1=0.2$. As such it is intimately tied to the multilayer aspect.

\section{Conclusions}
In this work we reviewed all symmetry-allowed spin-singlet and spin-triplet superconducting states in monolayer, bilayer and trilayer graphene with different stacking. By allowing on-site pairing, as well as pairing between nearest neighbors and next-to-nearest neighbors, we captured all possible spin-singlet and -triplet order parameters, from $s$- and $d$-wave to $p$- and $f$-wave states, including the chiral $d+id\,'$- and $p+ip\,'$-wave states. To analyze the properties of these states, we calculated the low-energy band structure, as well as the density of states as a function of the chemical potential and the superconducting pairing strength. The different SC states can be classified in two large classes, the fully-gapped states, such as the $s$-, $f$-, $d+id\,'$-, and $p+ip\,'$-wave states, that exhibit a U-shaped DOS and the nodal ones, such as the $d_{xy}$-, $d_{x^2-y^2}$-, $p_x$-, and $p_y$-wave states that exhibit nodal points in the band dispersion and a V-shaped DOS. Moreover, we focused on the existence of gap closing points in the DOS when changing the physical parameters, and found that many of these points can be understood by a careful examination of the symmetry of the intraband superconducting order parameter in monolayer graphene. For bilayer and trilayer graphene, we found that the interlayer coupling splits the nodal points in the band structure, as well as the gap closing points. We also analyzed the effect of the trigonal warping present in bilayer and trilayer graphene on the formation of gap closing points. By distinguishing between nodal and fully gapped superconducting states our work provides an experimentally viable tool to differentiate between different superconducting symmetries in all carbon-based superconductors. In future works we will discuss the topology and the formation of edge states in these systems, as well as the formation of Shiba states, in the hope to provide additional tools to further distinguish experimentally between various order parameters arising in superconducting graphene materials.

\acknowledgements We thank Miguel Alvarado and Tomas L\"othman for useful discussions.
We acknowledge financial support from the Swedish Research Council (Vetenskapsr\aa det Grant No.~2018-03488) and the Knut and Alice Wallenberg Foundation through the Wallenberg Academy Fellows program. 
\appendix

\onecolumngrid

\section{Derivation of the momentum space Hamiltonian}
\label{appendix1}

In this Appendix we provide additional details for the derivation of the BdG Hamiltonian in momentum space.

\subsection{Normal state Hamiltonian}

For the tight-binding term in momentum space we have
\begin{align}
H_\text{TB}(k)
  = \sum \limits_{k,\sigma} \left( h_0(k) a^\dagger_{k,\sigma} b_{k,\sigma} +  h_0^*(k) b^\dagger_{k,\sigma} a_{k,\sigma}\right).
\label{annex1}
\end{align}
In the BdG form, and also taking into account the doubling of the number of modes to include separately the electron and the hole spectrum for both up and down spins, this becomes
\begin{align}
H_\text{TB}(k)= \sum \limits_{k,\sigma} h_0(k)\left( a^\dagger_{k \sigma} b_{k \sigma} - b_{-k,\sigma}a^\dagger_{-k,\sigma}\right) 
+ \sum \limits_{k,\sigma} h_0^*(k)\left( b^\dagger_{k \sigma} a_{k \sigma} - a_{-k,\sigma}b^\dagger_{-k,\sigma}\right),
\end{align}
with a form factor 
\begin{equation}
\begin{aligned}
h_0(k)
=-t\left[e^{-ik_y} +2e^{\frac{i}{2}k_y}\cos\left(\frac{\sqrt{3}}{2}k_x\right)\right]. \\
\end{aligned}
\end{equation}
For the chemical potential term we have 
\begin{align}
H_\mu
= -\mu \sum \limits_{k \sigma} \left(a^\dagger_{k,\sigma} a_{k,\sigma} + b^\dagger_{k,\sigma} b_{k,\sigma}\right).
\end{align}
In the BdG form this becomes
\begin{align}
H_\mu=& -\mu \sum \limits_{k \sigma} \left( a^\dagger_{k,\sigma} a_{k,\sigma} - a_{-k,\sigma} a^\dagger_{-k,\sigma} + b^\dagger_{k,\sigma} b_{k,\sigma} - b_{-k,\sigma} b^\dagger_{-k,\sigma}\right).
\end{align}

\subsection{On-site pairing}
In real space the on-site pairing term takes the following form
\begin{equation}
H_\text{ON}= 2 \Big[\sum \limits_{i} \Delta_\text{ON} a^\dagger_{i\up} a^\dagger_{ i \down}  + \sum \limits_{i} \Delta_\text{ON} b^\dagger_{i\up} b^\dagger_{i\down}  + h.c.\Big],
\end{equation}
where the factor of 2 come from the doubling of the number of modes to take into account separately the electrons/holes with spin up and spin down.
The Fourier transform of this term is
\begin{equation}
H_\text{ON}= 2 \Big[ \sum \limits_{k \sigma}  \Delta_\text{ON} a^\dagger_{k\up} a^\dagger_{ -k \down} + \sum \limits_{k\sigma} \Delta_\text{ON} b^\dagger_{k\up} b^\dagger_{-k\down} + h.c.\Big],
\end{equation}
which can be written in the BdG form as
\begin{equation}
    H_\text{ON}= \sum \limits_{k\sigma} \Delta_\text{ON}\left (a^\dagger_{k \up} a^\dagger_{-k \down} -  a^\dagger_{-k \down} a^\dagger_{k \up} \right)
    + \sum \limits_{k\sigma} \Delta_\text{ON} \left(b^\dagger_{k \up} b^\dagger_{-k \down} - b^\dagger_{-k \down} b^\dagger_{k \up}\right)+h.c.
\end{equation}

\subsection{NN pairing}
For NN pairing in the spin-singlet channel ($\eta=0$) we have
\begin{equation}
H^0_\text{NN} = \sum \limits_{\langle i,j \rangle} \Delta^{\eta =0}_{ij} \left(a^\dagger_{i\up} b^\dagger_{j\down}  - a^\dagger_{i\down}b^\dagger_{j\up}\right) + h.c.
\end{equation}
while in the spin-triplet channel we find
\begin{align}
H^x_\text{NN}=&\sum \limits_{\langle i,j \rangle}  \Delta_{ij}^{\eta=x} \left(a^\dagger_{i\up} b^\dagger_{j\up} -  a^\dagger_{i\down} b^\dagger_{j\down}\right) + h.c., \\
H^y_\text{NN}=&\sum \limits_{\langle i,j \rangle}  \Delta_{ij}^{\eta=x}\left (a^\dagger_{i\up} b^\dagger_{j\up} + a^\dagger_{i\down} b^\dagger_{j\down} \right)+ h.c., \\
H^z_\text{NN}=& \sum \limits_{\langle i,j \rangle} \Delta^{\eta =z}_{ij} \left(a^\dagger_{i\up} b^\dagger_{j\down}  + a^\dagger_{i\down}b^\dagger_{j\up}\right) + h.c.
\end{align}
We can now perform the Fourier transform as before. In the spin-singlet channel we obtain
\begin{equation}
H_\text{NN}= \sum \limits_{k}h_\text{NN}^{0} \left( a^\dagger_{k\up} b^\dagger_{-k\down}  - a^\dagger_{k\down}b^\dagger_{-k\up}\right) + h.c.
\end{equation}
and in the spin-triplet channel
\begin{align}
H^x_\text{NN}=&\sum \limits_{k} h_\text{NN}^x \left(a^\dagger_{k\up} b^\dagger_{-k\up} -  a^\dagger_{k\down} b^\dagger_{-k\down} \right)+ h.c., \\
H^y_\text{NN}=& i\sum \limits_{k}h_\text{NN}^y\left( a^\dagger_{k\up} b^\dagger_{-k,\up}+ a^\dagger_{k,\down} b^\dagger_{-k\down}\right) + h.c.,  \\
H^z_\text{NN}=& \sum \limits_{k}h_\text{NN}^z\left(a^\dagger_{k\up}b^\dagger_{-k\down}+ a^\dagger_{k\down}b^\dagger_{-k\up}\right)+ h.c. 
\end{align}
Here $h_\text{NN}^{\eta}$ are the form factors, whose expressions depend on both the spin channel and the symmetry of the order parameter. Their general expression is:
\begin{equation}
h_\text{NN}^{\eta}=\Delta_\text{NN}^{\eta,d=1}e^{-ik_y}+\Delta_\text{NN}^{\eta,d=2} e^{\frac{i}{2}k_y -\frac{\sqrt{3}i}{2}k_x} 
+ \Delta_\text{NN}^{\eta,d=3} e^{\frac{i}{2}k_y + \frac{\sqrt{3}i}{2}k_x}.
\end{equation}
Here, $d=1,2,3$ correspond to the three NNs, following the convention of Fig.~\ref{figure2a}. The values of $\Delta^{\eta, d}$ are also detailed in Tables \ref{Table1} and \ref{Table2} for each symmetry. By replacing these values we obtain the form factors for NN pairing ($h_\text{NN}^{0, s_\text{ext}}$, $h_\text{NN}^{0, d_{xy}}$, $h_\text{NN}^{0, d_{x^2-y^2}}$, $h_\text{NN}^{\eta, p_x}$, $h_\text{NN}^{\eta,p_y}$, $h_\text{NN}^{\eta,f_x}$) in Tables \ref{Table3} and  \ref{Table4}.

We subsequently express the above NN pairing term in the BdG form
\begin{equation}
\begin{aligned}
H^{0}_\text{NN}=& \frac{1}{2} \sum \limits_{k}h_\text{NN}^{0}(k)\left( a^\dagger_{k\up} b^\dagger_{-k\down} - b^\dagger_{-k\down}a^\dagger_{k\up}  - a^\dagger_{k\down}b^\dagger_{-k\up} + b^\dagger_{-k\up} a^\dagger_{k\down}\right) + h.c.,\\
=& \frac{1}{2} \sum \limits_{k} \left(h_\text{NN}^{0}(k) a^\dagger_{k\up} b^\dagger_{-k\down} -  h_\text{NN}^{0}(-k) b^\dagger_{k\down} a^\dagger_{-k\up}- h_\text{NN}^{0}(k) a^\dagger_{k\down}b^\dagger_{-k\up} +h_\text{NN}^{0}(-k) b^\dagger_{k\up} a^\dagger_{-k\down}\right) + h.c.
\end{aligned}
\end{equation}
for the singlet channel and
\begin{equation}
\begin{aligned}
H^x_\text{NN}=&\frac{1}{2}\sum \limits_{k} \left(h_\text{NN}^{x}(k) a^\dagger_{k\up} b^\dagger_{-k\up} - h_\text{NN}^x(-k) b^\dagger_{k\up}a^\dagger_{-k\up} - h_\text{NN}^{x}(k) a^\dagger_{k\down} b^\dagger_{-k\down} + h_\text{NN}^{x}(-k) b^\dagger_{k\down} a^\dagger_{-k\down}\right)+ h.c., \\
H^y_\text{NN}=& \frac{i}{2} \sum \limits_{k} \left(h_\text{NN}^{y}(k) a^\dagger_{k\up} b^\dagger_{-k\up} - h_\text{NN}^{y}(-k) b^\dagger_{k\up} a^\dagger_{-k\up}+h_\text{NN}^{y}(k) a^\dagger_{k\down} b_{-k\down}- h_\text{NN}^{y}(-k) b^\dagger_{k\down} a^\dagger_{-k\down}\right) + h.c.,  \\
H^z_\text{NN}=& \frac{1}{2} \sum \limits_{k} \left(h_\text{NN}^{z}(k) a^\dagger_{k\up}b^\dagger_{-k\down} -h_\text{NN}^{z}(-k) b^\dagger_{k\down} a^\dagger_{-k\up} + h_\text{NN}^{z}(k) a^\dagger_{k\down}b^\dagger_{-k\up} -h_\text{NN}^{z}(-k) b^\dagger_{k\up}a^\dagger_{-k\down}\right)+ h.c.
\end{aligned}
\end{equation}
for the triplet channel.

\subsection{NNN pairing}
For NNN pairing in the spin-singlet channel ($\eta=0$)  we have
\begin{equation}
H^0_\text{NNN} = \sum \limits_{\langle\langle i,j \rangle\rangle} \Delta^{\eta =0}_{ij} \left(a^\dagger_{i\up} a^\dagger_{j\down}  - a^\dagger_{i\down}a^\dagger_{j\up}+b^\dagger_{i\up} b^\dagger_{j\down}  - b^\dagger_{i\down}b^\dagger_{j\up}\right) + h.c.,
\end{equation}
and in the spin-triplet channel
\begin{align}
H^x_\text{NNN} = \sum \limits_{\langle\langle i,j \rangle\rangle} \Delta^{\eta =0}_{ij} \left(a^\dagger_{i\up} a^\dagger_{j\up}  -a^\dagger_{i\down}a^\dagger_{j\down}+b^\dagger_{i\up} b^\dagger_{j\up}  - b^\dagger_{i\down}b^\dagger_{j\down}\right) + h.c.,\\
H^y_\text{NNN} = \sum \limits_{\langle\langle i,j \rangle\rangle} \Delta^{\eta =0}_{ij} \left(a^\dagger_{i\up} a^\dagger_{j\up}  +a^\dagger_{i\down}a^\dagger_{j\down}+b^\dagger_{i\up} b^\dagger_{j\up}  +b^\dagger_{i\down}b^\dagger_{j\down}\right) + h.c.,\\
H^0_\text{NNN} = \sum \limits_{\langle\langle i,j \rangle\rangle} \Delta^{\eta =0}_{ij}\left (a^\dagger_{i\up} a^\dagger_{j\down}  + a^\dagger_{i\down}a^\dagger_{j\up}+b^\dagger_{i\up} b^\dagger_{j\down}  +b^\dagger_{i\down}b^\dagger_{j\up}\right) + h.c.,
\end{align}
where the sums are now performed over the NNN pairs.
Exactly as before, performing the Fourier transform gives
\begin{equation}
H^0_\text{NNN}= \frac{1}{2} \sum \limits_{k} h^{\eta=0}_\text{NNN}(k) \left(a^\dagger_{k\up}a^\dagger_{-k\down} - a^\dagger_{k\down}a^\dagger_{-k\up}+b^\dagger_{k\up}b^\dagger_{-k\down} - b^\dagger_{k\down}b^\dagger_{-k\up}\right) +h.c.,
\end{equation}
in the spin-singlet channel and
\begin{subequations}
\begin{align}
H^x_\text{NNN}=& \frac{1}{2} \sum \limits_{k} h^x_\text{NNN}(k) \left(a^\dagger_{k\up} a^\dagger_{-k\up} - a^\dagger_{k\down} a^\dagger_{-k\down}+b^\dagger_{k\up} b^\dagger_{-k\up} - b^\dagger_{k\down} b^\dagger_{-k\down}\right) + h.c., \ \ \ \ \  \\
H^y_\text{NNN}=& \frac{i}{2} \sum \limits_{k} h^y_\text{NNN}(k) \left(a^\dagger_{k\up} a^\dagger_{-k\up} + a^\dagger_{k\down} a^\dagger_{-k\down}+b^\dagger_{k\up} b^\dagger_{-k\up} + b^\dagger_{k\down} b^\dagger_{-k\down}\right) + h.c., \ \ \ \ \ \\
H^z_\text{NNN}=& \frac{1}{2} \sum \limits_{k} h^z_\text{NNN}(k)\left (a^\dagger_{k\up}a^\dagger_{-k\down} + a^\dagger_{k\down}a^\dagger_{-k\up}+b^\dagger_{k\up}b^\dagger_{-k\down} + b^\dagger_{k\down}b^\dagger_{-k\up}\right) + h.c., \ \ \ \ \ 
\end{align}
\end{subequations}
in the spin-triplet one. Note that we have a $\frac{1}{2}$ factor appearing since one takes twice into account the links between the next-nearest neighbors when summing over all the atoms. Following the NNN conventions of Fig.~\ref{figure2a} we write
\begin{align}
h_\text{NNN}^{\eta}=&  \Delta_\text{NNN}^{\eta,d=1}e^{-\frac{\sqrt{3}i}{2}k_x + \frac{3i}{2}k_y} + \Delta_\text{NNN}^{\eta,d=2}e^{-i\sqrt{3}k_x} + \Delta_\text{NNN}^{\eta,d=3}e^{-\frac{\sqrt{3}i}{2}k_x - \frac{3i}{2}k_y}\nonumber \\ 
&+ \Delta_\text{NNN}^{\eta,d=4}e^{\frac{\sqrt{3}i}{2}k_x - \frac{3i}{2}k_y}+ \Delta_\text{NNN}^{\eta,d=5}e^{i \sqrt{3} k_x}+\Delta_\text{NNN}^{\eta,d=6}e^{\frac{\sqrt{3}i}{2}k_x + \frac{3i}{2}k_y}.
\end{align}
The explicit expressions of the resulting NNN form factors ($h_\text{NNN}^{0, s_\text{ext}}$, $h_\text{NNN}^{0, d_{xy}}$, $h_\text{NNN}^{0, d_{x^2-y^2}}$, $h_\text{NNN}^{\eta, p_x}$, $h_\text{NNN}^{\eta,p_y}$, $h_\text{NNN}^{\eta,f_x}$)  are given in Tables \ref{Table3} and  \ref{Table4}.

\section{Spin-triplet pairing Hamiltonian}
\label{Appendix2}
In this Appendix we derive the Hamltonian for spin-triplet pairing with $\eta = y,z$. The result for $\eta=x$ is given in the main text.
For the $\eta=y$ spin-triplet channel, the Hamiltonian reads
\begin{equation}
\begin{pmatrix}
\mu & h_0(k) & 0 & 0 & -i\tilde{h}_\text{NNN}^{\eta}(-k) & i  \tilde{h}_\text{NN}^{\eta}(k) & 0 & 0 \\
h_0^*(k) & \mu & 0 & 0 &-i\tilde{h}_\text{NN}^{\eta}(-k) & -i\tilde{h}_\text{NNN}^{\eta}(-k)& 0 & 0 \\
0 & 0 & \mu & h_0(k) & 0 & 0 & -i\tilde{h}_\text{NNN}^{\eta}(-k) & i\tilde{h}_\text{NN}^{\eta}(k) \\
0 & 0 & h_0^*(k) & \mu & 0 & 0& -ih_\text{NN}^{\eta}(-k) &-i\tilde{h}_\text{NNN}^{\eta}(-k)  \\
i\tilde{h}_\text{NNN}^{\eta*}(-k) & i\tilde{h}_\text{NN}^{\eta*}(-k) & 0 & 0 & -\mu & -h_0(k) & 0 & 0 \\
-i\tilde{h}_\text{NN}^{\eta*}(k) & i\tilde{h}_\text{NNN}^{\eta*}(-k) & 0 & 0&- h_0^*(k) & -\mu & 0 & 0 \\
0 & 0 & i\tilde{h}_\text{NNN}^{\eta*}(-k) & i\tilde{h}_\text{NN}^{\eta*}(-k) & 0 & 0 & -\mu &-h_0(k) \\
0 & 0& -i\tilde{h}_\text{NN}^{\eta*}(k) & i\tilde{h}_\text{NNN}^{\eta*}(-k) & 0 & 0 &- h_0^*(k) & -\mu
\end{pmatrix},
\end{equation}
while for the $\eta=z$ spin-triplet channel we find
\begin{equation}
\begin{pmatrix}
\mu & h_0(k) & 0 &0 & 0 & 0 &0 + \tilde{h}_\text{NNN}^{\eta}(-k) & -\tilde{h}_\text{NN}^{\eta}(k) \\
h_0^*(k) & \mu & 0 & 0 & 0 & 0 & \tilde{h}_\text{NN}^{\eta}(-k) &  \tilde{h}_\text{NNN}^{\eta}(-k) \\
0 & 0 & \mu & h_0(k) & \tilde{h}^{\eta}_\text{NNN}(-k) & -\tilde{h}_\text{NN}^{\eta}(k) & 0 & 0  \\
0 & 0 & h_0^*(k) & \mu &\tilde{h}_\text{NN}^{\eta}(-k) & \tilde{h}_\text{NNN}^{\eta}(-k) &0 &0  \\
0 & 0 &\tilde{h}^{\eta*}_\text{NNN}(-k) & \tilde{h}_\text{NN}^{\eta*}(-k) & -\mu & -h_0(k) & 0 & 0 \\
0 & 0 & -\tilde{h}_\text{NN}^{\eta*}(k) & \tilde{h}_\text{NNN}^{\eta*}(-k) & -h_0^*(k) & -\mu & 0 & 0 \\
\tilde{h}_\text{NNN}^{\eta*}(-k) & \tilde{h}_\text{NN}^{\eta*}(-k) & 0 & 0 & 0 & 0 & -\mu & -h_0(k) \\
-\tilde{h}_\text{NN}^{\eta*}(k) & \tilde{h}_\text{NNN}^{\eta*}(-k) & 0 & 0 & 0 & 0 & -h_0^*(k) & -\mu
\end{pmatrix}.
\end{equation}

\clearpage

\section{Lowest energy band for $\mu=1$}
\label{appendix3}
\begin{figure}[tbh]

\includegraphics[width=5cm]{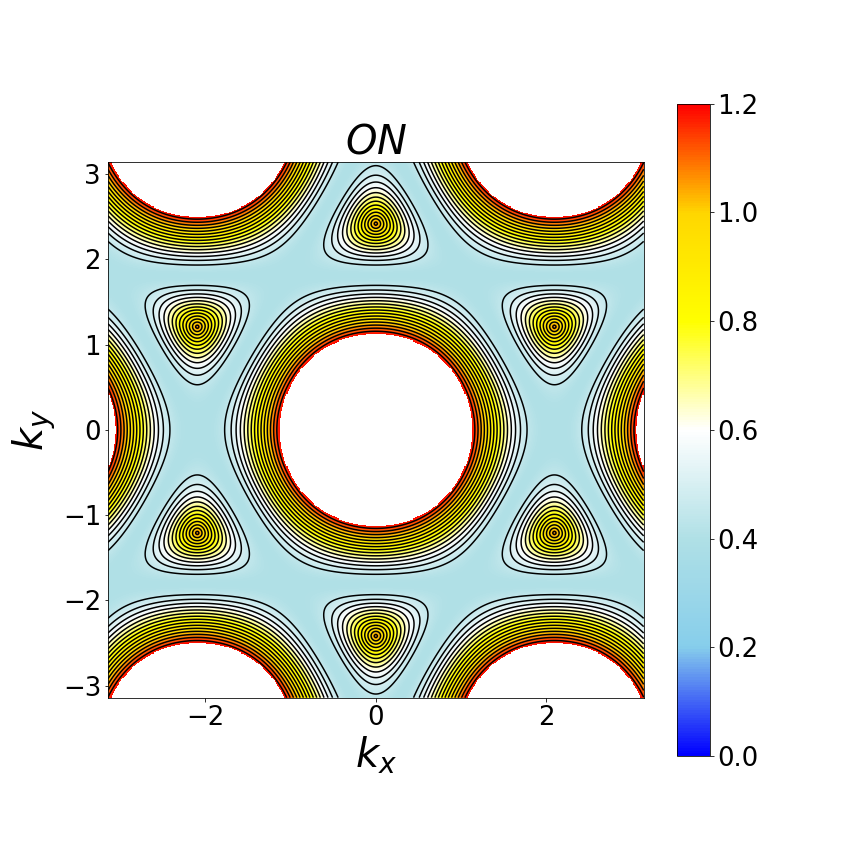}
\includegraphics[width=5cm]{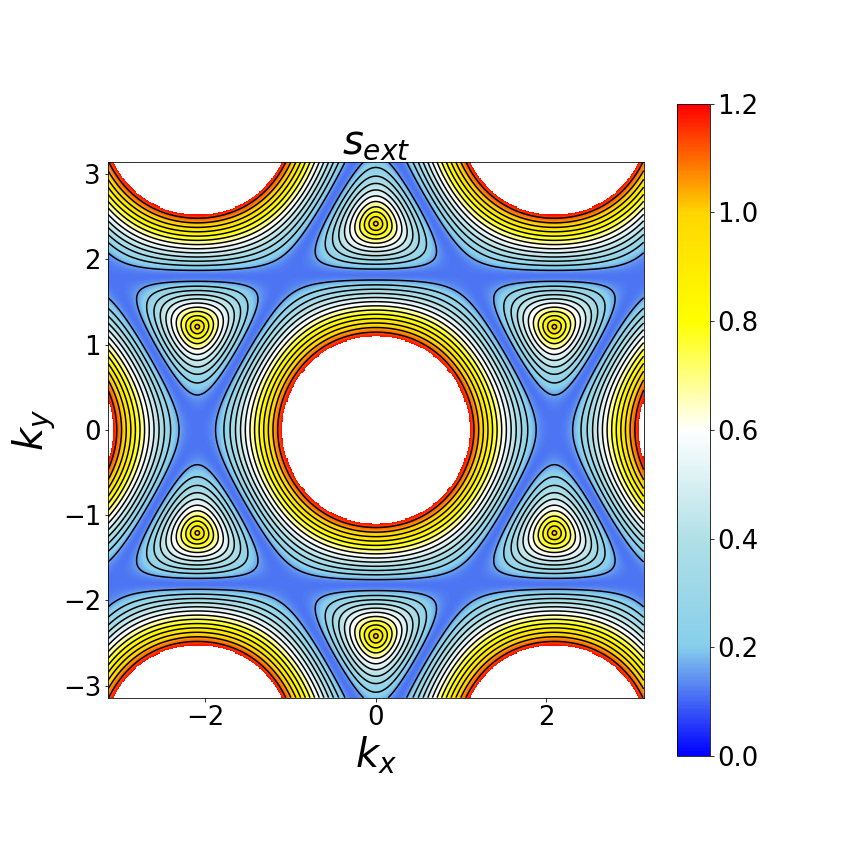}
\includegraphics[width=5cm]{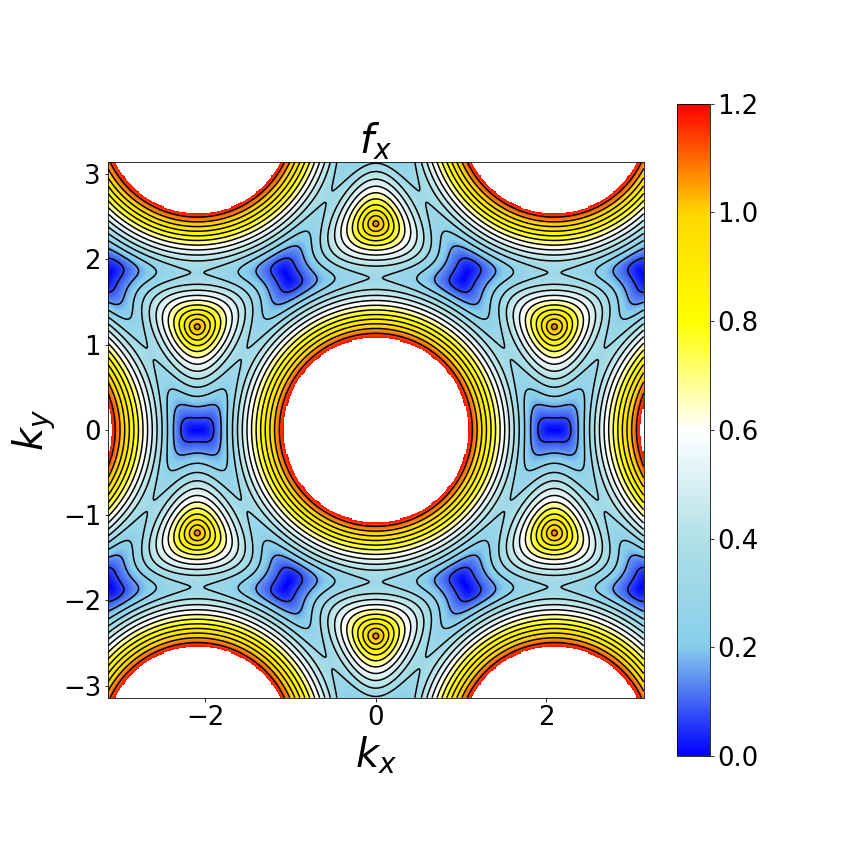}

\vspace{-0.5cm}

\includegraphics[width=5cm]{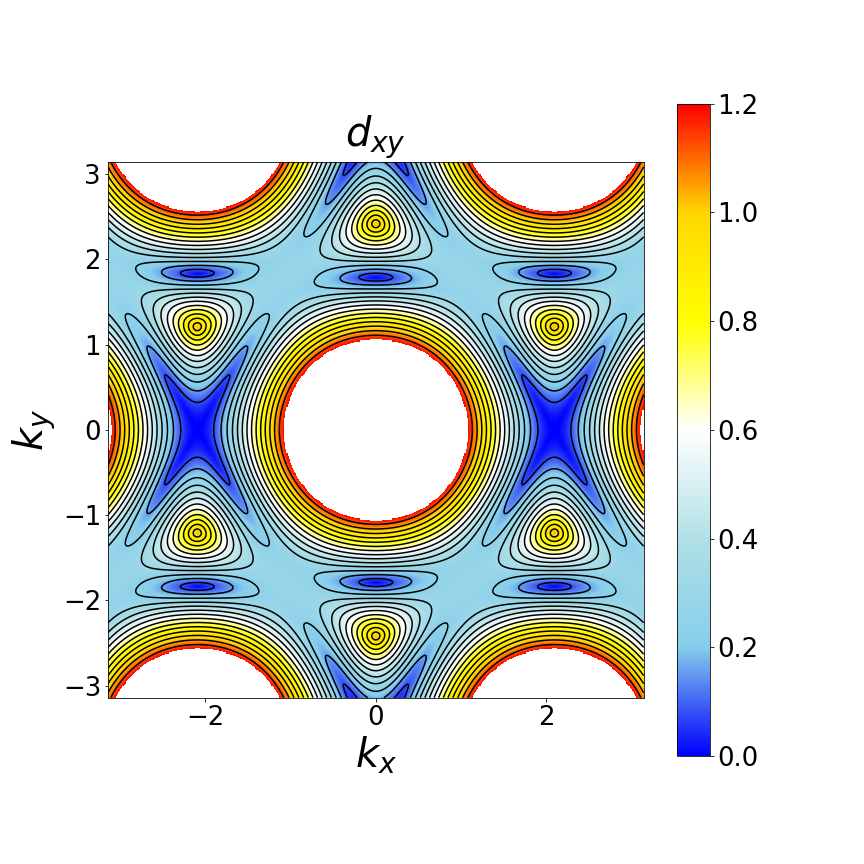}
\includegraphics[width=5cm]{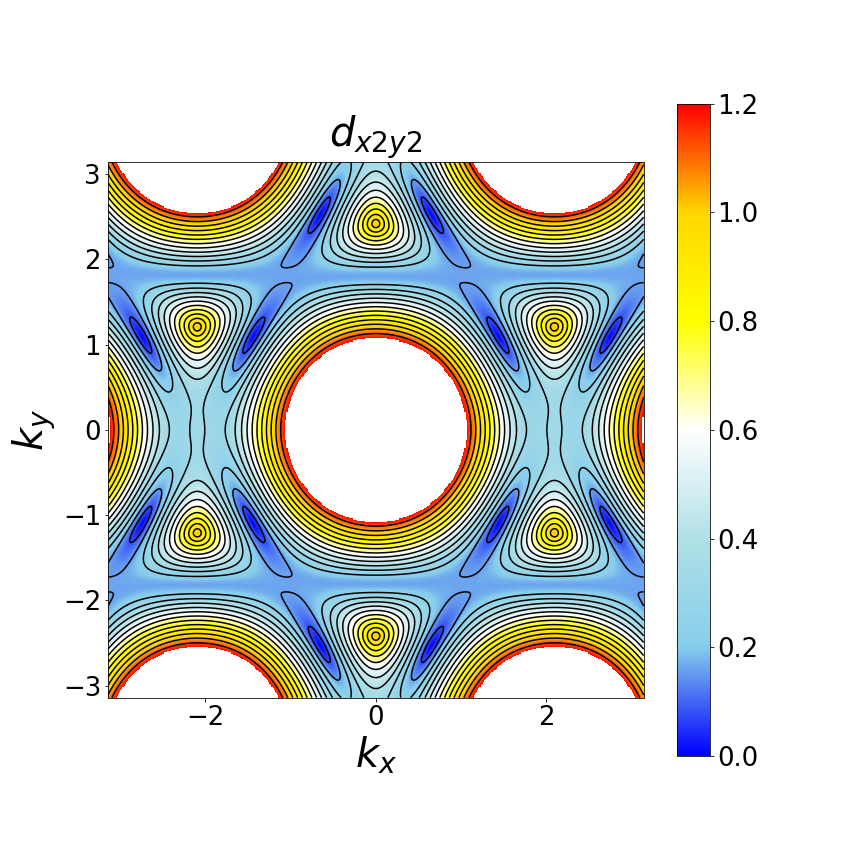}
\includegraphics[width=5cm]{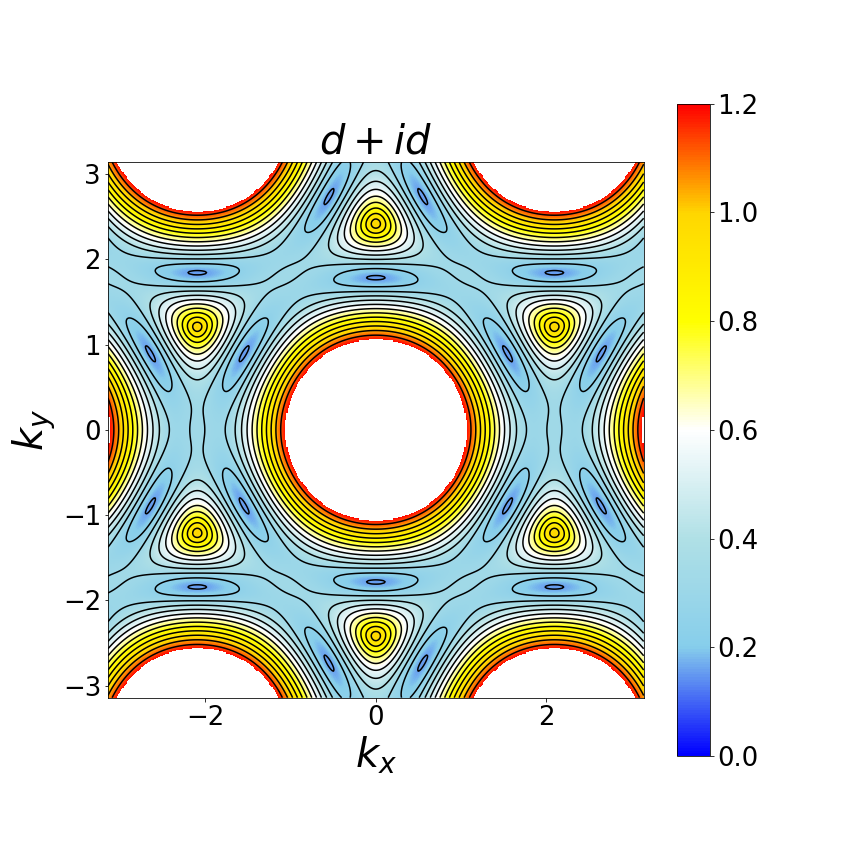}

\vspace{-0.5 cm}

\includegraphics[width=5cm]{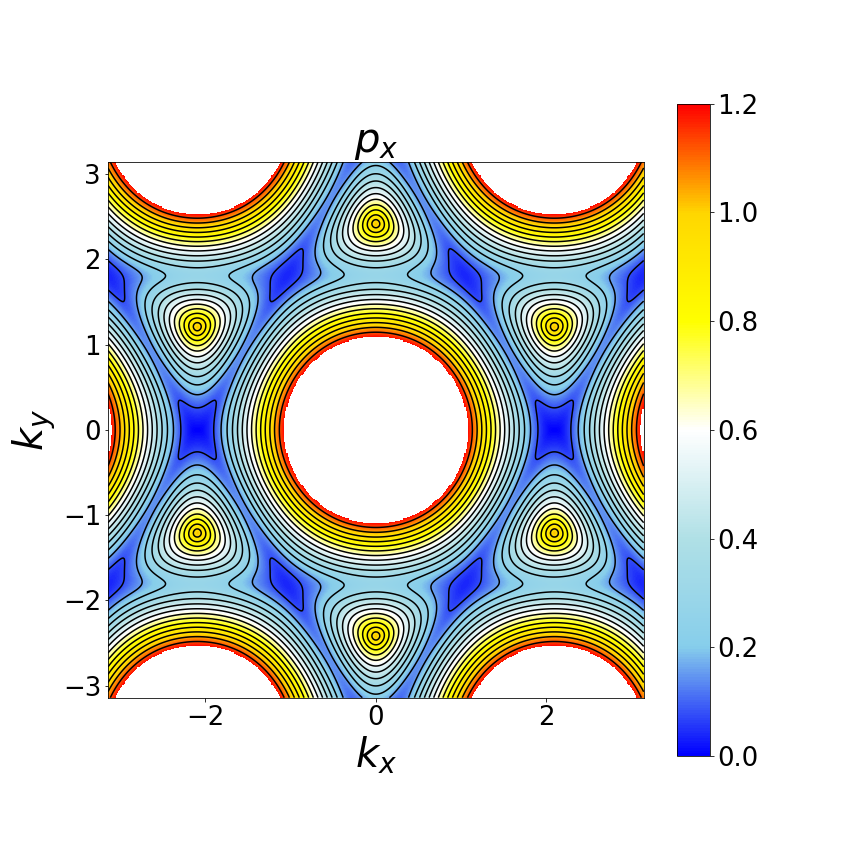}
\includegraphics[width=5cm]{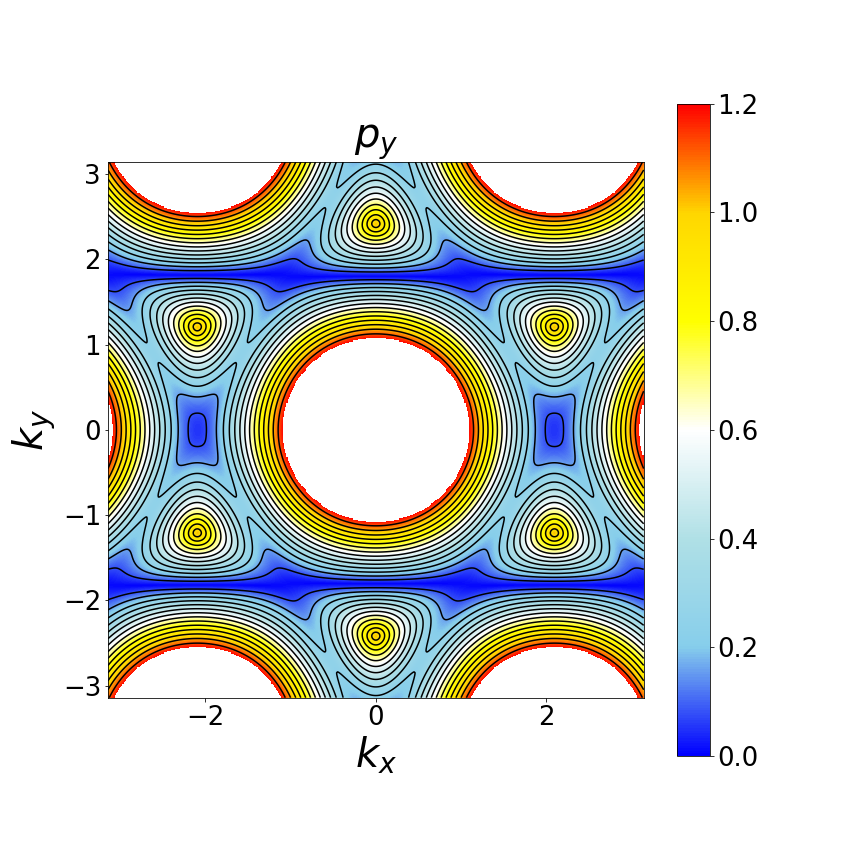}
\includegraphics[width=5cm]{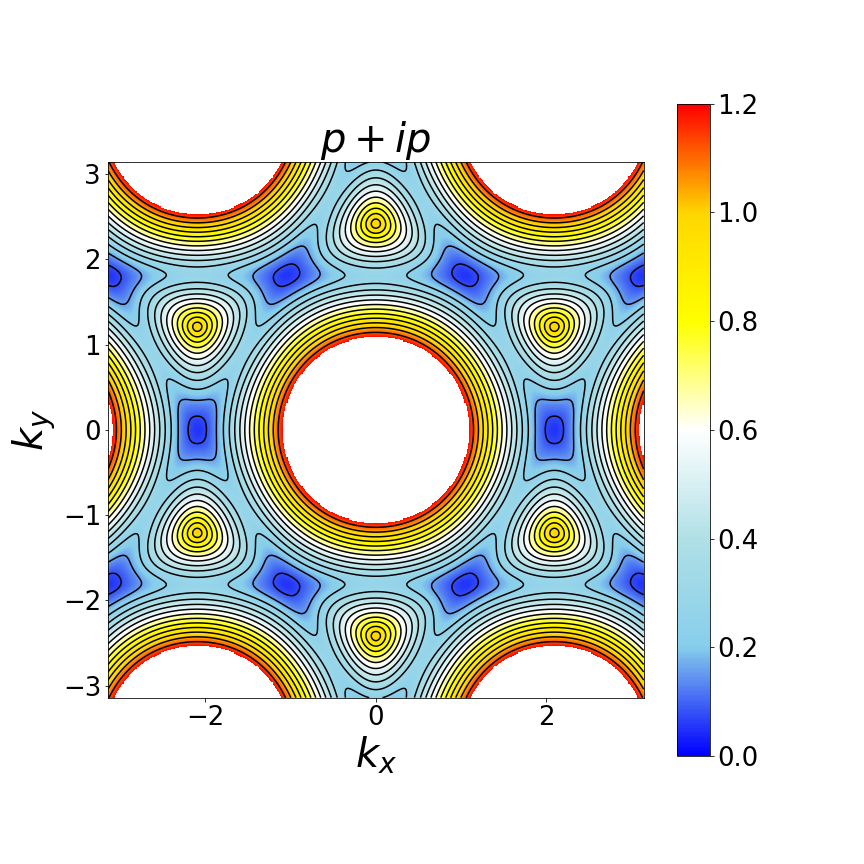}

\caption{Lowest energy band structure for $\mu=1$, $\Delta_0=0.4$, and for SC states with ON $s$-wave, NN $s_\text{ext}$-, $d_{xy}$-, $d_{x^2-y^2}$-, $d+id\,'$-, $p_x$-, $p_y$-, $p+ip\,'$-wave, as well as  NNN $f_x$ symmetries. Dark blue corresponds to zero energy.}
\label{figurea17}
\end{figure}

In this Appendix we provide plots of the lowest energy bands at $\mu=1$ in Fig.~\ref{figurea17}. This is to be compared to the results for $\mu=0.4$ given in Fig.~\ref{figurebands} in the main text.
For a doping close to the van Hove singularity, $\mu=1$, we thus plot the lowest energy band as a function of $k_x$ and $k_y$. We note that in this case the bands for some of the SC symmetries reach zero energy in the vicinity of the $M$ points (corresponding to the dark blue regions). This gives rise to a gap closing point in the DOS around the value of  $\mu=1$.

\clearpage

\section{Band structure for NNN pairing}
\label{appendix4}
\begin{figure}[tbh]
\centering
\includegraphics[width=4.8cm]{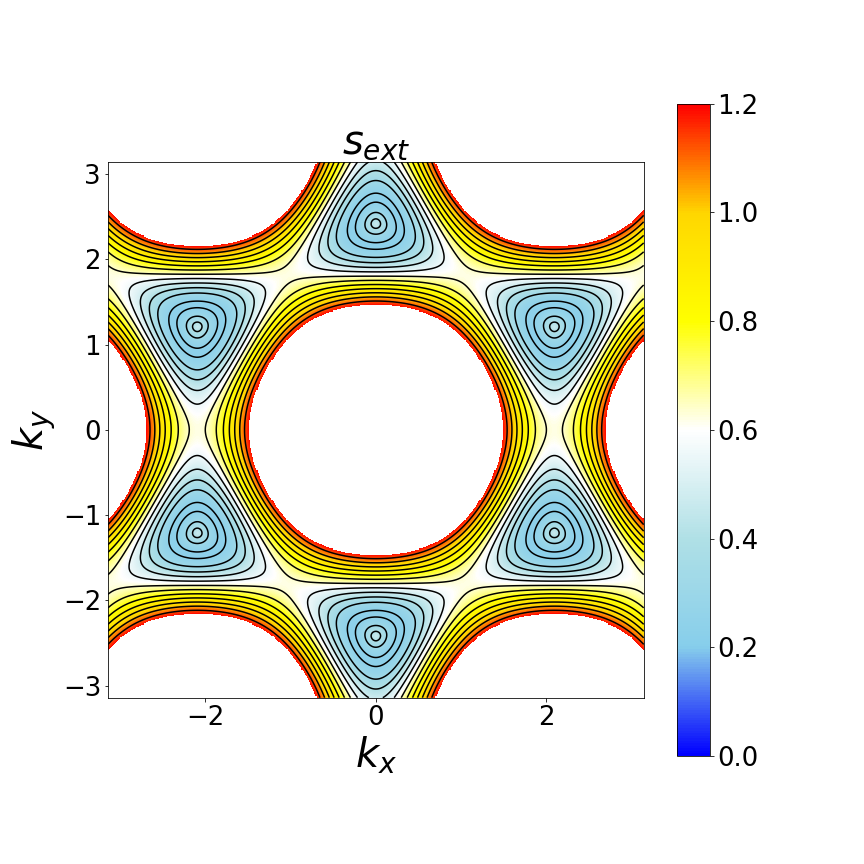}
\includegraphics[width=4.8cm]{bands_fx_mu_04.png}

\vspace{-0.5cm}

\includegraphics[width=4.8cm]{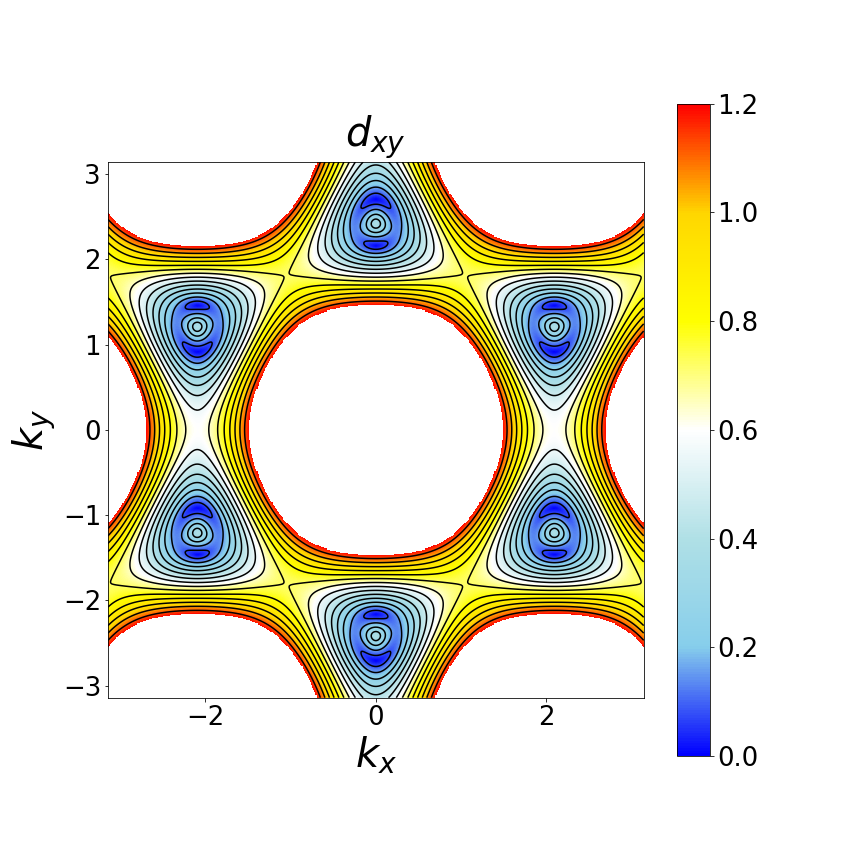}
\includegraphics[width=4.8cm]{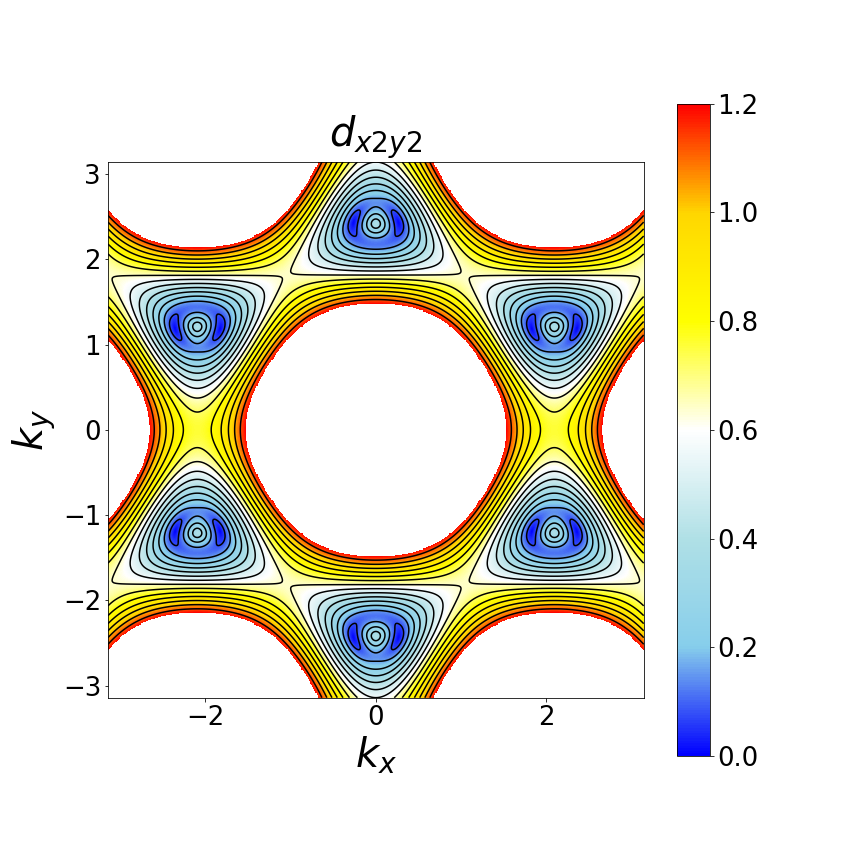}
\includegraphics[width=4.8cm]{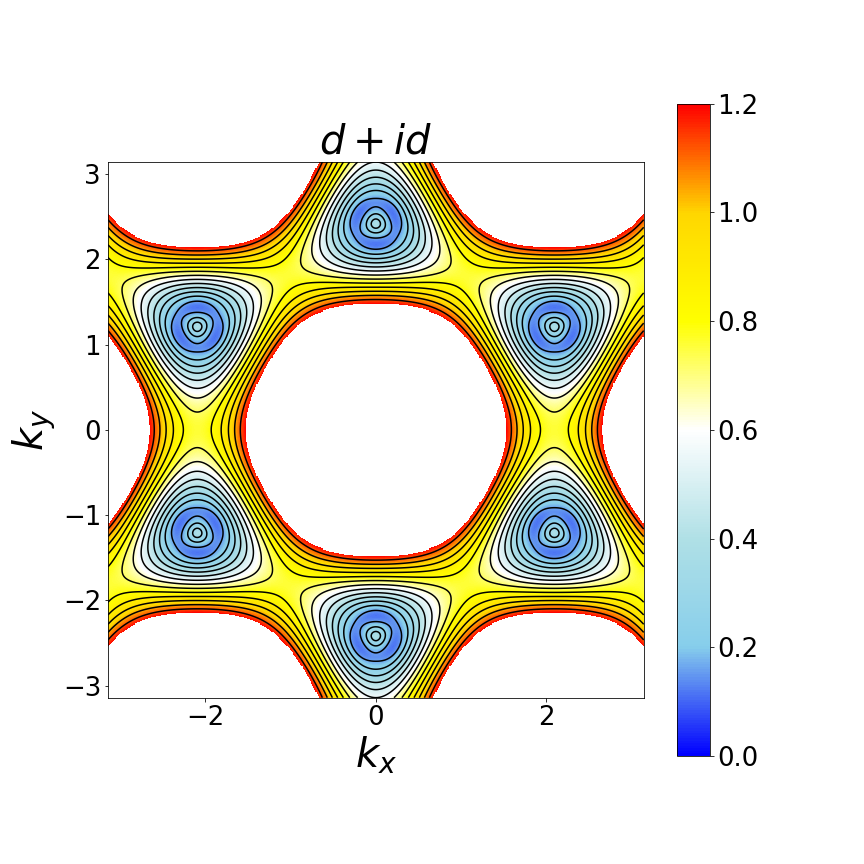}

\vspace{-0.5 cm}

\includegraphics[width=4.8cm]{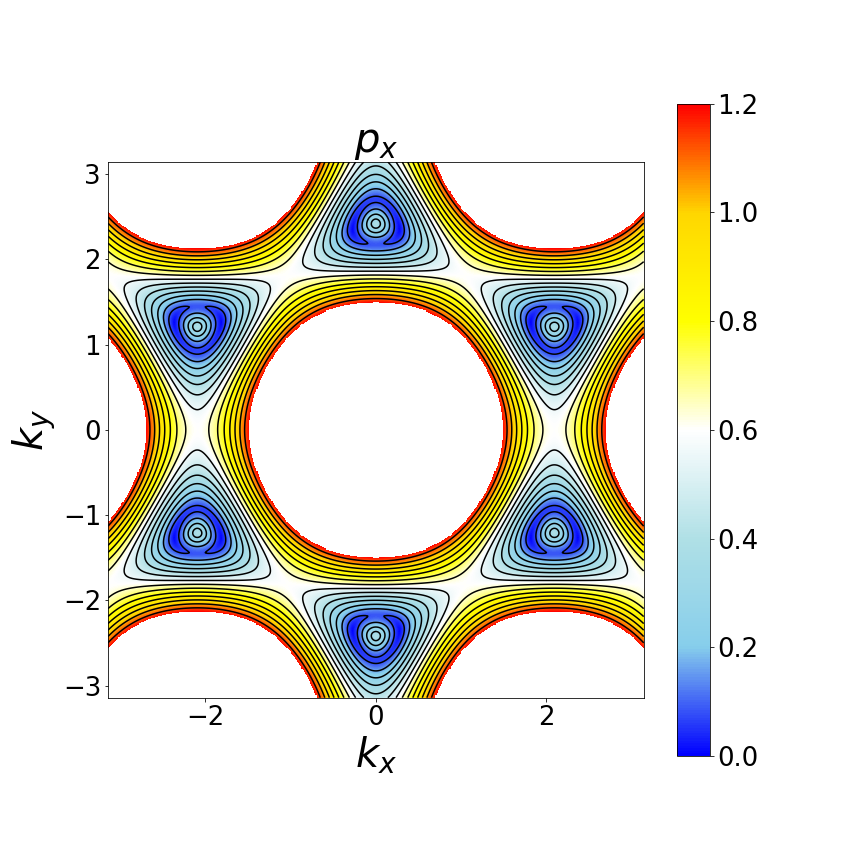}
\includegraphics[width=4.8cm]{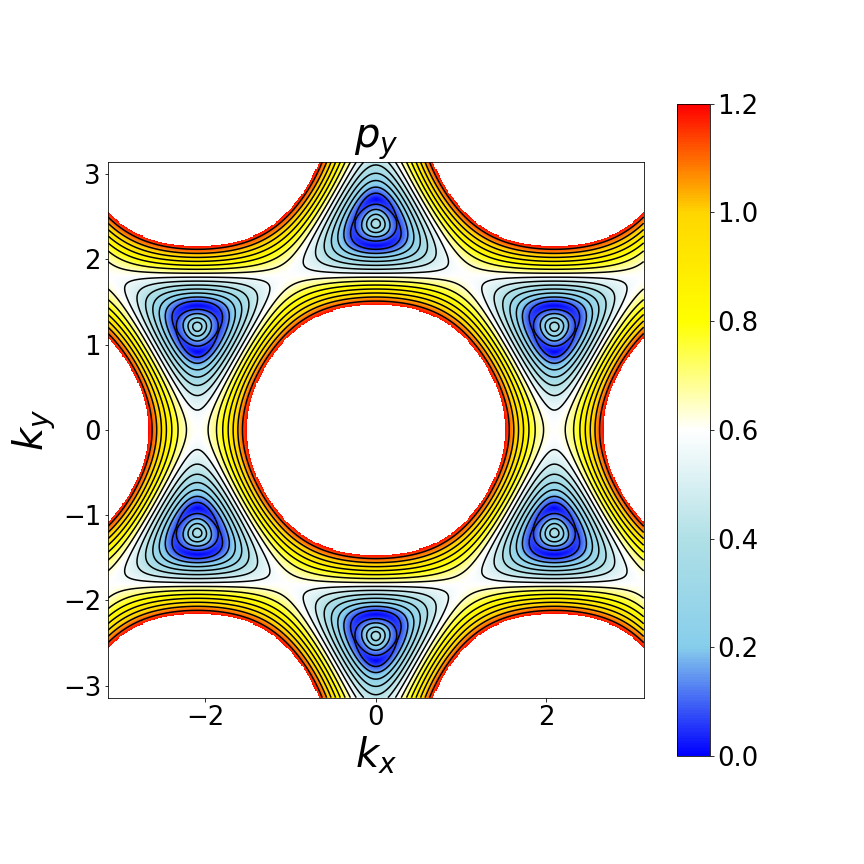}
\includegraphics[width=4.8cm]{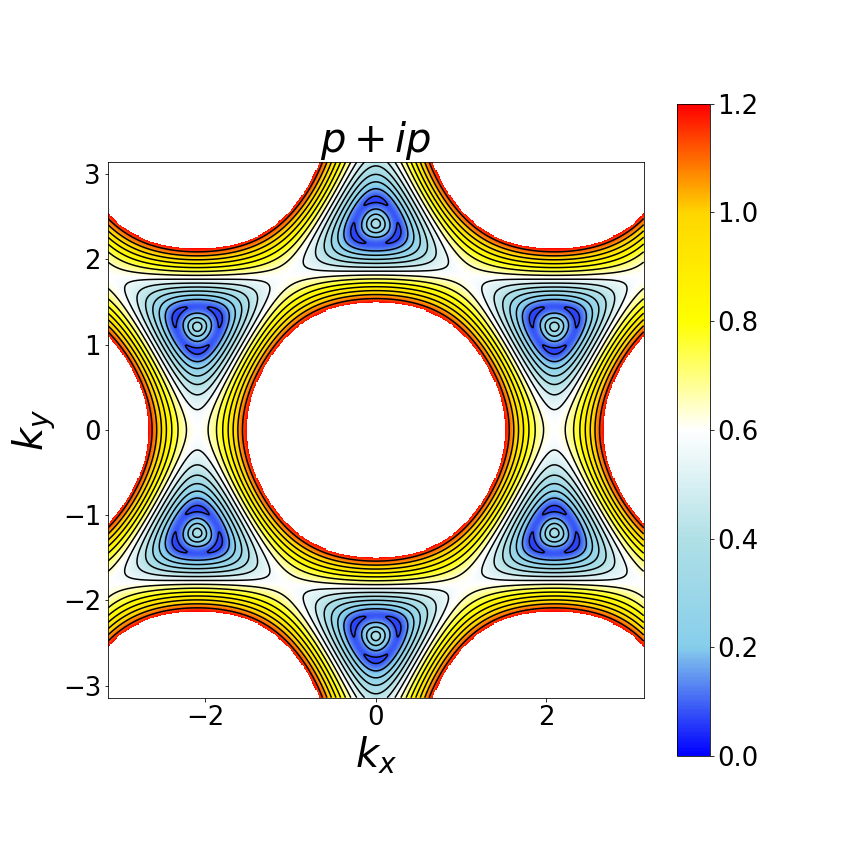}
\caption{Lowest energy band structure for $\mu=0.4$, $\Delta_0=0.4$, and for SC states with NNN pairing with   $s_\text{ext}$-, $d_{xy}$-, $d_{x^2-y^2}$-, $d+id\,'$-, $p_x$-, $p_y$-, $p+ip\,'$-, and $f_x$-wave symmetries. Dark blue corresponds to zero energy.}
\label{figure18}
\end{figure}

In this Appendix we provide plots of the lowest energy bands for NNN pairing in Fig.~\ref{figure18}. This is to be compared to results for NN pairing given in Fig.~\ref{figurebands} in the main text. We thus consider both spin-singlet and -triplet NNN pairing with various symmetries. We note a similarities with the bands resulting from NN pairing Fig.~\ref{figurebands}, such as the nodal points in the structure for $d_{xy}$-, $d_{x^2y^2}$-, $p_x$-, and $p_y$-wave (depicted in dark blue) and a fully gapped structure for the rest.

\clearpage

\section{DOS for NNN pairing}
\label{appendix5}
\begin{figure}[tbh]
\includegraphics[width=5.5cm]{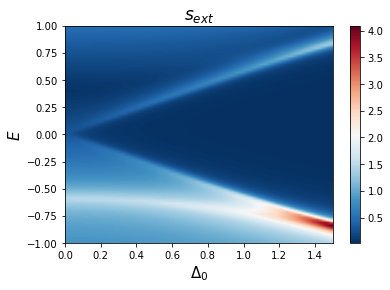}
\includegraphics[width=5.5cm]{graphene_pho_delta_fx.png}
\hspace*{5.5cm} \\
\includegraphics[width=5.5cm]{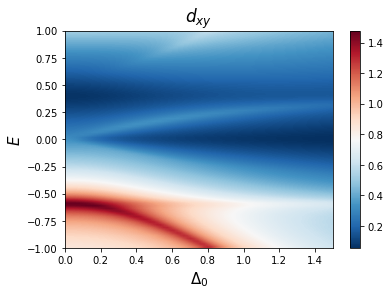} 
\includegraphics[width=5.5cm]{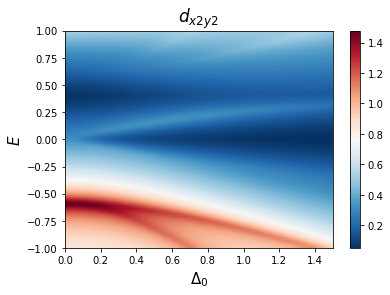}
\includegraphics[width=5.5cm]{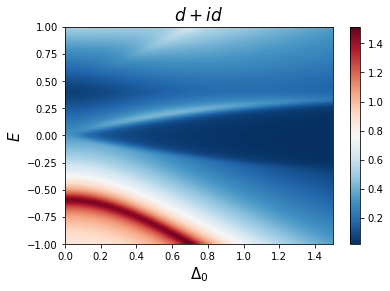} \\
\includegraphics[width=5.5cm]{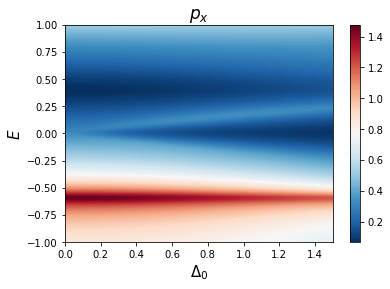}
\includegraphics[width=5.5cm]{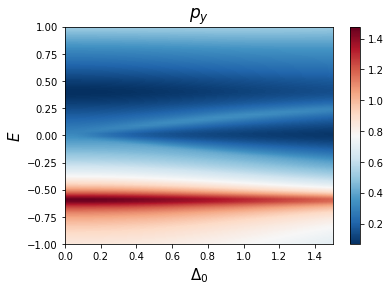}
\includegraphics[width=5.5cm]{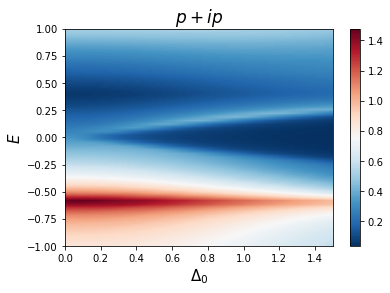}
\caption{DOS as a function of SC amplitude $\Delta_0$ for $\mu=0.4$ for SC states with NNN pairing with $s_\text{ext}$-, $d_{xy}$-, $d_{x^2-y^2}$-,  $d+id\,'$-, $p_x$-, $p_y$-, $f_x$- and $p+ip\,'$-wave symmetries. }
\label{figure19}
\end{figure}

\begin{figure}[tbh]
\includegraphics[width=5.5cm]{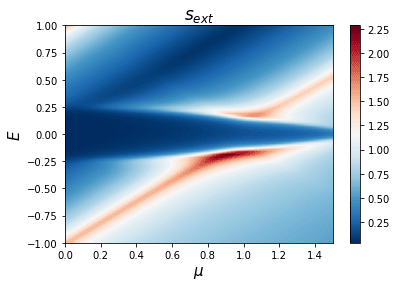}
\includegraphics[width=5.5cm]{graphene_pho_mu_fx.png}
\hspace*{5.5cm} \\
\includegraphics[width=5.5cm]{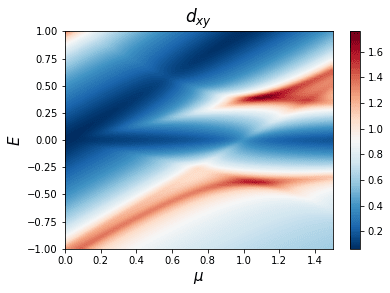} 
\includegraphics[width=5.5cm]{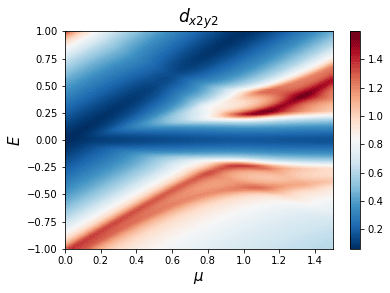}
\includegraphics[width=5.5cm]{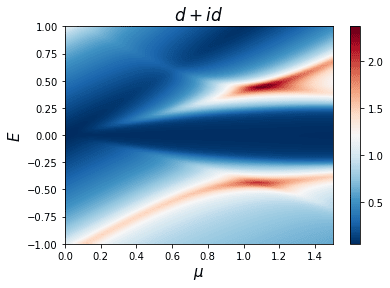} \\
\includegraphics[width=5.5cm]{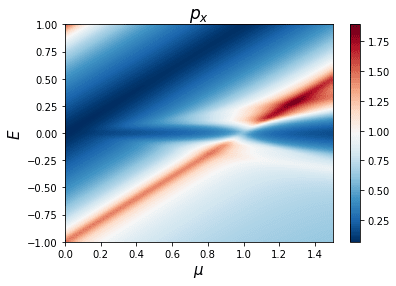}
\includegraphics[width=5.5cm]{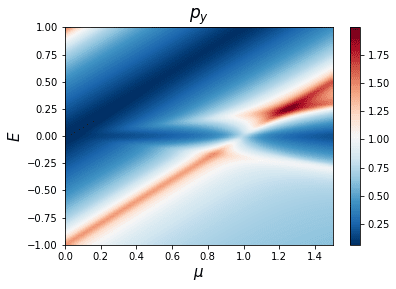}
\includegraphics[width=5.5cm]{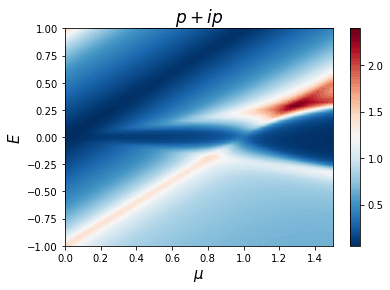}
\caption{DOS as a function of $\mu$ for $\Delta_0=0.4$ for SC states with NNN pairing with $s_\text{ext}$-, $d_{xy}$-, $d_{x^2-y^2}$-,  $d+id\,'$-, $p_x$-, $p_y$-, $f_x$- and $p+ip\,'$-wave symmetries.}
\label{figure20}
\end{figure}

\begin{figure}[tbh]
\begin{center}
\includegraphics[width=6cm]{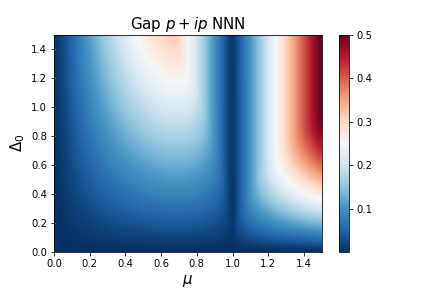}
\caption{Gap value as a function of the chemical potential $\mu$ and the SC amplitude $\Delta_0$ for the NNN $p+ip\,'$-wave state.}
\label{figure26}
\end{center}
\end{figure}

In this Appendix we provide information on the DOS for NNN pairing. First we plot the DOS for NNN pairing as a function of energy and SC pairing strength in Fig.~\ref{figure19} for the same parameter values as those considered for the NN pairing described in the main text and in Fig.~\ref{figure3}. We note that the main difference from the NN coupling case is that the dependence with $\Delta_0$ is roughly linear, and in contrast to the NN case, there is no critical value of $\Delta_0$ for which we observe additional gap closings, for any of the symmetries.
The dependence of the DOS with energy and the chemical potential is very similar to that for the NN couplings as seen when comparing Fig.~\ref{figure20} with Fig.~\ref{figure5} in the main text.
Furthermore, as mentioned in the main text, the gap closing for the NNN $p+ip\,'$-wave state occurs at $\mu=1$, independent of the value of the SC coupling, as seen in Fig. \ref{figure26}.

\clearpage

\section{DOS for NN $p+ip\,'$-wave pairing in multilayer graphene}
\label{appendix6}
\begin{figure}[tbh]
\centering
\hspace{-1.8cm}
\includegraphics[width=5.2cm]{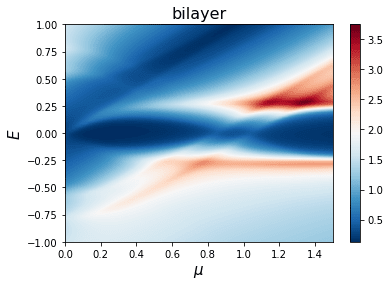}
\hspace{-0.1cm}
\includegraphics[width=5.2cm]{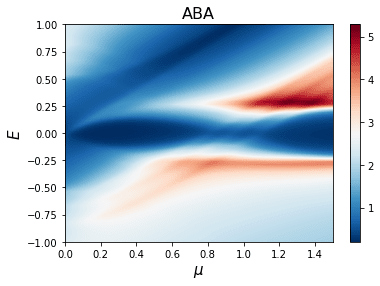}
\hspace{-0cm}
\includegraphics[width=5.2cm]{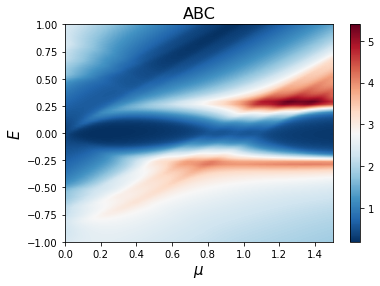}
\caption{DOS as a function of energy and $\mu$ for $\Delta_0=0.4$, $\gamma_1=0.2$, $\gamma_3=0$ in bilayer, trilayer ABA, and trilayer ABC graphene for the NN $p+ip\,'$-wave SC state .}
\label{figure21}
\end{figure}

\begin{figure}[tbh]
\centering
\hspace{-1.8cm}
\includegraphics[width=5.2cm]{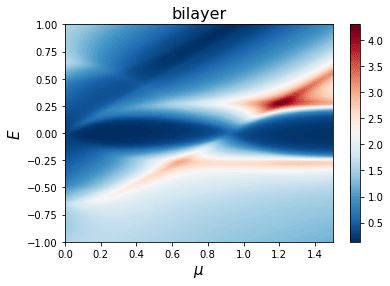}
\hspace{-0.1cm}
\includegraphics[width=5.2cm]{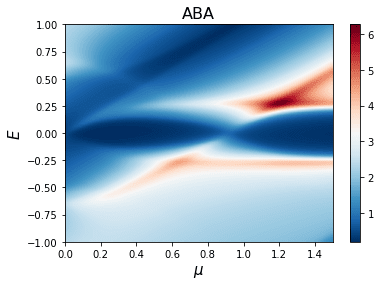}
\hspace{-0cm}
\includegraphics[width=5.2cm]{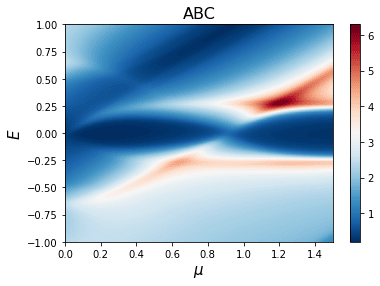}
\caption{DOS as a function of energy and $\mu$ for $\Delta_0=0.4$, $\gamma_1=0.2$, $\gamma_3=0.2$ in bilayer, trilayer ABA,  and trilayer ABC graphene for the NN $p+ip\,'$-wave SC state.}
\label{figure22}
\end{figure}

In this Appendix we provide additional information of the DOS for NN $p+ip\,'$-wave pairing in multilayer graphene.
For a vanishing trigonal warping, $\gamma_3=0$, we obtain the DOS as a function of energy and chemical potential in Fig.~\ref{figure21} for different multilayer configurations.
We note that these results are very similar to monolayer graphene, except for the number of gap closing points. As discussed in the main text, this number is also strongly affected by the trigonal warping, which we illustrate in Fig.~\ref{figure22} where we plot the DOS in the presence of non-zero trigonal warping $\gamma_3=0.2$.

\begin{figure}[tbh]
\centering
\hspace{-1cm}
\includegraphics[width=5.5cm]{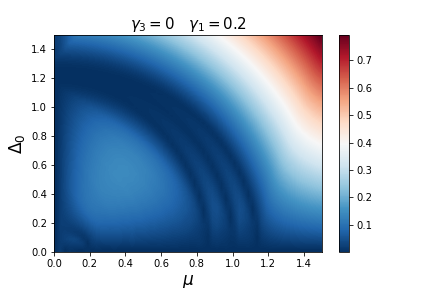}
\hspace{-0.7cm}
\includegraphics[width=5.5cm]{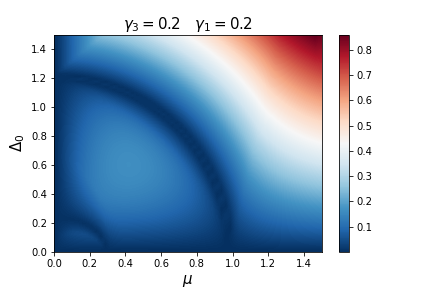}
\hspace{-0.7cm}
\includegraphics[width=5.5cm]{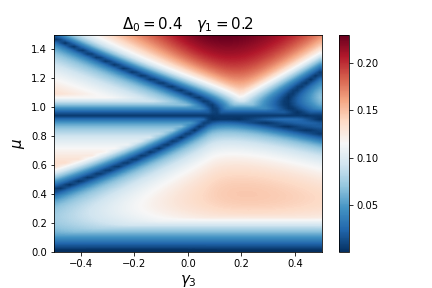}
\caption{DOS as a function of the chemical potential $\mu$ and the SC amplitude $\Delta_0$ for the NN $p+ip\,'$-wave state in ABA trilayer graphene for  $\gamma_3=0$ (left) and $\gamma_3 = 0.2$ (middle) with $\gamma_1 = 0.2$. (Right): DOS as a function of $\gamma_3$ and $\mu$ for $\Delta_0 = 0.4$ and $\gamma_1 = 0.2$. Dark blue corresponds to a vanishing DOS and to the gap closing points.}
\label{gapclosingABA}
\end{figure}

\begin{figure}[tbh]
\centering
\hspace{-1cm}
\includegraphics[width=5.5cm]{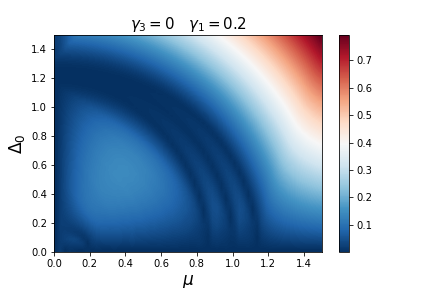}
\hspace{-0.7cm}
\includegraphics[width=5.5cm]{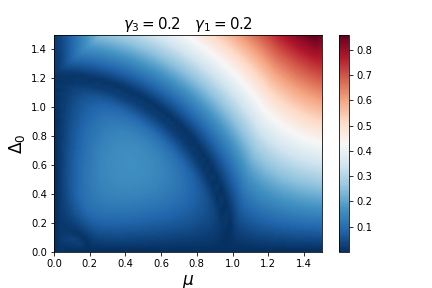}
\hspace{-0.7cm}
\includegraphics[width=5.5cm]{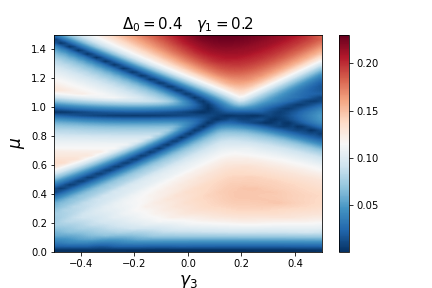}
\caption{DOS as a function of the chemical potential $\mu$ and the SC amplitude $\Delta_0$ for the NN $p+ip\,'$-wave state in ABC trilayer graphene for $\gamma_3=0$ (left) and  $\gamma_3 = 0.2$ (middle) with  $\gamma_1 = 0.2$. (Right): DOS as a function of $\gamma_3$ and $\mu$ for $\Delta_0 = 0.4$ and $\gamma_1 = 0.2$. Dark blue corresponds to a vanishing DOS and to the gap closing points.}
\label{gapclosingABC}
\end{figure}

As opposed to the monolayer case for which we have a single gap closing point as a function of the chemical potential, for trilayer graphene we find that this number oscillates between one and three. This is illustrated for ABA trilayer graphene in Fig.~\ref{gapclosingABA} and for ABC trilayer graphene in Fig.~\ref{gapclosingABC}. 
We find that the effect of trigonal warping is quite pronounced by modifying the gap closing points.

\twocolumngrid

\end{document}